\pdfoutput=1
\documentclass[PAPER, atlasdraft=false, coverpage=false, UKenglish, cernpreprint, texmf, orcidlogo]{atlasdoc}
\usepackage[backend=biber]{atlaspackage}
\usepackage[backref=true]{atlasbiblatex}

\usepackage[jetetmiss=true]{atlasphysics}
\let\oldalign\align
\let\oldendalign\endalign

\renewenvironment{align}{\linenomathNonumbers\oldalign}{\oldendalign\endlinenomath}

\graphicspath{{./logos/},{./figures/}}

\usepackage{ANA-JETM-2024-01-PAPER-defs}
\usepackage{tabularx}
\usepackage{multirow}
\usepackage{stackengine}
\usepackage{multirow}

\AtlasTitle{Precision calibration of calorimeter signals in the ATLAS experiment using an uncertainty-aware neural network}

\AtlasAbstract{%
The ATLAS experiment at the Large Hadron Collider explores the use of modern neural networks for a multi-dimensional calibration of its calorimeter signal defined by clusters of topologically connected cells (topo-clusters).
The Bayesian neural network (BNN) approach not only yields a continuous and smooth calibration function that improves performance relative to the standard calibration but also provides uncertainties on the calibrated energies for each topo-cluster.
The results obtained by using a trained BNN are compared to the standard local hadronic calibration and to a calibration provided by training a deep neural network.
The uncertainties predicted by the BNN are interpreted in the context of a fractional contribution to the systematic uncertainties of the trained calibration.
They are also compared to uncertainty predictions obtained from an alternative estimator employing repulsive ensembles.
}

\AtlasRefCode{JETM-2024-01}

\PreprintIdNumber{CERN-EP-2024-317}

\AtlasJournal{SciPost\ Phys.}
\AtlasJournalRef{SciPost\ Phys. 19, 155 (2025)}
\AtlasDOI{10.21468/SciPostPhys.19.6.155}

\usepackage{authblk}
\author{The ATLAS Collaboration%
\thanks{The full author list can be found at:\newline
\url{https://atlas.web.cern.ch/Atlas/PUBNOTES/ATL-PHYS-PUB-2024-XXX/authorlist.pdf}}
}
\AtlasCoverEgroupAnalysisTeam{atlas-jetm-2024-01-analysis-team@cern.ch}

\hypersetup{pdftitle={ATLAS document},pdfauthor={The ATLAS Collaboration}}

\begin{document}

\maketitle

\tableofcontents

\section{Introduction}
\label{sec:intro}
The principal signals of the non-compensating calorimeters employed by the ATLAS experiment~\cite{PERF-2007-01} at the Large Hadron Collider (LHC) are clusters of topologically connected cell signals (\topos) that reconstruct locally deposited energy into a three-dimensional signal object~\cite{PERF-2014-07}.
The calorimeter signals are calibrated to correctly measure the energy deposited by electromagnetic showers, thus providing no compensation for the fraction of deposited energy not contributing to the signal in the complex development of hadronic showers.
This leads to on average smaller observed signals for hadrons depositing the same energy as electrons, positrons, or photons.

In ATLAS, the standard approach to compensate for the hadronic signal deficiencies at the level of individual \topos is to apply a \emph{local hadronic calibration} as part of a full calibration chain that also includes corrections for other effects.
This chain is implemented as a four-step sequence of (1) a cluster-by-cluster signal classification determining the likelihood of the \topo to be of electromagnetic origin, (2) a hadronic calibration employing a local cell-signal weighting technique giving the chain its name (LCW), (3) an out-of-cluster correction, and (4) a correction for energy losses in inactive material in the proximity of the \topo, as detailed in \RefRef{PERF-2014-07}.
All classification probabilities and calibration and correction scale factors used in this sequence are retrieved from multi-dimensional lookup tables that are calculated from single particle response simulations.
These tables provide averages in bins of observables sensitive to the nature of the energy deposit and to effects on the signal introduced by the detector geometry, and its signal formation and extraction characteristics at the \topo location.
In addition to bin edge effects potentially introducing discontinuities of scale factors, the correlations between the observables are reduced to average relations in this approach.
This is expected to introduce limitations to the precision of the LCW calibration, as the specific relation between the observables used as inputs is lost for any given \topo.

Modern machine-learning (ML) methods~\cite{Plehn:2022ftl} can be used to learn multi-dimensional, continuous calibration functions without step-like transitions.
These functions preserve, and even use, the underlying correlations between observables, thus promising a more accurate \topo calibration at higher precision.
A set of \topo observables commonly referred to as \emph{features} provides the inputs to the corresponding training, and to the inference of the trained model.
This set includes physics-motivated data representations encompassing measures of the \topo signal characteristics, the underlying nature of the energy deposit, and the signal environment at the LHC.

A one-step training procedure omitting the dedicated \topo classification step was designed in a previous approach to determine such a function.
The results obtained by training a deep neural network (DNN) to learn the response of individual \topos from which the signal calibration is derived are presented in \RefRef{ATL-PHYS-PUB-2023-019}.
The performance of the DNN-trained calibration with respect to signal linearity and resolution was found to yield significant improvements when compared with the standard hadronic calibration provided in the LCW sequence.
Other machine-learning-based approaches to calibrate \topos use cells in \topos in deep learning networks employing images, graphs, and point clouds.
Those were successfully trained to calibrate single particles in full Monte Carlo (MC) simulations of the ATLAS detector~\cite{ATL-PHYS-PUB-2020-018,ATL-PHYS-PUB-2022-040}.

In this paper, results from the application of a Bayesian neural network (BNN)~\cite{Mackay:1995qgm,10.5555/525544,bnn_early3,kendall2017uncertainties,Bollweg:2019skg,Kasieczka:2020vlh,Badger:2022hwf}
for the calibration of the \topos obtained with full MC simulations of multijet final states of the proton--proton (\pp)
collisions at the LHC \RunTwo (2015--2018) centre-of-mass energy of \sqrts{13} are
presented.
These simulations encompass physics modelling of the particle-level final state and a detailed detector simulation.
The detector simulation models the signal contributions from \pu introduced by the high inelastic \pp cross-section, in combination with the high proton beam intensities observed in \RunTwo.

The BNN is trained to not only predict the response and thus the calibration function cluster-by-cluster, but to also learn a statistical and a systematic uncertainty component associated with this prediction.
The naming of these uncertainty components in the context of this study reflects expectations for their respective behaviour and may differ from typical interpretations in calibrations that are derived using measures of centrality in samples of signal objects, rather than directly determining them for a single object (\topo) alone.
Also not considered here are possible other (systematic) uncertainty contributions not directly associated with the application of the learned calibration model, including those arising from residual discrepancies between experimental data and MC simulations.
A example of a complete determination of systematic uncertainties can be found in the jet calibration in ATLAS, with its approach for LHC \RunTwo discussed in detail  in \RefRef{JETM-2018-05}.

The predicted statistical uncertainty component captures the dependence of the prediction on the size of the dataset used for training the BNN, while the systematic uncertainty component provides a measure for the accuracy of the prediction that cannot be improved by a larger training dataset or more training cycles (\emph{epochs}).
Depending on the approach to the full local hadronic calibration of \topos, which, as mentioned before, potentially includes components introduced by the comparison of experimental
data to MC simulations, these uncertainties can provide important contributions to the overall systematic uncertainty associated with the application of the trained BNN model in the course of such calibration.

The individual uncertainty predictions are available for all \topos with a calibration learned by the BNN, including those that are not part of the signal of a calibrated physics object like jets or particles.
They can be employed to improve the input signal quality by \topo selections prior to the reconstruction of particles and jets.
In addition to these selections, uncertainty propagations can be applied in \emph{bottoms-up} calibration schemes used in the context of  the reconstruction of more complex and softer energy flows that are potentially needed for the measurements of the missing transverse momentum, the soft hadronic recoil in vector boson production, and other hadronic event shapes, as well as in jet substructure reconstruction and analysis.

To increase confidence in the uncertainty predictions from the BNN, they are compared to an alternative uncertainty estimate determined with \emph{repulsive ensembles} (RE) \cite{DAngelo2021RepulsiveDE,Rover:2024pvr}.
Observed large predicted uncertainties for certain \topos are interpreted in the context of the signal characteristics associated with complex transition regions in the ATLAS calorimeter system.

The structure of this paper is as follows.  In \SectRef{sec:detector}, the ATLAS detector and the used datasets are presented.
Introduction to the neural-network-based calibration, the features it uses, and the BNN and RE network architecture and configurations are given in \SectRef{sec:datasets}.
In \SectRef{sec:performance}, the performance of the calibration is compared to the LCW and the DNN-derived calibrations.
\SectRef{sec:uncertainties} discusses the validity of the uncertainties predicted by the BNN in a comparison with the corresponding RE predictions and in the context of detector signal features.
The paper concludes with \SectRef{sec:conclusion}.

\section{Experimental setup and simulations}
\label{sec:detector}

\newcommand{\AtlasCoordFootnote}{%
ATLAS uses a right-handed coordinate system with its origin at the nominal interaction point (IP) in the centre of the detector and the \(z\)-axis along the beam pipe.
The \(x\)-axis points from the IP to the centre of the LHC ring, and the \(y\)-axis points upwards.
Polar coordinates \((r,\phi)\) are used in the transverse plane, \(\phi\) being the azimuthal angle around the \(z\)-axis.
The pseudorapidity is defined in terms of the polar angle \(\theta\) as \(\eta = -\ln \tan(\theta/2)\) and is equal to the rapidity $y = \frac{1}{2} \ln \left( \frac{E + p_z}{E - p_z} \right)$ in the relativistic limit.
The pseudorapidity \etadet in the detector frame of reference follows the definition of $\eta$ with reference to the nominal collision vertex at $(x,y,z) = (0,0,0)$.
Angular distance is measured in units of \(\Delta R \equiv \sqrt{(\Delta y)^{2} + (\Delta\phi)^{2}}\).}

The results presented here are solely obtained with MC simulations that model
the setup of the ATLAS experiment and the signal formation
and extraction for the various sub-detectors in great detail.
The emphasis is on the central region of the calorimeter system.
It is the most relevant detector component for the studies presented in this paper, as the \topos used to learn the response and validate the obtained calibration are collected from calorimeter jets in this region.
The only other detector systems contributing here are the inner tracking detector for the reconstruction of the collision vertices and the dedicated forward detectors comprising the luminosity measurement system~\cite{DAPR-2021-01} used for the reconstruction of the number of \pu collisions.

\subsection{The ATLAS detector}
\label{sec:detdat:detector}

The ATLAS detector at the LHC covers nearly the entire solid angle around the \pp collision vertex.\footnote{\AtlasCoordFootnote}
It consists of an inner tracking detector system (ID) surrounded by a thin superconducting solenoid, electromagnetic and hadronic calorimeters, and a muon spectrometer (MS) incorporating three large superconducting air-core toroidal magnets.

The calorimeter system provides near full absorption coverage in the pseudorapidity range $|\etadet|<4.9$.
Electromagnetic calorimetry is provided up to $|\etadet|<3.2$ by highly
granular barrel (EMB) and \EndCap (EMEC) lead/liquid-argon (\LAr)
calorimeters.
An additional thin \LAr \PreSampler covers $|\etadet|<1.8$ to correct for energy losses in material in front of the calorimeters.

Hadron calorimetry up to $|\etadet|<1.7$ is provided by the steel/scintillator-tile (\Tile) calorimeter with three barrel structures (one barrel Tile and two extended barrel Tile calorimeters).
The two hadronic \EndCap (HEC) copper/\LAr calorimeters provide coverage for $1.5<|\etadet|<3.2$.
Scintillating counters are installed between the barrel and the extended Tile calorimeters to measure energy losses in this transition region within $0.8<|\etadet|<1.6$ (Tile gap scintillators).

The two forward calorimeters (FCal) cover $3.1<|\etadet|<4.9$.
Each consists of a copper/\LAr module followed by two tungsten/\LAr modules for a total of three longitudinal segments, a setup optimised for electromagnetic and hadronic energy measurements.

The ATLAS calorimeters feature highly granular readout, with a \PreSampler followed by six longitudinal readout segments in the barrel region $|\etadet| < 1.5$, a \PreSampler up to $|\etadet| = 1.8$ and six to seven longitudinal segments in the \EndCap region, and three longitudinal segments in the forward region.
The calorimeter has a depth of at least $\UNIT{10}{\lambda}$ everywhere, with $\lambda$ being the hadronic absorption length.
Overall, it has nearly $200\,000$ independent readout cells, with the highest lateral granularities found in the EMB and the EMEC\@.
The software suite described in \RefRef{ATL-SOFT-PUB-2021-001} is used for the reconstruction and analysis of the MC simulations.
It is in large parts identical to the one used for experimental data.

\subsection{Monte Carlo simulations}
\label{sec:detdat:jets}

The fully simulated jet samples were produced according to the description given in \RefRef{ATL-PHYS-PUB-2019-017}.
The hard-scatter \pp interactions were generated at $\sqrt{s} = \UNIT{13}{\TeV}$ using \PYTHIA[8.230]~\cite{Sjostrand:2014zea} with the A14~\cite{ATL-PHYS-PUB-2014-021} set of tuned parton shower and underlying event parameters and the \NNPDF[2.3lo]~\cite{Ball:2012cx} set of parton distribution functions (PDF).

To simulate the detector signals, the stable particles\footnote{Here particles are considered stable if their proper lifetime $\tau$ is $\tau > \UNIT{10}{\ps}$.} from the generated final state were tracked through the \ATLAS detector using the \GEANT \cite{Agostinelli:2002hh} toolkit.
The energy deposited in the calorimeters was collected in the detector simulation and is available for each calorimeter cell and in regions of inactive material.
The signal formation in the various detector systems is modelled following the respective signal extractions and digitisations in the experiment~\cite{SOFT-2010-01}.
The effect of multiple \pp interactions in the same (in-time \pu) and neighbouring bunch crossings (out-of-time \pu) was modelled by overlaying the simulated signals of the hard-scattering event with signals from inelastic \emph{minimum bias} \pp collisions generated with \PYTHIA[8.186]~\cite{Sjostrand:2007gs} using the \NNPDF[2.3lo] set of PDFs and the A3 set of tuned parameters~\cite{ATL-PHYS-PUB-2016-017}.
The number of overlaid \pu interactions is sampled from the distribution measured in the data during LHC \RunTwo.

\subsection{Calorimeter signal formation and features}
\label{sec:detdat:clusters}

The formation of \topos principally collects calorimeter cells starting from seed cells with highly significant signals and growing by following cell signal significance patterns within and across calorimeter modules in three dimensions.
The initial clustering is followed by an algorithm splitting large \topos between local signal maxima, again in three dimensions.
The full procedure and the reconstructed observables associated with the final cluster are detailed in \RefRef{PERF-2014-07}.
This reference also introduces the LCW calibration that uses signal characteristics, shapes and locations associated with the \topos.

\subsubsection{\Topo kinematics}
\label{sec:detdat:clusters:kine}

The reconstructed \topos have kinematics that are represented at two scales, the basic electromagnetic (\escale) scale and the calibrated \lscale scale.
At both scales, \topos are considered to be massless \emph{pseudo-particles}.
The four-momentum representation \fmomclusem of a \topo at the \escale scale is given by
\begin{align}
\label{eq:cluskine}
\fmomclusem = \eclusem\cdot\left(1,\,\cos\aclusem/\cosh\yclusem,\,\sin\aclusem/\cosh\yclusem,\,\tanh\yclusem\right)\;,
\end{align}
where \eclusem, \yclusem and \aclusem are the cluster energy, rapidity and azimuth reconstructed at the \escale scale, respectively.

The ML calibrations derived in this study are specific for each \topo and focus on the energy deposit it represents at its location in the calorimeter.
There is no attempt to include predictions for the correction of energy losses outside of the \topo.
Such corrections are included in the full LCW chain, though, see \AppRef{app:lcw} for more details.

The \emph{hadronically calibrated} cluster energy \eclushad is available from the LCW sequence, before the corrections for out-of-cluster and inactive material energy losses are applied.
It serves as a reference for performance comparisons with the ML calibrations.
The expectation is that it represents the energy \eclusdep deposited in the calorimeter cells collected into the \topo,
\begin{align*}
\eclusem
\:\:\xrightarrow{\:\substack{\text{hadronic}\\\text{calibration}}\:\:}\:\:
\underbracket{\eclushad =
\eclusdep}_{\mathclap{\text{expectation/goal}}}\;.
\end{align*}
Consequentially, \eclushad is referred to as the signal at the local hadronic LCW scale, as it is reconstructed in the context of this calibration but omits further corrections not related to the energy deposit within the volume of the \topo.
Due to the cell signal weighting applied in the hadronic calibration step of the LCW procedure, typically $\eclushad > \eclusem$, $\yclusem\neq\yclushad$ and $\aclusem\neq\aclushad$ are expected for all \topos not classified as purely electromagnetic (EM) based on an EM-likelihood $\emlike < 1$ measure provided in the classification step of this procedure.
The machine-learning-based calibration applies a scale factor for \eclusem to represent \eclusdep, and as such does not change the \topo direction reconstructed at \escale scale.

\subsubsection{Simulation data content}\label{sec:detdat:clusters:truth}

The detector simulation provides the relevant \emph{truth information} that assesses the energy \eclusdep deposited inside the \topo.
This is the sum of all energies \ecelldep deposited in the \ncellclusall clustered cells during the shower simulation of \emph{primary particles} emerging directly from the generated hard-scatter interaction,
\begin{align*}
\eclusdep = \sum_{i=1}^{\ncellclusall} w_{\text{cell},i}^{\text{geo}}\ecelldepi{i}\;.
\end{align*}
The geometrical weight $0 < w_{\text{cell},i}^{\text{geo}} \leq 1$ reflects the fractional contribution of \ecelldepi{i} to the \topo if the cell $i$ is shared with another cluster, with $w_{\text{cell},i}^{\text{geo}} = 1$ for cells $i$ exclusively collected into one \topo.

Within this approach \eclusdep is the simulated \emph{detector truth} that is generated by the energy flow carried by primary particles into the calorimeter and collected within the constraints introduced by the \topo algorithm.
Thus, cells that have energy deposits from these particles, but are not clustered, do not contribute to the detector truth of a given final state.
Also not included in \eclusdep are energies deposited by particles from in-time \pu collisions, or any energies deposited by particle generated in nearby bunch crossings contributing to out-of-time \pu.
Any target for a machine-learned calibration, or any other, which is constructed from \eclusdep thus reflects only the energy deposit generated by the \pp interaction of interest, and not from \pu interactions.
This principally allows determining calibration functions that locally, at the level of each \topo, mitigate the effect of \pu on the \topo signal \eclusem with these simulations.

Each \ecelldep, and thus by extension the \eclusdep it contributes to, comprises visible and invisible energy losses within the particle shower.
The visible energy deposits typically consist of losses through ionisation and other interactions that potentially contribute to the cell signal \ecell, and thus the \topo signal \eclusem.
Invisible losses do not have a signal contribution.
They include the rest energy of stopped particles, particles absorbed or decaying without visible energy release, and energy invested in (slow) nuclear processes following an inelastic interaction between incoming shower particles and the matter they traverse.
In addition, all energy invested in invisible particles like neutrinos, which can escape the detector without further trace, is included.\footnote{Within the ATLAS detector simulation, all energy released in the development of particle showers, and not invested into the production of  secondary particles that can be further tracked through the detector setup, is collected at the location where it is produced. For the energy deposits discussed here, this is the calorimeter cell volume in which the interaction happened. The concept of stopped particles is implemented in the interaction modelling within \geant for particles with energies or ranges below the corresponding configured thresholds for secondary particle tracking and thus does not necessarily reflect physical behaviour at the interaction level.}
The invisible losses mostly appear in the development of hadronic showers, and thus are the reason for the non-compensating signal character of the ATLAS calorimeter.

The composition of \ecelldep, and therefore also \eclusdep, is preserved in fractional contributions depending on the nature of the primary particles that cause the energy release into a cell volume in their interactions with the matter of the calorimeter.
In particular, \eclusdep can be characterised by the fraction of energy \fclusdepem deposited by showers from primary electrons/positrons ($e^{\pm}$) and photons ($\gamma$).
Within the context of the studies presented in this paper, the following nomenclature is used to categorise \topos according to \fclusdepem:
\begin{mldescription}
\item[\EmTopos]\mbox{}\\ have at least \perCent{90} of \eclusdep coming from primary $e^{\pm}$ or $\gamma$, $\fclusdepem > 0.9$.
\item[\HadTopos]\mbox{}\\ have less than \perCent{10} of \eclusdep deposited by primary $e^{\pm}$ and $\gamma$, $\fclusdepem < 0.1$.
\item[\CompTopos]\mbox{}\\ have varying fractional contributions to \eclusdep from various primary particles such that $0.1 \leq \fclusdepem \leq 0.9$.
\end{mldescription}
This categorisation is not directly used in the training of the machine-learning-based calibrations, as it is not available for \topos in the experiment.
It is useful when evaluating the sensitivity of observables to the cluster signal origin.
More sensitive observables are expected to better reflect the relation between
\eclusdep and the signal \eclusem it generated, which, among other
considerations, suggests to include them as inputs for the calibration
training, as further discussed in \SectRef{sec:datasets:features}.

\subsection{Event observables}
\label{sec:detdat:event}

The MC simulation samples include \pu at levels observed in LHC \RunTwo, which affects the \topo signal \eclusem and thus the calibration.
Measurable indicators of the \pu activity in a given event are therefore considered for inclusion into the training of the cluster calibration network.
The two relevant observables in data and MC simulations are the number of reconstructed primary vertices \npv~\cite{ATL-PHYS-PUB-2015-026} and the (measured) actual number of \pu interactions in the event \npu.
Generally both \npv and \npu are correlated, with \npv providing a direct measure for in-time \pu activity of each bunch crossing while \npu is also indicative of the out-of-time \pu activity.\footnote{For MC simulations, \npu is sampled from the distribution of \mean{\mu}, which is the number of interactions measured over a luminosity block at the LHC (duration about \SI{30}{s} to \SI{60}{s}) divided by the number of proton bunch crossings in this interval (average over all bunches). Due to this time-integrated construction, \mean{\mu} provides an average measure for the out-of-time \pu activity. For data, \npu is determined from the number of \pu interactions averaged over all bunches at a fixed position in the bunch train, again in a luminosity block~\cite{DAPR-2021-01}.}

\section{Datasets and network setup}
\label{sec:datasets}
\subsection{Dataset}
\label{sec:datasets:topo}

The dataset consists of selected \topos reconstructed in MC simulations of full \pp collision events at \sqrts{13} with multijet final states.
The events are generated and simulated as described in \SectRef{sec:detdat:jets}.

The \topos are collected from fully reconstructed calorimeter jets in the central detector region of ATLAS, as provided by the MC simulations.
Each jet establishes a calorimeter volume containing its \topo constituents.
With this, they not only provide a reference phase space for a group of clusters for future comparisons with experimental data, but also an environment characterised by varying \topo densities covering many spatial cluster distributions also expected in data.

\subsubsection{Final-state objects and phase space}
\label{sec:dataset:object}

Calorimeter jets are formed from \topos with signals at the \escale scale using the \antikt algorithm~\cite{Cacciari:2008gp} in \fastjet~\cite{Fastjet} with a radius parameter $R=0.4$.
They are calibrated using the procedures described in \RefRef{JETM-2018-05}, including \pu corrections, the MC-simulations-derived jet energy scale (JES) calibration, and direction corrections.
The considered phase space for these jets is motivated by typical analysis selections and to restrict the data used for the machine-learning calibration to \topos located in the high-granularity region $|\etadet|<2.5$ of the ATLAS calorimeters, where the \emph{expressiveness} of their associated observables is expected to be the highest.
It is defined using reconstructed calorimeter jet kinematics,
\begin{align}
\ptjes > \SI{20}{\GeV}
\qquad\mand\qquad
\abs{\yjes} < 2.0 \;,
\end{align}
where \ptjes is the fully calibrated transverse momentum of the jet, and \yjes is its rapidity after all corrections.

In the context of this study particle jets in MC simulations are considered as the {physics truth} generating the detector truth introduced in \SectRef{sec:detdat:clusters:truth}.
They are clustered from stable particles in the generated particle-level final state of the hard-scatter interaction, with the exception of muons and neutrinos, using the \antikt algorithm with the same radius parameter used for the calorimeter jets.
Only calorimeter jets within an angular distance $\Delta R< 0.4$ to the nearest particle jet are accepted, to assure a significant amount of true signal contribution to the \topos they are clustered from.
No \pu-jet tagging or removal is applied to these simulated and matched calorimeter jets.

The \topos are collected from the simulated calorimeter jets passing the selection above.
Given the jet definition, they are found to be approximately within $\abs{\yclusem-\yjes}\lesssim 0.4$, with \yclusem being the cluster rapidity reconstructed at the EM signal scale, as introduced in \EqRef{eq:cluskine} of \SectRef{sec:detdat:clusters:kine}.
To remove clusters generated by \pu, all clusters considered for training, testing and validation of the calibration networks are required to have a true energy
content of $\eclusdep>\SI{300}{\MeV}$.
Most \topos not satisfying this selection were found to have a high level of randomness in their signal and other associated observables, with only a weak or no relation to \eclusdep at all.\footnote{Per ATLAS convention, and following the discussion in \SectRef{sec:detdat:clusters:truth}, the \topos generated purely by \pu alone have $\eclusdep = 0$. A meaningful target for calibrations requires $\eclusdep > 0$, at least.}
They are excluded as they are not significantly contributing to the hadronic final state (and jet) signals in the context of this study.

The collected sample of \topos is randomised and all information from the jet context is dropped, as it is not relevant for training the networks and the performance evaluation of the learned calibration.

\subsubsection{Feature selection for the calibration training}
\label{sec:datasets:features}

In addition to the kinematics represented by \eclusem and \yclusem, each \topo is associated with a set of observables that are canonically referred to as \emph{cluster moments}.
These are part of the general analysis data model in ATLAS.
The moments were designed to provide signal characteristics including its timing and significance as well as selected measures indicative of the nature of the energy deposit in the \topo, together with other measures that have impact on a cluster classification by signal sources and a local hadronic calibration.

The \topo moments that are expected to be useful for a local calibration are those that establish an observable space that represents the relation between the signal \eclusem and its source \eclusdep.
This relation can only be directly established, within the limitations of the models applied, in MC simulations.
When using a selection of these observables for a machine-learning-based multi-dimensional \topo calibration, not only the individual observables but also the correlations between them are important.

In machine-learning, the inputs to a classification or regression network are often referred to as features.
For learning the calibration, each \topo is represented by a \emph{feature set} \fsetclus that serves as input to a learned calibration network.
There are several important aspects to be considered when collecting some of the features summarised in \TabRef{tab:observables} into \fsetclus.
These mostly relay to
\begin{enumerate}
\item the expectation for the achievable accuracy of the calibration not only for the MC simulations but also for the application to experimental data; \label{enum:precision}
\item the expressiveness of each feature with respect to the calorimeter signal source and its reflection in the signal definition given by the topological cell clustering; \label{enum:expressive}
\item and a meaningful feature reconstruction with sufficient resolution for almost all \topos, if not all, reconstructed to measure the final state of the \pp collisions in the ATLAS calorimeters.\label{enum:coverage}
\end{enumerate}
Both aspects captured in item~\ref{enum:precision} and item~\ref{enum:expressive}, respectively, are tightly connected.
Features with little sensitivity to the \topo signal source are irrelevant as inputs and are thus omitted.
Similarly, the selection of expressive but highly correlated features may lead to an unnecessarily large numerical system decreasing the computing performance and potentially introducing a higher level of \emph{stochasticity} (noise).

The application of the \topo calibration learned from MC simulations to experimental data requires that the individual features and their correlations are well modelled.
This is generally achievable for individual features that are lower order \topo moments, as documented in \RefRef{PERF-2014-07}.
Several highly expressive second-order moments are not considered, as they are too sensitive to the (hadronic) shower modelling and the signal calculation in the simulation including \pu.
In particular, the modelling of \pu contributions to these complex observables requires minimum bias physics and detector simulations to be accurate at often small scales that, at the present state-of-art, have not been achieved in all aspects.
The second order moments that provide measurements of the longitudinal and lateral \topo signal dispersion (\cluslong, \cluslat) are included in \fsetclus, because they were found to be largely insensitive to modelling issues and \pu \cite{PERF-2014-07}.
In addition, the signal compactness measure \clusptd and the energy-squared-weighted variance of the cell time distribution \tclusvar inside the \topo, both calculated as described in \AppRef{app:features:construction}, are included because by construction they are expected to have little sensitivity to potential (small scale) modelling issues.

The modelling of the feature correlations can only be evaluated in the context of a full performance study in which the trained calibration is applied to data, with well defined metrics.
This study is out-of-scope for this method paper.

\begin{table}[tt!]
\centering
\caption[]{The \topo features used in the training of the \topo calibration networks, as summarised in \EqRef{eq:observables}.
The variables are defined in \RefRef{PERF-2014-07}, except for \clusptd
and \tclusvar defined in \AppRef{app:features}. Selected features used in the LCW calibration approach are indicated.
}
\renewcommand{\arraystretch}{1.25}
\newlength{\fcolw}\settowidth{\fcolw}{Signal strength\ }
\newlength{\scolw}\settowidth{\scolw}{\ \tclusvar\ }
\newlength{\lcolw}\settowidth{\lcolw}{\ LCW\ }
\newcolumntype{N}{>{\raggedright\arraybackslash}p{\fcolw}}
\newcolumntype{V}{>{\centering\arraybackslash}p{\scolw}}
\newcolumntype{L}{>{\centering\arraybackslash}p{\lcolw}}
\newcolumntype{R}{>{\raggedright\arraybackslash}X}
\label{tab:observables}
\begin{tabularx}{0.98\textwidth}{@{}NVLR@{}}
\hline\hline
Category & Symbol & LCW & Comment \\ \hline
\multirow[t]{2}{*}{Kinematics} & \eclusem & yes & Signal at the electromagnetic energy scale    \\
& \yclusem & yes & Rapidity at the electromagnetic energy scale \\ \hline
\multirow[t]{2}{*}{\Shortunderstack[l]{{Signal strength,} {time structure}}} & \sclus & no & Signal significance  \\
& \tclusvar & no & Variance of \tcell distribution \\ \hline
\multirow[t]{6}{*}{\Shortunderstack[l]{{Shower depth,} {shower shape,} compactness}} & \cluslam   & yes & Distance of the \cog from the calorimeter front face (along principal cluster axis)\\
& \cluscog         & no & Distance of the \cog from the nominal vertex\\
& \clusfemc       & no & Fraction of energy in the electromagnetic calorimeter\\
& \clusdensfirst &  yes & Cluster signal density measure\\
& \cluslong        & no & Normalised energy dispersion along the principal cluster axis\\
& \cluslat           & no & Normalised energy dispersion perpendicular to the principal cluster axis\\
& \clusptd          & no & Signal compactness measure (inspired by~\RefRef{CMS-PAS-JME-13-002})\\\hline
Topology & \clusfiso          & yes & Cluster isolation measure\\ \hline
\multirow[t]{3}{*}{\PU}
& \tclus               & no & Signal timing \\
& \npv                 & no & Number of reconstructed primary vertices\\
& \npu                 & no & Number of \pu interactions per bunch crossing\\
\hline \hline
\end{tabularx}
\end{table}

The final important aspect noted, in item~\ref{enum:coverage} in the list above, must be considered when composing \fsetclus.
All chosen features need to be reconstructed for each cluster in a large sample of \topos in a meaningful way, with the highest possible resolution.
This requires least impact from the \topo formation algorithm and cell-level detector inefficiencies due to, e.g., insufficient granularity and noise, on their expressiveness for the calibration.
For this reason, the jet phase space, and consequently the \topo phase space, covered by this proof-of-principle study of the applicability of a ML calibration and the understanding of its predictions, is restricted to the ATLAS calorimeter region with the highest readout granularity, as mentioned in \SectRef{sec:dataset:object}.

Even within the chosen phase space, very few \topos in the used dataset may still not have enough cells or suffer from other deficiencies introduced by the detector or the collision environment such that it is not possible to reconstruct some or all of the features with the highest possible accuracy and precision.
Nevertheless, and mostly to avoid selection biases and thus predict the most universal calibration function for all \topos in the illuminated calorimeter region, these clusters are represented as well in both training and testing of the calibration network.

The final composition of \fsetclus contains the 15 features described in \TabRef{tab:observables},
\begin{align}
\begin{split}
& \fsetclus = \label{eq:observables}\\
& \bigl\{
\underbracket{\vphantom{\cluslong}\eclusem,\yclusem}_{\mathclap{\text{kinematics}}},
\overbracket{\vphantom{\cluslong}\sclus}^{\mathclap{\text{signal
relevance}}},
\underbracket{\vphantom{\cluslong}\tclusvar,\cluslam,\cluscog,\clusdensfirst,\cluslong,\cluslat,\clusptd,\clusfemc}_{\mathclap{\text{shower
nature (position, compactness, signal density and internal time
structure)}}},
\overbracket{\vphantom{\cluslong}\clusfiso}^{\mathclap{\text{topology
(isolation)}}},
\underbracket{\vphantom{\cluslong}\tclus,\npv,\npu}_{\mathclap{\text{\pu}}}
\bigr\}
\end{split}
\;.
\end{align}
It reflects the guidance discussed above.\footnote{The DNN trained in the context of the studies presented in \RefRef{ATL-PHYS-PUB-2023-019} employs the same feature set.}
Included are the kinematic variables \eclusem and \yclusem, both basic inputs for any calorimeter calibration.
Also included are all of the \topo observables employed by the classification and the hadronic calibration in the LCW sequence, thus removing the need for an explicit cluster classification but preserving the information entering this step.
In addition, the event-level \pu measures \npv and \npu and the \topo isolation measure \clusfiso are included to represent topological (varying cluster densities in jets) and environmental (\pu) effects on the cluster signal.
All members of \fsetclus are available for both experimental data and MC simulations.

\subsection{\Topo calibration training}
\label{sec:data_cal}

The hadronic calibration in the LCW sequence aims at the (true) deposited energy \eclusdep as a calibration target.
The determination of the local \topo calibration using machine learning techniques introduces a set of novel aspects.
Rather than learning \eclusdep with a typical value ranging from
$\mathcal{O}(\SI{100}{\MeV})$ to $\mathcal{O}(\SI{1}{\TeV})$ directly, the
response \rclusem as a function of the feature set \fsetclus from
\EqRef{eq:observables} is learned for each \topo in a multi-dimensional
regression fit.
The target \rclusem is defined as the ratio of the cluster signal \eclusem at \escale scale and the deposited energy in the cluster \eclusdep,
\begin{align}
\label{eq:target-response}
\rclusem = \eclusem / \eclusdep\;.
\end{align}
It is available in MC simulations for each \topo extracted from the jet and selected according to the criteria introduced in \SectRef{sec:dataset:object}.
The significantly reduced value range for \rclusem lowers the computational effort in the training of the network by reducing the number of epochs, with the promise of a more accurate and precise calibration.

Both the BNN and the RE learn uncertainties related to the trained calibration for each \topo individually.
These uncertainties allow for a better understanding of (1) detector signal features, (2) possible signal quality issues in the data, and (3) possible fundamental limitations and other issues associated with the network training.
As further discussed in \SectRef{sec:uncertainties:predictions}, they are learned for each \topo to provide a measure of the accuracy achieved with the trained calibration network.

\subsubsection{Training target}
\label{sec:data_cal:target}
\begin{figure}[tt!]
\centering
\subfloat[]{\includegraphics[width=1.00\figwidth]{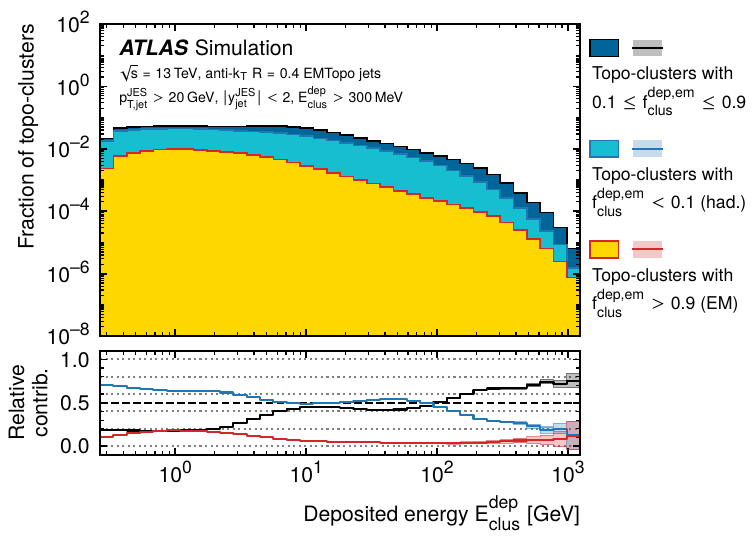}\label{fig:composition:deposit}}\quad
\subfloat[]{\includegraphics[width=1.00\figwidth]{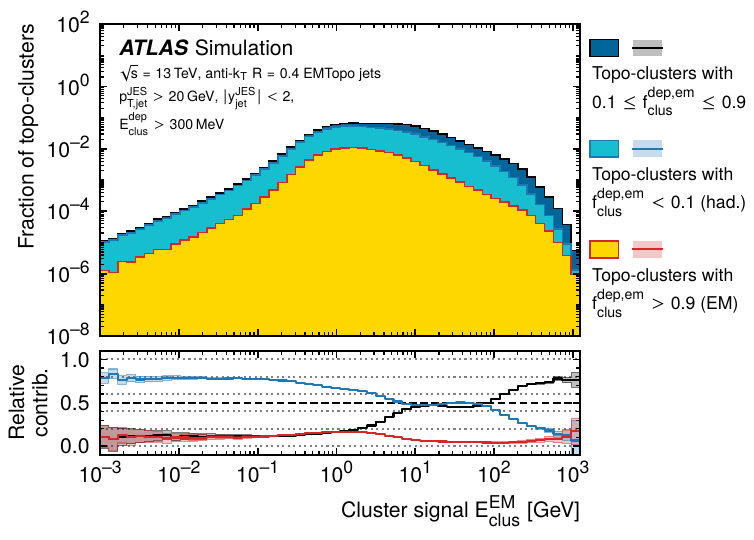}\label{fig:composition:signal}}\\
\subfloat[]{\includegraphics[width=1.00\figwidth]{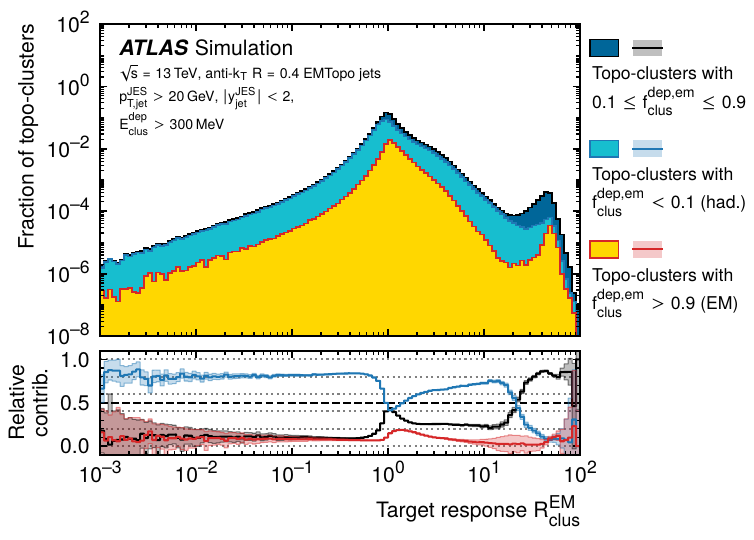}\label{fig:composition:response}}
\caption[]{The distributions of the energy \eclusdep deposited in \topos are shown in \subref{fig:composition:deposit} in a stack, for the clusters categorised as electromagnetic, hadronic or composite by applying the classifications introduced in \SectRef{sec:detdat:clusters:truth}.
In \subref{fig:composition:signal}, the corresponding stacked distributions for the basic cluster signal \eclusem at EM scale are shown.
The distributions of the machine-learning training target, the \topo response \rclusem constructed from \eclusem and \eclusdep according to \EqRef{eq:target-response}, are shown stacked in
\subref{fig:composition:response} for the same three categories.
The distributions are filled from the sample of \topos collected from fully simulated calorimeter jets, as described in \SectRef{sec:dataset:object}.
The respective lower panels show the relative contributions from the three categories to the inclusive spectrum. The shaded areas indicated the statistical errors.
}
\label{fig:composition}
\end{figure}

In addition to using the \topo response \rclusem as a calibration target, instead of the deposited energy \eclusdep that is the direct target of the LCW hadronic calibration, the dedicated classification step characteristic for the LCW procedure is omitted when training the BNN and the RE network.
Instead, the features  \eclusem, \yclusem, \cluslam and $\clusdensfirst \equiv \clusrho$ used in the LCW classification are included, as presented in \TabRef{tab:observables} and \EqRef{eq:observables}.
The result of the BNN(RE) training is the individual response prediction $\rclusbnnre(\fsetclus)$ for a given \topo with a feature set \fsetclus, with
\begin{align}
\label{eq:calib_goal}
\rclusbnnre(\fsetclus) \overset{\text{trained}}{=}\;\rclusem(\fsetclus)\;,
\end{align}
with \rclusem from \EqRef{eq:target-response}.
Both networks are independently trained to encode the respective representation \rclusbnnre of the target \rclusem as a function of \fsetclus given in \EqRef{eq:observables}.
The derivation of the corresponding calibrated \topo energy \eclusbnnre then employs \rclusbnnre such that
\begin{align}
\label{eq:energy-recons}
\eclusbnnre
= \frac{\eclusem}{\rclusbnnre(\fsetclus)}\;.
\end{align}
\FigRef{fig:composition} shows the complex distributions for \eclusdep, \eclusem and \rclusem for the three \topo categories defined in \SectRef{sec:detdat:clusters:truth}.
The particular distribution shapes observed for both \eclusdep and \eclusem are driven by the multijet final state simulation configuration that determines the incoming flow of particles from the selected jets in the MC simulations, with its calorimeter representation subjected to the \topo selection discussed in \SectRef{sec:dataset:object}.
The distributions provide an integral view on regional hadronic signal characteristics, like the varying levels of non-compensation across the illuminated \topo rapidity range.

The contribution of \emTopos to the \eclusem distribution in \FigRef{fig:composition:signal} is near constant for $\eclusem\lesssim\SI{2}{\GeV}$.
The lower \eclusem regime is largely populated by \hadTopos, with their relative contribution dropping to match the one from \compTopos for $\SI{10}{\GeV}\lesssim\eclusem\lesssim\SI{100}{\GeV}$.
\CompTopos then dominate in the high signal regime $\eclusem > \SI{100}{\GeV}$, indicating a dominant contribution from clusters generated by merged showers and shower fragments.

The resulting distributions of \rclusem shown on \FigRef{fig:composition:response} for the \topo categories depict the complexity of the calibration task.
Small $\rclusem \ll 1$ indicate that the contribution of the deposited energy \eclusdep to the \topo signal \eclusem is insignificant.
This is the case for \eclusdep predominantly arising from stopped particles with invisible decay products like neutrinos together with a small fraction of \eclusdep generating the signal, or in deep inelastic hadronic interactions where a large amount of invisible energy is produced.
In addition, the shower-generated signal reflecting \eclusdep may be suppressed because the total signal \eclusem of the \topo is dominated by negative cell signals from out-of-time \pu noise~\cite{PERF-2014-07}.

Many of the \topos with large responses  $\rclusem \gg 1$  have a \cog located in the Tile gap scintillators~\cite{TCAL-2021-01},
where the response reflects detector signal characteristics that are not only related to \pu.
These counters are not calorimeters in that they have no internal absorber structures generating showers.
Their signal \eclusem at the EM scale is calibrated to (on average) correct for energy losses in the inactive materials around them.
Consequently, with the definition of \rclusem given in \EqRef{eq:target-response}, the deposited energy \eclusdep being small in the absence of an absorber,
and the signal \eclusem by design representing energies much larger than \eclusdep, a large \rclusem is expected.

The energy loss of particles produced in in-time \pu interactions, or the secondary particles they generate by showering in the material around the Tile gap scintillators, is not accounted for in \eclusdep that is conceptionally defined in \SectRef{sec:detdat:clusters:truth} as the detector truth generated by the final state of the hard scatter \pp interaction.
Due to the \topo selection presented in \SectRef{sec:dataset:object}, each cluster has a finite part of its signal \eclusem generated by $\eclusdep > \SI{300}{\MeV}$, with the remainder generated by a contribution from \pu that is subject to considerable fluctuations.
The peak of the \rclusem distribution around $\rclusem \approx 50$ manifests these characteristic signal features.
It hints at a direct proportionality between \eclusem and \eclusdep for the bulk of the \topos population under it, which is expected for the signal from ionisation-only energy deposits in a scintillating counter.\footnote{The energy \eclusdep in a Tile gap scintillator only represents the sum of all ionisation losses from charged particles emerging from the hard scatter interaction and passing through this detector. With the ionisation loss in the thin scintillator being about the same for each of those particles, within narrow Landau distributions \cite{Landau:1944fvs,Vavilov:1957zz}, \eclusdep is expected to be proportional to the number of charged particles  passing through it. This charged particle multiplicity in turn is an indirect measure of energy losses in inelastic interactions occurring in the inactive material located around the scintillator. Thus, \eclusdep is approximately proportional  to these losses when induced by particles from the hard scatter final state, but does not capture energy losses induced by charged particles directly emerging from in-time \pu interactions or generated in showers generated by these particles.}

Fluctuating levels of \pu, in terms of energy flow density and particle types in both prompt emissions from the interactions in the proton bunch crossing and in the shower development in the materials in proximity of the Tile gap scintillators caused by those, are dominant contributions to the width and shape of this peak, aside from possible modulations introduced by the \topo selection applied in this study.
These \pu-induced response fluctuations arise from the \pp bunch-crossing-to-bunch-crossing variations of the \pu activity, in combination with the amount of material located  between the Tile gap scintillators and the collision vertex region that comprises calorimeter modules as well as non-instrumented mechanical structures.
This upstream material acts as shielding that, when folded with the particle flow from the vertex region contributing to the reconstructed jets used for this study, introduces fluctuations of \eclusem (and thus \rclusem) of varying magnitude for \topos between the lower rapidity edge of the Tile gap scintillator coverage  at $|\yclusem| \approx 0.8$  (largest amount of upstream material) and the higher rapidity edge at $|\yclusem|\approx1.8$ (lowest amount of upstream material).
Some of these observations are revisited in the context of the predicted uncertainties in \SectRef{sec:uncertainties:detector}.

Larger \rclusem are also observed for some \topos located in the electromagnetic calorimeters, which are closest to the collision region and thus exposed to the highest flow of particles from \pu collisions.
Generally, and in extension to the interpretations of the \eclusdep and \eclusem spectra given above, the complexity of the \rclusem distributions reflect the incoming particle
flow, the collision environment, and the varying detector geometries and, consequently, the varying calorimeter signal characteristics.

\subsubsection{Feature and target transformations and projections}
\label{sec:data_preproc}

To limit the numerical value range of the network inputs and to create more appropriate (peaked) feature distributions for the training whenever possible, the features are transformed following three different procedures.
If the spectrum of a feature $x\in\fsetclus$ is smoothly falling and $x > 0$ always, $x$ is replaced by a \emph{standardised logarithmic representation},
\begin{align}
\label{eq:log-and-standardisation}
x \to \frac{\logten x -\mean{\logten x}}{\std(\logten x)}\;.
\end{align}
Here $\mean{\logten x}$ is the statistical average of the $\logten x$ distribution from the training sample, and $\std(\logten x)$ denotes the corresponding standard deviation of the transformed features $x$.

\begin{table}[tt!]
\centering
\caption[]{Summary of the transformations applied to the various observables collected into the feature set \fsetclus from \EqRef{eq:observables}, and the target response \rclusem from \EqRef{eq:target-response}.}
\label{tab:transforms}
\renewcommand{\arraystretch}{1.25}
\newlength{\faaaa}\settowidth{\faaaa}{Logarithmic (\logten) standardisation (\EqRef{eq:log-and-standardisation})}
\newcolumntype{N}{>{\raggedright\arraybackslash}p{\faaaa}}
\newcolumntype{R}{>{\raggedright\arraybackslash}X}
\begin{tabularx}{0.98\textwidth}{@{}NR@{}}
\hline\hline
Transformation (preprocessing) & Features \\ \hline
Logarithmic (\logten) standardisation (\EqRef{eq:log-and-standardisation})  & \eclusem, \sclusem, \tclusvar, \cluslam, \clusdensfirst \\
Linear standardisation (\EqRef{eq:standardisation})               & \yclusem, \cluscog, \clusfemc, \cluslong, \cluslat, \clusptd, \clusfiso, \npv, \npu \\
Maximum-absolute normalisation (\EqRef{eq:cluster-time})     & \tclus \\
Logarithmic (\logten) transformation (\EqRef{eq:log-r})                            & \rclusem \\
\hline\hline
\end{tabularx}
\end{table}

For other distribution shapes, $x$ is standardised without additional transformation (\emph{linear standardisation}),
\begin{align}
\label{eq:standardisation}
x \to \frac{x-\mean{x}}{\std{x}}\;,
\end{align}
where $\mean{x}$ and $\std{x}$ are the mean and the standard deviation of the distribution of $x$ from the training sample.
An exception is the cluster time \tclus, which is normalised to its maximum absolute value $\max(\vert\tclus\vert)$ found in the training sample, to keep its sparsity (\emph{maximum-absolute normalisation}),
\begin{align}
\tclus \to \frac{\tclus}{\max(\vert\tclus\vert)}\;.
\label{eq:cluster-time}
\end{align}
For the regression target, a simple logarithmic transform is applied to reduce its numerical range further,
\begin{align}
\rclusem \to \logten \rclusem\;.
\label{eq:log-r}
\end{align}
The transformations of the individual features listed in
\TabRef{tab:observables} and the regression target from
\EqRef{eq:target-response} are summarised in \TabRef{tab:transforms}.

\begin{table}[ttt!]
\centering
\caption[]{The architecture and hyper-parameters for both the BNN and the
RE network with the three-mode Gaussian mixture likelihood. Both networks
feature a decaying learning rate (\lrate) implemented by the learning-rate
scheduler \StepLR from the \pytorch \cite{Paszke:2019xhz} software tool kit, with
\lrate reduced by a factor $\gamma$ at each of the epochs within the range
specified in braces below.}
\renewcommand{\arraystretch}{1.25}
\begin{tabular}{ll}
\hline\hline
Hyper-parameter & Network architecture and setup \\
\hline
Likelihood model & Gaussian mixture \\
Number of mixture modes \nmix & $3$ \\
Number of hidden layers & $4$ \\
Number of nodes per layer (input, hidden, and output) &
$\braces{15, 64, 64, 64, 64, 9}$ \\
Activation function (inner layers) & rectified linear unit
(ReLU) \\
Activation function (last layer) & none \\
Central-value prediction (mean or maximum) & maximum of the
likelihood \\
\hline
Optimizer and learning rate (\lrate) & \adam \cite{Kingma:2014vow} with $\lrate = 10^{-4}$ \\
Learning-rate scheduler & \StepLR, epochs $\braces{25, 100}$, $\gamma=0.1$ \\
Number of training epochs & $150$ \\
Training (and inference) batch size & $4096$ ($512$) \\
\hline
Dataset size (all selected \topos)           & $14.5\times10^{6}$ \\
Data sample size for training (\DSsizeTrain)   & $8.7\times10^{6}$ (\perCent{60.0}) \\
Data sample size for validation & $5.0\times10^{5}$ ($\perCent{3.4}$)    \\
Data sample size for testing     & $5.3\times10^{6}$ ($\perCent{36.6}$)     \\
Sampling at inference on the test dataset (BNN only, $N$) & $50$ times \\
Number of ensemble members (RE only, $N$)      & 50 \\
\hline\hline
\end{tabular}
\label{tab:hyperparameters}
\end{table}
\subsection{Models and network setup}
\label{sec:data_net}

To learn the response given in \EqRef{eq:target-response} as a function of \fsetclus, both a Bayesian neural network and a repulsive ensemble are employed.
Both are configured with the same network model, with the hyper-parameters presented in \TabRef{tab:hyperparameters}.
They differ significantly in the implementation of the variational approach to determine the uncertainties, and thus in the set of learned network parameters and the respective loss functions, as further discussed in \SectsRef{sec:data_net:bnn_loss}{sec:data_net:re_loss}.

\subsubsection{Principle design of a Bayesian neural network}
\label{sec:data_net:bnn_design}

\begin{figure}[tt!]
\centering
\includegraphics[width=1.00\textwidth]{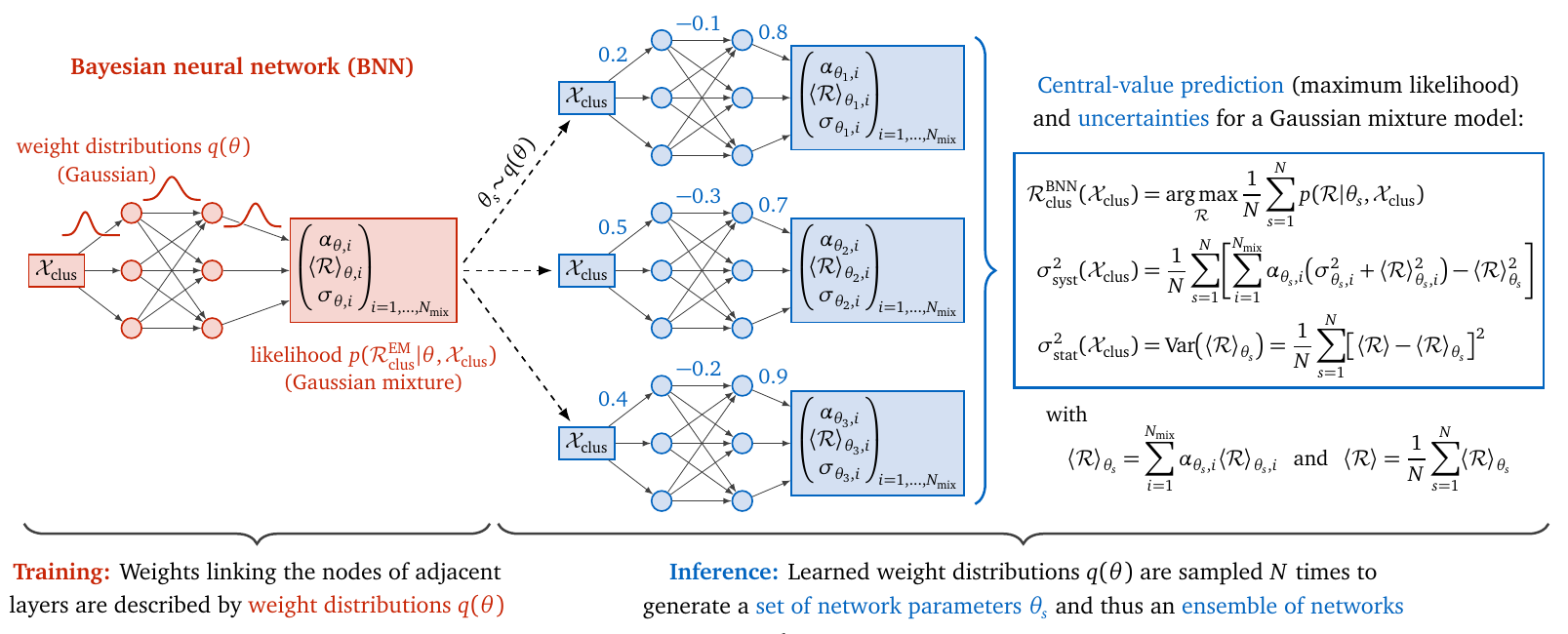}
\caption[]{The principal design of the Bayesian neural network (BNN)
employed for the regression fits and the corresponding uncertainty
predictions for the \topo calibration.
The network on the left is designed such that the resulting weights linking the nodes of adjacent layers are described by weight functions $q(\theta)$ with learned averages and widths.
The inference of the trained model, represented by a set of learned network parameters $\theta$, is shown on the right,
where an ensemble of $N$ networks is sampled from $q(\theta)$ to
generate $N$ predictions from the model.
The central prediction $\rclusbnn$ is the mode (most probable value) of the average of $N$ probability density functions $p(\mathcal{R}|\theta_{s},\fsetclus)$, where each individual $p(\mathcal{R}|\theta_{s},\fsetclus)$ represents the probability that the network describes the target $\mathcal{R} = \rclusem$ for a given \topo with feature set \fsetclus.
Each of the $p(\mathcal{R}|\theta_{s},\fsetclus)$ is defined as the mixture of $i = 1 \ldots \nmix$ Gaussian distributions with learned means $\langle \mathcal{R}\rangle_{\theta_{s},i}$, widths $\sigma_{\theta_{s},i}$ and coefficients $\alpha_{\theta_{s},i}$, as introduced in \EqRef{eq:bnn_gmix}, with $\nmix = 3$.
The averaged function is the normalised sum of all probability density functions obtained by sampling weights from $q(\theta)$ $N$ times, with $N=50$.
It is composed of a total of $N\times\nmix$ Gaussian distributions.
The weighted average response $\langle \mathcal{R}\rangle_{\theta_{s}}$ is the sum of $\alpha_{\theta_{s},i}\langle \mathcal{R}\rangle_{\theta_{s},i}$ from the \nmix distributions for each sample $s$.
It is needed to predict the systematic ($\uncbnnsyst(\fsetclus)$) and statistical ($\uncbnnstat(\fsetclus)$) uncertainties, which are further discussed in \SectRef{sec:uncertainties:components}.
It is calculated for each of the $N$ sampled sets of network parameters $\theta_{s}$ to obtain $\langle \mathcal{R}\rangle$, the mean response averaged over the $N$ corresponding $\langle \mathcal{R}\rangle_{\theta_{s}}$.
This mean is needed exclusively for the calculation of \uncbnnstat.
All calculations are individually performed for any given \topo at inference of the trained BNN model.
The numbers written above some of the edges (links) between nodes are weights sampled from $q(\theta)$ at network inference.
They are for illustration only and show the sampling character of the predictions.
}
\label{fig:bnn_sketch}
\end{figure}

As previously mentioned, Bayesian neural networks do not only predict the target of regressions like the one exercised in the context of the \topo calibration, but also provide uncertainty predictions associated with the central value prediction.
\FigRef{fig:bnn_sketch} illustrates the principle setup for the training of the BNN and its inference on the (test) data.

The most significant deviation of the BNN model from the usual deep neural network design is that the weights connecting the nodes in adjacent layers are not trained as fixed values, rather they are represented by learned average values and widths of probability distributions $q(\theta)$.\footnote{The distributions $q(\theta)$ are chosen to be Gaussian functions. This is not required but enhances computational performance without observable effects on the accuracy of the predictions \cite{Bollweg:2019skg}.}
Here $\theta$ are the trained network parameters.
When the BNN is inferred on test data, these distributions are sampled a fixed number of times, denoted by $N$ and set to $N = 50$ for this result.
This is the same as collecting the predictions from $N$ different networks trained using the same data, without the need to do this training $N$ times.
The central prediction \rclusbnn is then constructed from the $N$ sampled predictions such that it reflects the prediction with the highest probability to represent the target \rclusem, as indicated in \FigRef{fig:bnn_sketch}.
The uncertainties are constructed as indicated in the same figure.
They are further discussed in \SectRef{sec:uncertainties}.

\subsubsection{BNN loss function}
\label{sec:data_net:bnn_loss}

The loss function employed in the BNN training is designed to maximise the likelihood \probbnn of the network parameters $\theta$ to describe the training dataset \trainds.
This dataset is represented by a vector of target and feature set pairs
$\left\{ \rclusem, \fsetclus\right\}_{j}$ for $j = 1,\ldots,\DSsizeTrain$
\topos,
\begin{align}
\label{eq:trainds}
\trainds = \left(
\left\{\rclusem,\fsetclus\right\}_{1},\,
\left\{\rclusem,\fsetclus\right\}_{2},\,\ldots,\,
\left\{\rclusem,\fsetclus\right\}_{\DSsizeTrain} \right)\;.
\end{align}

The form of \probbnn is not known a priori.
Therefore, and following Bayes' theorem,  the posterior \probbayes that provides the likelihood that \trainds describes $\theta$, is maximised instead of \probbnn, as discussed in \RefRef{Plehn:2022ftl}.
Assuming a Gaussian likelihood for \probbayes is well motivated, but in case of the \topo-response predictions, a mixture of $\nmix = 3$ Gaussian functions is chosen to accommodate possible deviations from this shape,
\begin{align}
\label{eq:bnn_gmix}
\probbayes
= p(\{\rclusem\} | \theta, \{\fsetclus\})
= \prod_{j=1}^{\DSsizeTrain} \sum_{i=1}^{\nmix} \awi\,\gauss(
\r_{\text{clus},j}^{\text{EM}} |
\mean{\r}_{\theta,i},
\sigma_{\theta,i} ) \;.
\end{align}
Here \awi are the weights with which the Gaussian functions $\gauss(\rclusem |
\mean{\r}_{\theta,i}, \sigma_{\theta,i} )$ with mean $\mean{\r}_{\theta,i}$
and
standard
deviation $\sigma_{\theta,i}$ contribute to \probbayes, with
\begin{align*}
\awi \geq 0
\qquad\mand\qquad
\sum_{i=1}^{\nmix}\awi = 1\;.
\end{align*}
All \awi, $\mean{\r}_{\theta,i}$ and $\sigma_{\theta,i}$ are predicted by the
BNN and are thus functions of $\fsetclus$.
The mixture weights \awi are forced to be positive and to add up to unity by applying a \emph{softmax}~\cite{Bridle:1989a,Bridle:1990a} function.

To find the maximum of \probbayes, the negative logarithmic likelihood $-\log\probbayes$ is minimised.
As the uncertainties $\sigma_{\theta,i}$ are learned as well, rather than being fixed \emph{a-priori}, this approach implements a \emph{heteroscedastic} loss function that regularises the otherwise prevalent trend of the minimisation to prefer larger values of $\sigma_{\theta,i}$ in the exponents of the Gaussian functions.
Therefore, trivial solutions $\sigma_{\theta,i} \rightarrow \infty$ are avoided and the values of $\sigma_{\theta,i}$ are generally restricted to meaningful ranges.
The introduction of a (reverse) Kullback-Leibler divergence $\kl[q(\theta),\pprior(\theta)]$ \cite{KL_1951} in the BNN loss function \lbnn acts as a ridge (L2) regularisation \cite{Hastie_2020} that controls over-fitting and focuses the fit on regions of relevant data,
\begin{align}
\label{eq:bnn_loss}
\lbnn = \kl[q(\theta),\pprior(\theta)] -
\big\langle\log\probbayes\big\rangle_{\underbracket{\theta\!\sim\!q(\theta)}_{\mathclap{\substack{\theta\text{\
sampled}\\\text{from \ }q(\theta)}}}}\;.
\end{align}
The $\kl[q(\theta),\pprior(\theta)]$ term vanishes when the distribution of learned network parameters $q(\theta)$ matches the prior $\pprior(\theta)$.
Both $q(\theta)$ and $\pprior(\theta)$ are chosen to be Gaussians so that
$\kl[q(\theta),\pprior(\theta)]$ can be calculated analytically. The prior
mean and standard deviation are set to zero and one, respectively.
The likelihood \probbayes is taken from \EqRef{eq:bnn_gmix} and the network
parameters $\theta$ are sampled once from $q(\theta)$ to calculate the average
negative logarithmic likelihood term in \EqRef{eq:bnn_loss}.
More details of the motivation and the design of \lbnn are given in \AppRef{app:loss:bnn}.

Compared to a regression training without uncertainty predictions, the network output in case of a loss defined by a single Gaussian function, configured by using $\nmix =1$ in  \EqRef{eq:bnn_gmix}, is doubled.
It consists of a local central value function as a representation of the regression target and a local uncertainty map.
Generally, the number of network outputs increases by a factor of $3\nmix-1$ for a mixture of \nmix Gaussian distributions.

The predictions from the BNN are a novel approach to learn uncertainty contributions to calorimeter signal calibrations.
They can represent relevant contributions to the nuisances characterising the overall systematic uncertainty associated with the application of the learned local (\topo by \topo) calibration in data and MC simulations.
The learned uncertainties also serve as indicators of the detector signal expressiveness in terms of the measurement of the true energy deposit, as shown in \SectRef{sec:uncertainties:detector}.
They can thus be used as a base for \topo selections to improve the accuracy and precision of the energy measurement for reconstructed particles inside and outside of jets, and thus the general energy flow in the final state of the \pp collisions at the LHC.

\subsubsection{Repulsive ensembles}
\label{sec:data_net:re_design}

\begin{figure}[tt!]
\centering
\includegraphics[width=1.00\textwidth]{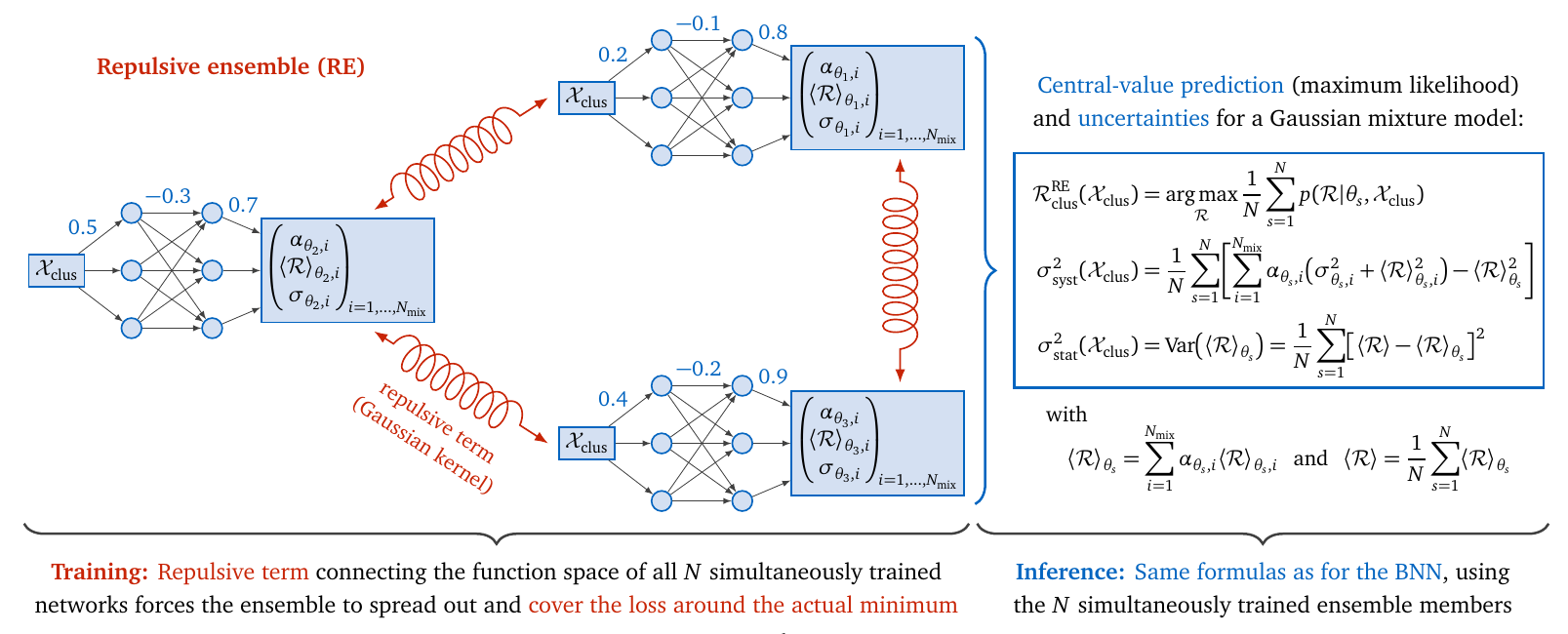}
\caption[]{Schematics describing the design of the repulsive ensemble (RE)
model employed for the regression fits and the corresponding uncertainty
predictions of the \topo calibration. An ensemble of $N$ networks with
identical configurations is configured. These $N$ networks are trained simultaneously with interconnections between their loss functions acting as repulsive forces between them (shown as spring connectors in the schematic).
The predictions from the $N$ networks are collected for each \topo in each processed batch during training.
A central prediction and the uncertainties are calculated in the same way as
illustrated in \FigRef{fig:bnn_sketch} for the BNN,  with the same nomenclature introduced there.
The numbers written above some of the edges (links) between nodes illustrate the learned weights for one of the $N$ repulsive ensembles.
They are shown to illustrate that each network fits its own weights within the repulsive action of the loss function.
}
\label{fig:re_sketch}
\end{figure}

To gain confidence in the BNN uncertainty predictions for the local \topo calibration, an alternative and independent model for learning such uncertainties employing a network of repulsive ensembles is considered.
The basic idea of RE is to determine uncertainties for the learned parameters $\theta$, and consequently the functions they describe, by forcing ensembles of identically configured training networks to not predict the same best-fit network parameters, but rather explore the landscape around the corresponding minimum of the loss.
While some variation of the predictions between members of an ensemble of networks can be observed due to random network initialisation and random training batch selection, there is no explicit mechanism that guarantees that the full weight space encoding \probbayes is mapped out.
It can be shown that adding a repulsive force (network interaction) to the usual update rule that minimises $-\log\probbayes$ based on gradient descent, maps the weight space completely when applied to all
members of the network ensemble simultaneously at each update $t\rightarrow t+1$ evolving $\theta^{t}\rightarrow\theta^{t+1}$.
This procedure allows predicting uncertainties, see, \eg, the comprehensive discussion in \RefRef{Plehn:2022ftl}.
The basic design of the RE training model is sketched in \FigRef{fig:re_sketch}.
While the formal determination of the central prediction and the uncertainties shown in this figure is identical to the one for the BNN shown in \FigRef{fig:bnn_sketch}, the underlying sampling strategies are fundamentally different.
The BNN directly samples in weight space, while the RE samples in the function space establishing the repulsive force between networks.
Some relevant details of the complex theoretical foundation and the implementation of RE are given in \AppRef{app:loss:re_impl}.

\subsubsection{Repulsive ensemble loss function}
\label{sec:data_net:re_loss}

The loss function of the RE network is derived from the modified update rule.
Its principal form is a sum of $i=1,\ldots,N$ terms \lreterm for an ensemble of $N$ networks,
\begin{align}
\label{eq:re_loss}
\lre = \sum_{i=1}^{N} \lreterm\;.
\end{align}
Each \lreterm has three components comprising
\begin{enumerate}
\item the negative logarithm of a posterior likelihood \probbayes of Bayes' theorem representing \probbnn, the likelihood of the network parameters $\theta$ to describe the training dataset \trainds in \EqRef{eq:trainds}, similar to the approach in the design of the BNN loss function in~\SectRef{sec:data_net:bnn_loss};\label{enum:re_neglog}
\item a repulsive term connecting the loss function space of all $N$ networks with a given network $i\in[1,N]$ using Gaussian kernels; \label{enum:re_loss}
\item and  a Gaussian prior introduced by the application of Bayes' theorem when representing $\probbnn$ by $\probbayes$, again similar to what is introduced for the BNN in~\SectRef{sec:data_net:bnn_loss}.\label{enum:re_prior}
\end{enumerate}

The gradient of \lre is computed relative to all network parameters $\theta$ in the ensemble.
The repulsive term in item \ref{enum:re_loss} includes a \emph{stop-gradient operation} on all $\theta$ in the ensemble, to assure valid gradients at each update.
This term and the prior are also normalised by $1/\DSsizeTrain$ to avoid any scaling with the size of the training sample \DSsizeTrain.
More insight on the formalism of the RE loss function is given in \AppRef{app:loss:re_loss}.

\section{Performance evaluations}
\label{sec:performance}

The performance of the learned \topo responses is fully evaluated for the BNN
predictions (\rclusbnn) by applying the trained model to a test dataset that
has no content overlap with the training dataset.
Principal measures to determine the quality of this calibration are prediction power, the signal linearity and the local energy resolution.
The median is used as a measure of the central tendency of the analysed distributions, because, unlike the mode, it is analytically accessible without explicit determining (fitting) the shape of the distribution around the most probable value.
The median is also relatively stable with respect to sample statistics, which can affect the statistical mean when distributions are not symmetric.
Emerging tails in distributions of the prediction power and signal linearity, both introduced as evaluation metrics in \SectRef{sec:perf-metrics} below, with increasing sample statistics were observed in some cases when evaluating them in bins of certain features or truth-related variables.
These tails lead to observable changes of the means with increasing sample sizes in these cases, while medians were found to be stable once statistically meaningful samples were collected.

In addition, the performance of the BNN-derived \topo calibration is compared to the one learned by the RE\@.
Both calibrations are expected to produce the same central predictions and thus yield the same median signal linearity and energy resolution.
This is a necessary requirement for a meaningful comparison of the uncertainties predicted by these two models.

\subsection{Evaluation metrics}
\label{sec:perf-metrics}

Determining the performance of the ML calibration cluster-by-cluster includes
the evaluation of its accuracy or \emph{prediction power}, similar to a closure test of a
given functional model to predict the target in any calibration attempt,
including non-machine-learning-based ones.\footnote{Closure tests are usually
performed using the same dataset with which the calibration was derived. In
the context of the machine-learning-derived calibration, an independent
dataset is used for this test.}
In addition, the \emph{signal linearity} achieved for the ML-calibrated energies are compared with the corresponding ones at other (calibrated) energy scales by determining their respective proportionality to the energy \eclusdep deposited in the \topo.
Both prediction power and signal linearity are trivially connected for each \topo, but show different distribution shapes for the test variables defined in \SectsRef{sec:perf-metrics:pred}{sec:perf-metrics:linear} below, leading to different measures of both centrality and width.

The energy resolution for individual \topos is determined before any local calibration and after application of the ML- and the LCW-based calibrations.
These corresponding comparisons allow identifying the best approach to mitigate calorimeter- and collision-environment-induced signal fluctuations affecting the cluster signal quality.
The immediate expectation here is that calibration functions that are highly sensitive to the signal source yield the best resolution power.

The learned uncertainties are only available from the BNN and the RE network, and are further discussed in \SectRef{sec:uncertainties}.

\subsubsection{Accuracy of the learned response}
\label{sec:perf-metrics:pred}

To evaluate the accuracy of the ML calibrations, the test dataset is used to compare the outputs \rcluskap of the networks $\kappa$ for individual \topos with the respective training target \rclusem.
The deviation from an accurate prediction \delrkap is introduced as a measure of the prediction power of the trained networks, with
\begin{align}
\label{eq:delta-r}
\delrkap = \frac{\rcluskap}{\rclusem}-1
\quad\mwith\quad
\kappa\in\braces{\text{DNN}, \text{BNN}} \;.
\end{align}
It is calculated for each cluster in the test dataset with the expectation to approach zero for accurate predictions (highest prediction power).
The prediction power of the DNN is taken from \RefRef{ATL-PHYS-PUB-2023-019}, which uses the same datasets.

\subsubsection{Signal linearity}
\label{sec:perf-metrics:linear}

A fully calibrated calorimeter signal is expected to be directly proportional to a reference energy, which can be the incoming particle energy or, as in the case of the \topo calibration,
the locally deposited energy \eclusdep introduced in \SectRef{sec:detdat:clusters:truth}.
The signal linearity can be measured cluster-by-cluster by \delekap, which quantifies its deviation from perfect linearity by\footnote{The prediction power represented by \delr and the signal linearity represented by \dele are related by the response definition given in \EqRef{eq:target-response} and the reconstruction of the calibrated energy given in \EqRef{eq:energy-recons}, such that $\dele = - \delr / (\delr + 1)$. This generally leads to two different distributions with possibly different central values, $\median{\delr} \neq \median{\dele}$, and widths for any selected sample of \topos, and thus to differences in the performance. Only when the response is perfectly learned, indicated by $\delr = 0$, a perfect signal linearity $\dele = 0$ is achieved for a given \topo.}
\begin{align}
\label{eq:delta-e}
\delekap = \frac{\ecluskap}{\eclusdep}-1
\quad\mwith\quad
\kappa\in\braces{\text{EM}, \text{had}, \text{DNN}, \text{BNN}}\;.
\end{align}
The considered \topo energy scales $\kappa$ are the basic EM scale given in \EqRef{eq:cluskine}, the hadronic (had) energy scale provided
in the LCW sequence presented in \SectRef{sec:detdat:clusters:kine},
and the ML calibrations from the DNN~\cite{ATL-PHYS-PUB-2023-019} and the BNN, with the respective response predictions applied according to \EqRef{eq:energy-recons} in \SectRef{sec:data_cal}.
The cluster signal at EM scale \eclusem is not expected to be linear, due to the non-compensating character of the ATLAS calorimeters and the absence of corrections for that at this scale.
Larger deviations of $\deleem$ from zero are expected for small $\eclusdep$, where ionisations increasingly dominate the energy loss, and in regimes where pile-up significantly adds to \eclusem.
Therefore, \eclusem is not considered to be calibrated, unlike all other considered energy scales, in the context of the results presented here.
Nevertheless, it provides a baseline for the signal-related \topo performance.

Calibrated energy scales are supposed to provide signals directly proportional to \eclusdep.
A residual level of fluctuations can be expected after \topo calibration, due to the highly stochastic nature of its response.
Nevertheless, the chosen measure of centrality, the median deviation from linearity \median{\dele}, is expected to be zero after calibrations are applied.
Any deviation from this expectation can affect the measurement of final state objects like particles, jets and event shapes composed of \topos in terms of the introduction of kinematic biases, an increase of systematic uncertainties and a potential degradation of the resolution.

\subsubsection{Relative local energy resolution}
\label{sec:perf-metrics:reso}

The measure of the relative local energy resolution
$\relres(\ecluskap/\eclusdep)$ employs the \perCent{68} inter-quantile range
$\qwidthnorm$ and the median of the respective
$\delekap+1=\ecluskap/\eclusdep$ distribution,
\begin{align}
\label{eq:reso}
\relres(\ecluskap/\eclusdep)
= \frac{\qwidthnorm(\ecluskap/\eclusdep)}{2\median{\ecluskap/\eclusdep}}
\quad \mwith \quad
\kappa\in\braces{\text{EM}, \text{had}, \text{DNN}, \text{BNN}}
\;,
\end{align}
on the various energy scales $\kappa$.
Its construction is based on \EqsRef{eq:qwidthnorm}{eq:resolution} given in \AppRef{app:metric:perf}.
\subsection{Prediction power}
\label{sec:performance:prediction_power}

The prediction power of the ML-learned \topo calibration can be evaluated cluster-by-cluster as functions of various variables available from MC simulations and as functions of any of the features from the feature set \fsetclus used for the training, for both the BNN-derived and the RE-derived calibrations.
A qualitative analysis allows comparing distribution shapes for the predictions from the networks with the corresponding distributions for the target, without the need to introduce statistical measures that may not describe the overall distribution well.
The quantitative analysis introduces measures for the central values of these distributions that allow comparing them directly using, e.g., \delr from \EqRef{eq:delta-r}.

\subsubsection{Distribution shapes}

\begin{figure}[tt!]
\centering
\subfloat[]{\includegraphics[width=0.32\textwidth]{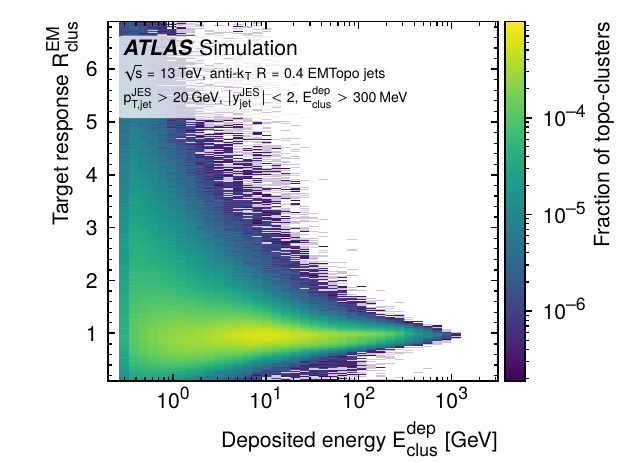}\label{fig:predpower:rems:edep}}
\subfloat[]{\includegraphics[width=0.32\textwidth]{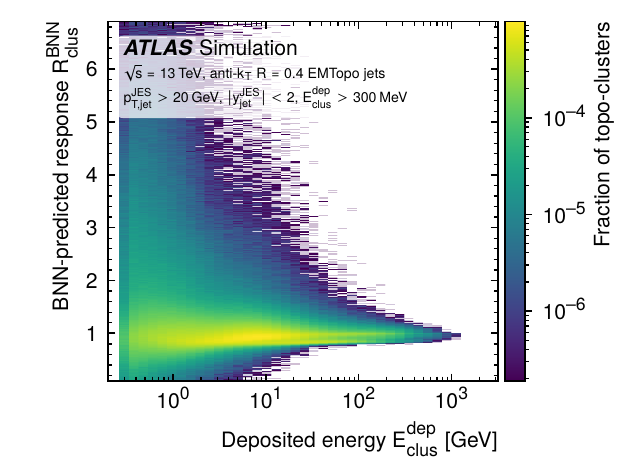}\label{fig:predpower:rbnn:edep}}
\subfloat[]{\includegraphics[width=0.32\textwidth]{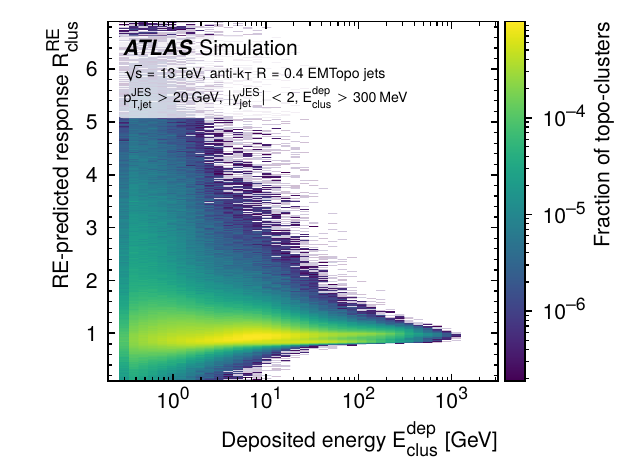}\label{fig:predpower:rrde:edep}}
\caption[]{The \topo response \rclus is evaluated cluster-by-cluster as a function of the energy \eclusdep deposited in the cluster. In \subref{fig:predpower:rems:edep},
the distribution of the response at EM scale \rclusem (the training target) is shown,
while \subref{fig:predpower:rbnn:edep} shows the distribution of the response \rclusbnn predicted by the trained BNN.
The corresponding distribution of the response \rclusre predicted by the RE is shown in \subref{fig:predpower:rrde:edep}.
The \topos are extracted from MC simulations, as described in \SectRef{sec:dataset:object}.}
\label{fig:predpower:truth}
\end{figure}

A qualitative comparison of the distributions of the training target \rclusem and the predictions from both the BNN and the RE, all as a function of the energy \eclusdep deposited in the \topo, is shown in \FigRef{fig:predpower:truth}.
The target distribution in \FigRef{fig:predpower:rems:edep} is reproduced well by both the BNN (\FigRef{fig:predpower:rbnn:edep}) and RE (\FigRef{fig:predpower:rrde:edep}) predictions.
Some small shifts in the populations between target and predictions can be observed in that the networks provide a sharper image of regions of higher population density.
These indicate that the learned \rclusem is dominated by the statistically better illuminated regions, thus giving a qualitative impression of the limitations of the networks to predict outliers.
With this respect, both the BNN and the RE network show very similar behaviour.

\begin{figure}[tt!]
\centering
\subfloat[]{\includegraphics[width=0.32\textwidth]{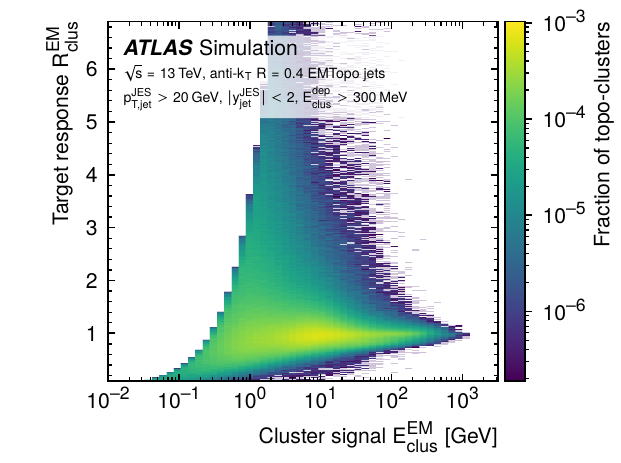}\label{fig:predpower:rems:eems}}
\subfloat[]{\includegraphics[width=0.32\textwidth]{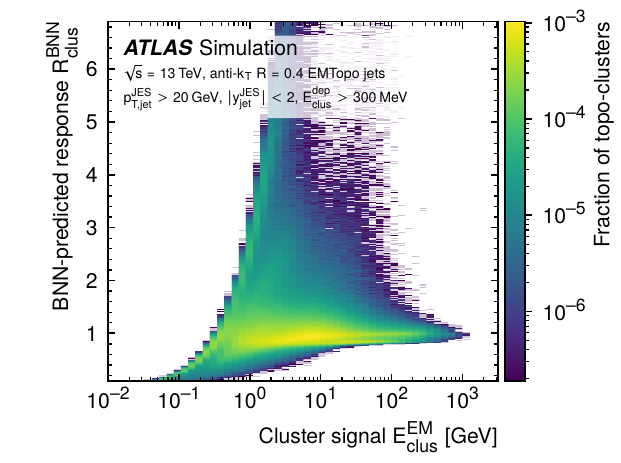}\label{fig:predpower:rbnn:eems}}
\subfloat[]{\includegraphics[width=0.32\textwidth]{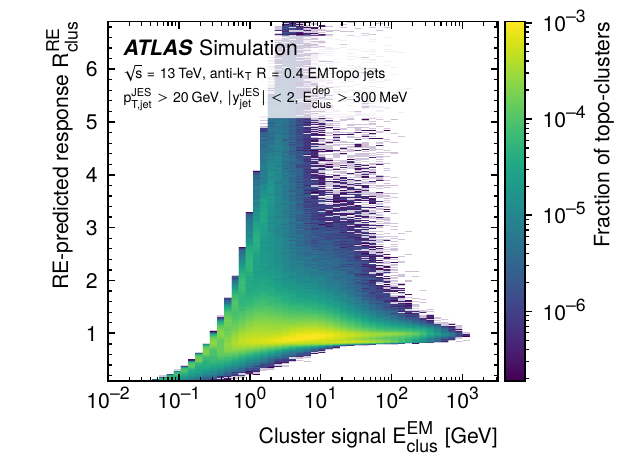}\label{fig:predpower:rrde:eems}}\\
\subfloat[]{\includegraphics[width=0.32\textwidth]{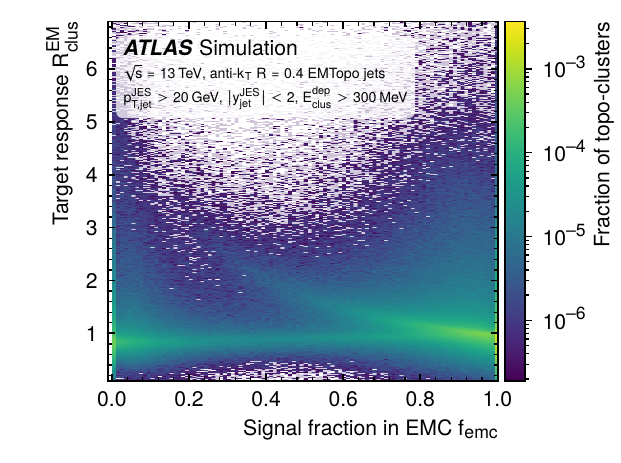}\label{fig:predpower:rems:mfem}}
\subfloat[]{\includegraphics[width=0.32\textwidth]{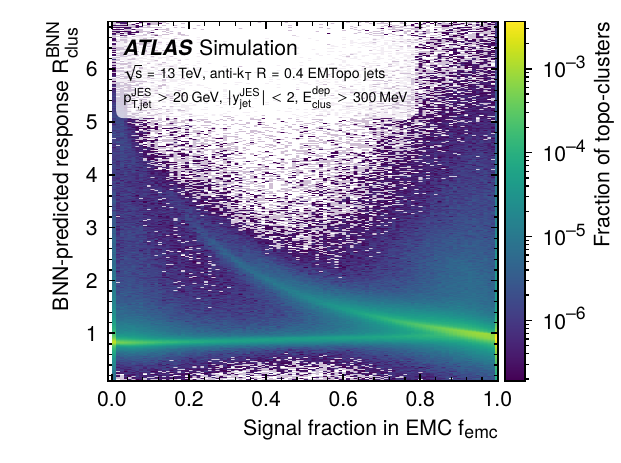}\label{fig:predpower:rbnn:mfem}}
\subfloat[]{\includegraphics[width=0.32\textwidth]{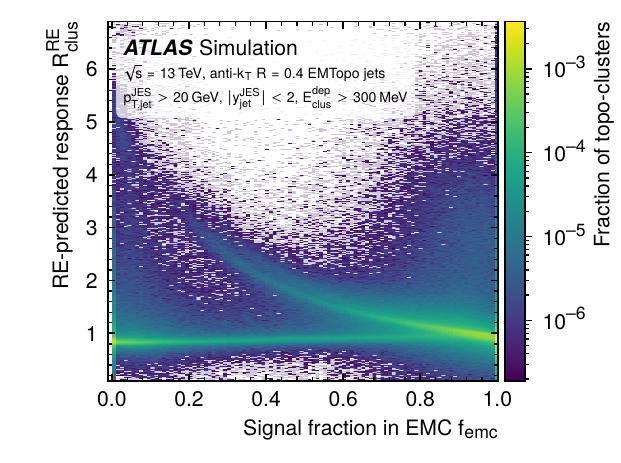}\label{fig:predpower:rrde:mfem}}\\
\subfloat[]{\includegraphics[width=0.32\textwidth]{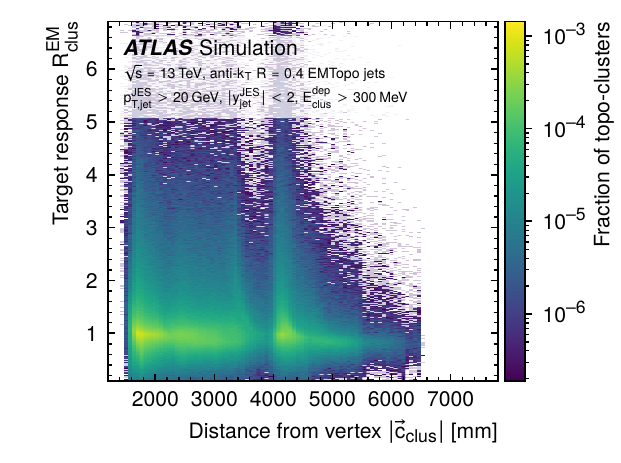}\label{fig:predpower:rems:mcog}}
\subfloat[]{\includegraphics[width=0.32\textwidth]{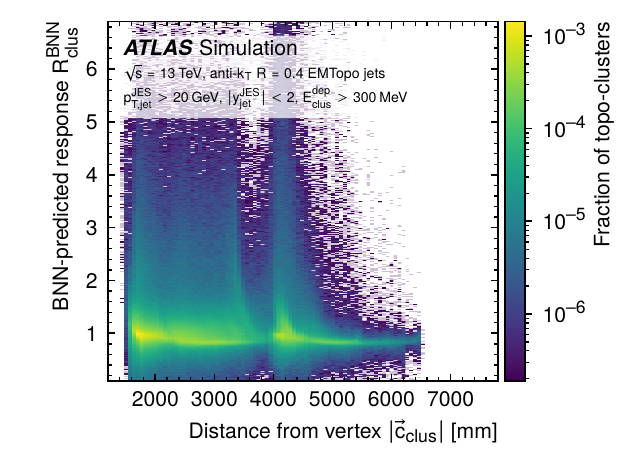}\label{fig:predpower:rbnn:mcog}}
\subfloat[]{\includegraphics[width=0.32\textwidth]{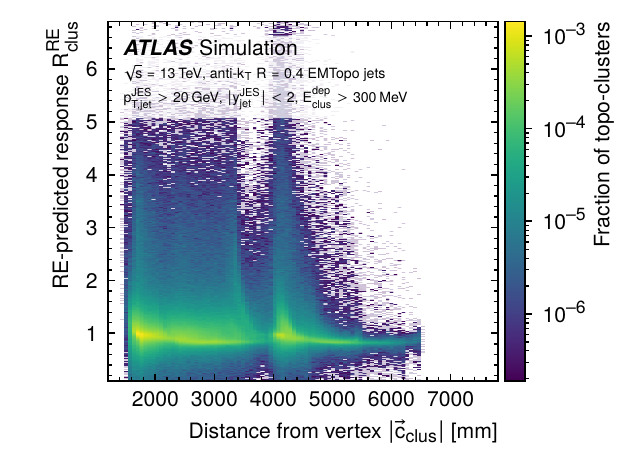}\label{fig:predpower:rrde:mcog}}
\caption[]{The distributions of the \topo response \rclusem and the corresponding predictions from the BNN (\rclusbnn) and the RE (\rclusre) are evaluated cluster-by-cluster as a function of selected features. The distributions of the training target
\rclusem are shown in
\subref{fig:predpower:rems:eems} as a function of the cluster signal \eclusem, in
\subref{fig:predpower:rems:mfem} as a function of the cluster signal fraction \clusfemc in the electromagnetic calorimeter, and in
\subref{fig:predpower:rems:mcog} as a function of the distance of the \topo \cog \cluscog.
The distributions of the predicted responses \rclusbnn and \rclusre as functions of the same features are respectively shown in
\subref{fig:predpower:rbnn:eems}, \subref{fig:predpower:rrde:eems} for \eclusem, in
\subref{fig:predpower:rbnn:mfem}, \subref{fig:predpower:rrde:mfem} for \clusfemc, and in
\subref{fig:predpower:rbnn:mcog}, \subref{fig:predpower:rrde:mcog} for \cluscog.
The \topos are extracted from MC simulations, as described in \SectRef{sec:dataset:object}. }
\label{fig:predpower:features}
\end{figure}

Further qualitative analyses of the BNN and RE predictions as functions of selected features used for the network training of the \topo response are collected in \FigRef{fig:predpower:features}.
As already observed in \FigsRef{fig:predpower:rbnn:edep}{fig:predpower:rrde:edep}, the learned responses show some differences at the edges of the densely populated ridges when compared with the corresponding target distributions and among themselves, and evaluated as a function of \eclusem.
This again is a display of \emph{statistical dominance} of the distribution regions with higher population densities in the derivation of the predictions.

The complex distributions of \rclusem as a function of the cluster signal fraction in the electromagnetic calorimeter \clusfemc shown in \FigRef{fig:predpower:rems:mfem} arise from the response characteristics and the calorimeter geometry.
The near horizontal ridge extending from $\rclusem\approx0.8$ at $\clusfemc = 0$ to $\rclusem\approx1.0$ for $\clusfemc=1$ shows the expected behaviour for a non-compensating calorimeter where a lower \clusfemc is indicative to an increased contribution from hadronic showers in \topos exclusively located in the hadronic calorimeter or extending from the electromagnetic to the hadronic calorimeter.
The upper ridge is populated by \topos with rising \rclusem for decreasing \clusfemc.
It is indicative of \topos with \eclusdep predominantly generated by ionisations in hadronic showers that increasingly extend into the Tile calorimeters.
The cluster signal \eclusem at EM scale overestimates \eclusdep because these ionizations generate larger signlsl than electromagnetic showers depositing the same \eclusdep in this calorimeter \cite{Adragna:2009zz,Ajaltouni:1996fx}.

\FigSubRangeRef{fig:predpower:rems:mcog}{fig:predpower:rrde:mcog} show the distributions of \rclusem, \rclusbnn and \rclusre as a function of the distance of the \topo \cog \cluscog from the nominal vertex in ATLAS\@.
The most prominent observed structure around $\cluscog\approx\SI{3900}{\mm}$ is introduced by the detector geometry.
It shows the effect of the transition from the central barrel to the \EndCap calorimeters.

The general observation from the comparison of the $\rclusem(\clusfemc)$ distribution in \FigSubRangeRef{fig:predpower:rems:mfem}{fig:predpower:rrde:mfem} and the $\rclusem(\cluscog)$ distribution in \FigSubRangeRef{fig:predpower:rems:mcog}{fig:predpower:rrde:mcog} is that the complex structures of the target distributions are well reproduced by both the BNN and the RE predictions, with some slightly sharpened edges around the ridges of higher \topo populations for the predictions due to the statistical dominance of these regions in the learning.

\subsubsection{Central prediction power}

\begin{figure}[tt!]
\centering
\subfloat[]{\includegraphics[width=1.00\figwidth]{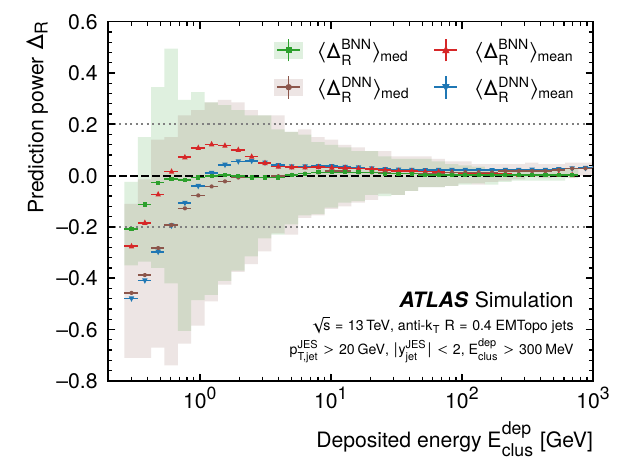}\label{fig:predpower:quantitative:edep}}\quad
\subfloat[]{\includegraphics[width=1.00\figwidth]{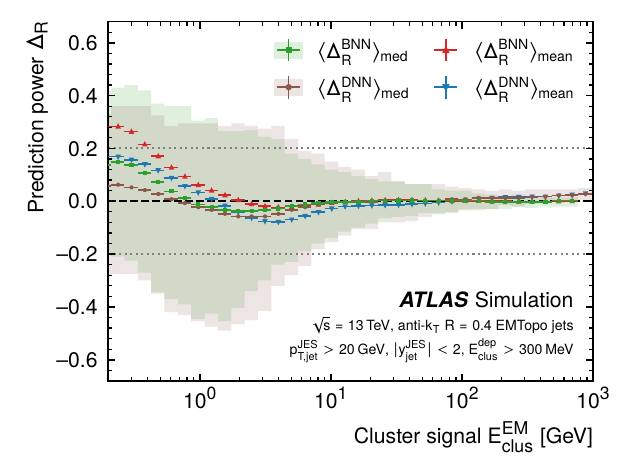}\label{fig:predpower:quantitative:eems}}\\
\subfloat[]{\includegraphics[width=1.00\figwidth]{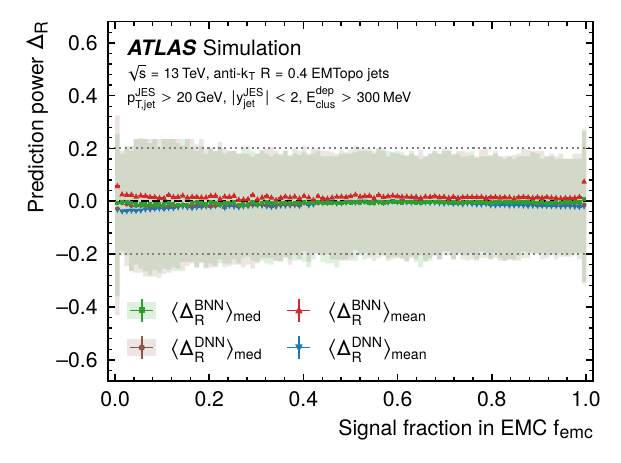}\label{fig:predpower:quantitative:mfem}}\quad
\subfloat[]{\includegraphics[width=1.00\figwidth]{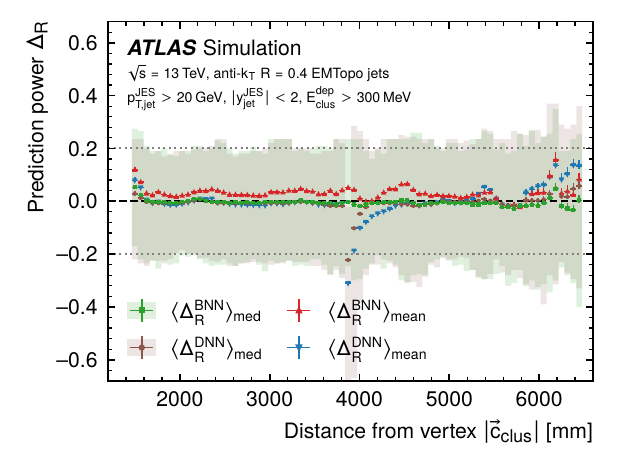}\label{fig:predpower:quantitative:mcog}}
\caption[]{The mean and median prediction power of the trained BNN (\delrbnn) and the trained DNN (\delrdnn) are compared as a function of
\subref{fig:predpower:quantitative:edep} the energy \eclusdep deposited in the \topo,
\subref{fig:predpower:quantitative:eems} the cluster signal \eclusem,
\subref{fig:predpower:quantitative:mfem} the cluster signal fraction in the electromagnetic calorimeter \clusfemc, and
\subref{fig:predpower:quantitative:mcog} the cluster \cog \cluscog.
The measures of centrality for the BNN are determined from the corresponding distributions shown in \FigRef{fig:predpower:rbnn:edep} and in
\FigRef{fig:predpower:rbnn:eems}, \subref{fig:predpower:rbnn:mfem} and \subref{fig:predpower:rbnn:mcog}, while the ones for the DNN are extracted from \RefRef{ATL-PHYS-PUB-2023-019}.
The \topos are collected from MC simulations, as described in \SectRef{sec:dataset:object}.
The shaded areas indicate the narrowest range of the respective \delr distributions containing \perCent{68} of all entries.
Statistical uncertainties are indicated by vertical bars if larger than the symbol size.
}
\label{fig:predpower:quantitative}
\end{figure}

Results from a quantitative analysis of the prediction power of the BNN \topo calibration model are collected in \FigRef{fig:predpower:quantitative}.
The metric used here is the relative difference \delr between learned \rclusbnn and the target \rclusem, as defined in \EqRef{eq:delta-r} in \SectRef{sec:perf-metrics:pred}.
It is measured cluster-by-cluster and analysed as functions of variables from MC simulations and the features in \fsetclus, using two different measures of centrality for the resulting \delr distributions.
For example, \FigRef{fig:predpower:quantitative} shows the mean and median prediction power of the BNN as function of \eclusdep, \eclusem, \clusfemc and \cluscog and compares those to the median and
mean prediction power of the DNN~\cite{ATL-PHYS-PUB-2023-019}.
The expectation from the particular (different) choices of loss functions for the BNN and the DNN training is that the mode $\mode{\delr} \approx 0$ for all \delr distributions in bins of features.
It is found that $\median{\delr} \approx \mode{\delr}$ for both networks and all evaluation scales, with the advantage that \median{\delr} is well defined and does not require a particular algorithm or peak shape fit to be determined.
The findings presented in \FigRef{fig:predpower:quantitative} show that the BNN has a similar if not stronger median prediction power than the DNN for all shown evaluation scales with the exception of \eclusem.
The mean prediction power is, as expected from the choice of the loss functions for the BNN and the DNN, not as strong as the median for all evaluation scales.

\subsubsection{Network training issues}
\label{sec:performance:issues}

About \perCent{3} of all \topos collected from the simulated calorimeter jets were found to be resistant to the BNN, RE, and DNN training.
These clusters are characterised by a target response $\rclusem \gg 1$.
This resistance potentially arises from a large \pu contribution that is not accounted for in \eclusdep but increases \eclusem significantly.
As the affected \topos are mostly observed in the hadronic calorimeters with considerably coarser readout geometry than the electromagnetic ones, this signal contribution from \pu at the cluster location may not be reflected in the features due to a limited (spatial) resolution.
The consequential lack of precision of the features can lead to a loss of expressiveness concerning the particular \topo response and thus a severely reduced prediction power.
In addition to the location, this interpretation is supported by the frequency of occurrence of these training-resistant \topos as a function of \pu, rising from about \perCent{0.2} of all clusters collected at $\npu \simeq 0$ (no \pu) to close to \perCent{10} of all clusters collected at $\npu \simeq 90$ (highest observed \pu level in the dataset).

A dedicated neural network was trained to tag the affected \topos and remove them from the training and test datasets for the results presented in this paper.
The ability to efficiently tag them, and to locate them in certain calorimeter regions under certain \pu conditions, supports the potential application of dedicated calibrations and corrections beyond the scope of the study presented here.
Further explorations are under way to address these issues.

\subsection{Central calibration performance}
\label{sec:performance:calib}

The prediction power discussed above is not a sufficient indicator for the accuracy and precision of the corresponding \topo calibration.
The deviation from signal linearity measure \dele introduced in \SectRef{sec:perf-metrics:linear} provides a direct scale to determine both, and thus the overall quality of the learned calibration.
This measure can be constructed at various energy scales.
For comparisons, the median deviation from linearity is evaluated for \topos from the test dataset as a function of \eclusdep and selected features for the basic EM scale (\deleem), the LCW hadronic scale (\delelcw),
the energy scale derived from the DNN calibration (\delednn), and the scale after applying the BNN-learned calibration (\delebnn).

\begin{figure}[tt!]
\centering
\subfloat[]{\includegraphics[width=1.00\figwidth]{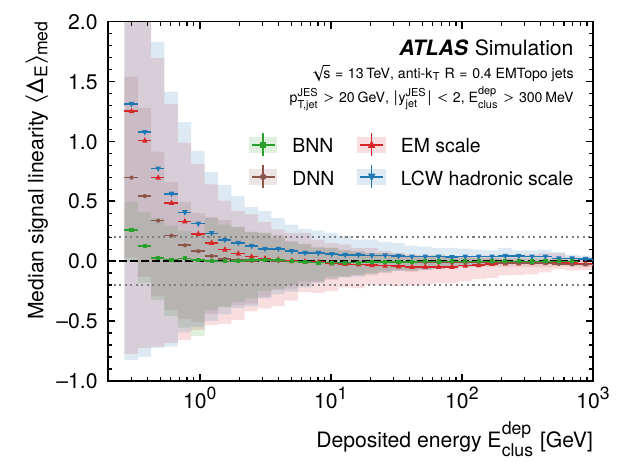}\label{fig:elinear:edep}}\quad
\subfloat[]{\includegraphics[width=1.00\figwidth]{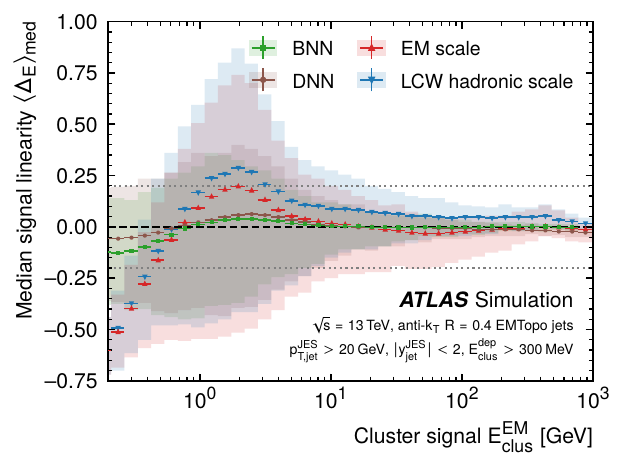}\label{fig:elinear:eems}}\\
\subfloat[]{\includegraphics[width=1.00\figwidth]{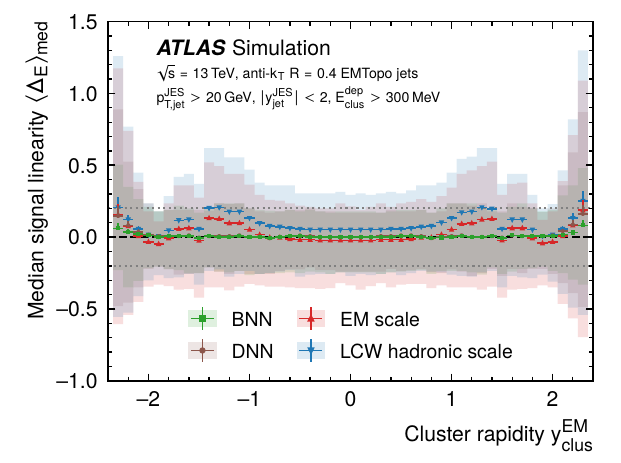}\label{fig:elinear:yems}}\quad
\subfloat[]{\includegraphics[width=1.00\figwidth]{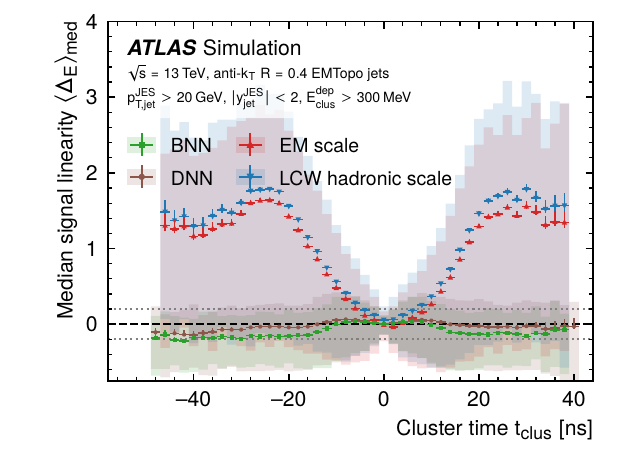}\label{fig:elinear:tems}}
\caption[]{The median deviation from linearity \median{\dele} for \topos at the basic EM, the LCW hadronic, the BNN-derived and the DNN-derived energy scales is shown as a function of
\subref{fig:elinear:edep} the energy \eclusdep deposited in the cluster,
\subref{fig:elinear:eems} the cluster signal \eclusem,
\subref{fig:elinear:yems} the cluster rapidity reconstructed at the basic EM scale \yclusem, and
\subref{fig:elinear:tems} the cluster signal time \tclus.
The \topos are extracted from MC simulations according to the description in \SectRef{sec:dataset:object}.
The shaded areas indicate the narrowest range of the respective \dele distributions containing \perCent{68} of all entries.
Statistical uncertainties are indicated by vertical bars if larger than the symbol size.
}
\label{fig:elinear:features:basics}
\end{figure}

\begin{figure}[tt!]
\centering
\subfloat[]{\includegraphics[width=1.00\figwidth]{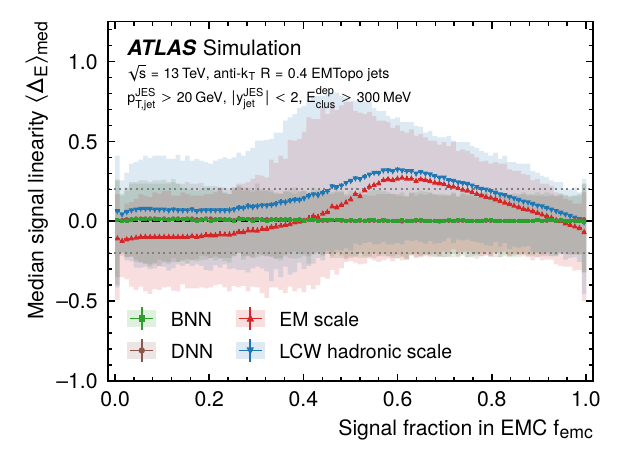}\label{fig:elinear:mfem}}\quad
\subfloat[]{\includegraphics[width=1.00\figwidth]{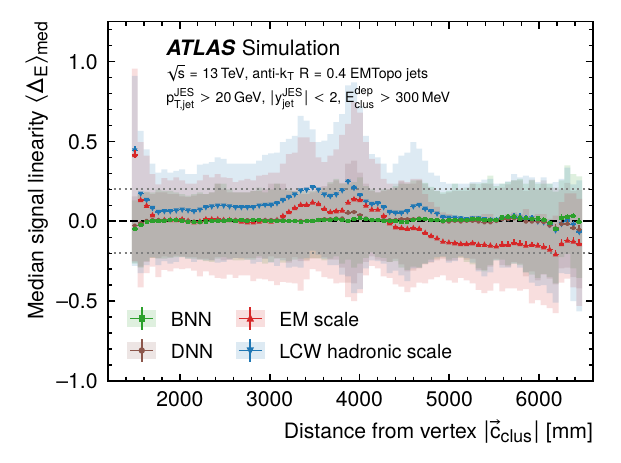}\label{fig:elinear:mcog}}\\
\subfloat[]{\includegraphics[width=1.00\figwidth]{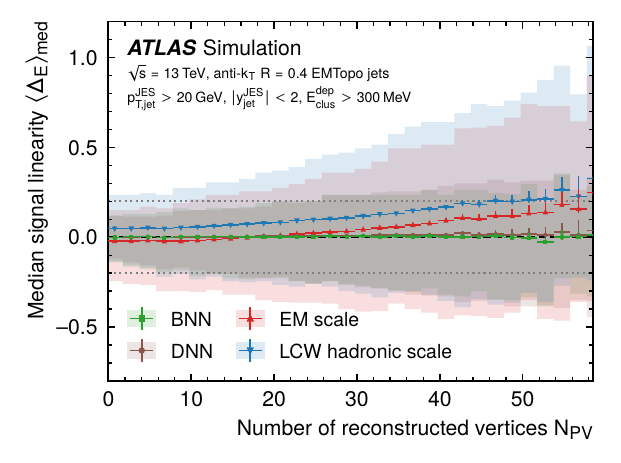}\label{fig:elinear:pnpv}}\quad
\subfloat[]{\includegraphics[width=1.00\figwidth]{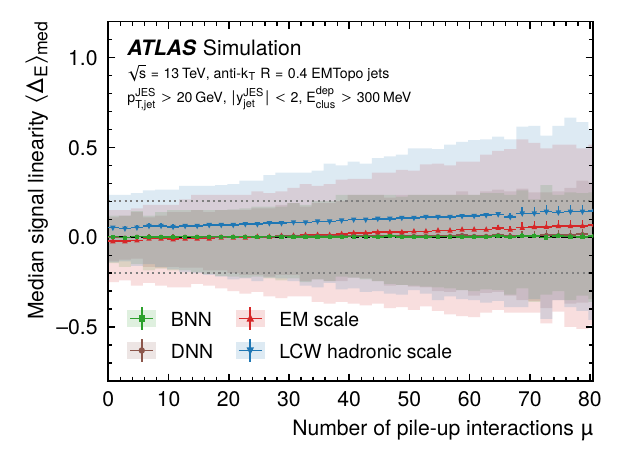}\label{fig:elinear:pmmu}}
\caption[]{The median deviation from signal linearity \median{\dele} for \topo signals at the basic EM, the LCW hadronic, the BNN-derived and the DNN-derived energy scales is shown as a function of
\subref{fig:elinear:mfem} the cluster signal fraction in the electromagnetic calorimeter \clusfemc,
\subref{fig:elinear:mcog} the distance of the cluster \cog from the nominal collision vertex \cluscog,
\subref{fig:elinear:pnpv} the number of reconstructed primary vertices \npv (in-time \pu measure), and
\subref{fig:elinear:pmmu} the number of \pu collisions in the bunch crossing \npu (out-of-time \pu measure).
The results are achieved with \topos from the test dataset, extracted as described in \SectRef{sec:dataset:object}.
The shaded areas indicate the narrowest range of the respective \dele distributions containing \perCent{68} of all entries.
Statistical uncertainties are indicated by vertical bars if larger than the symbol size.
}
\label{fig:elinear:features:all}
\end{figure}

\FigRef{fig:elinear:features:basics} compares \dele on the various energy scales, as a function of selected \topo features and \eclusdep.
As seen in \FigRef{fig:elinear:edep}, the BNN-derived calibration extends the
region of accurate energy reconstruction with $|\median{\delebnn}| \lesssim
\perCent{2}$ for $\eclusdep \gtrsim \SI{500}{\MeV}$ down from the next-best
calibration based on the DNN predictions, where a similar median deviation
from linearity is reached for $\eclusdep \gtrsim \SI{1}{\GeV}$.
This is a significant gain of accuracy in the low energy regime.

When evaluated as a function of the \topo signal \eclusem, the accuracy of both ML calibrations is significantly improved compared to the standard LCW-based hadronic calibration and the EM scale energy reconstruction, as is shown in \FigRef{fig:elinear:eems}.
The accuracy is not as high as observed in \FigRef{fig:elinear:edep}, which is likely due to a residual, unmitigated and highly stochastic contribution to smaller signals that is mostly introduced by \pu dominating \eclusem in this region.
Nevertheless, both machine-learning based calibrations show significant less deviation from linearity than the standard LCW hadronic energy scale, which is derived from a calibration that uses \eclusem as input as well.

Both \eclusbnn and \eclusdnn provide an accurate median measure of \eclusdep across the illuminated cluster rapidity \yclusem space, as shown in \FigRef{fig:elinear:yems}.
Some increase of both \median{\delebnn} and \median{\delednn} can be observed at the boundaries of this space at $|\yclusem|\approx 2.5$.
The \topos reconstructed at these edges can extend beyond the $|\etadet| = 2.5$ boundary, where the cell readout granularity of the ATLAS calorimeters is reduced by a factor of four.
The increase in cell size potentially reduces the expressiveness of some features for these clusters.
This leads to a loss of central accuracy, mainly due to a loss of sensitivity of the affected features that mitigate stochastic effects on \rclusem.
A larger deviation from linearity measured by $\median{\delelcw}$ at the hadronic LCW scale, already observed in \FigRef{fig:elinear:eems} for $\median{\delelcw}(\eclusem)$, is seen for
$\median{\delelcw}(\yclusem)$ in \FigRef{fig:elinear:yems} as well for $|\yclusem|\lesssim 2$, with maxima for \topos located in the complex calorimeter transition region at $|\etadet|\approx 1.5$.
Both BNN and DNN learned calibrations measure \eclusdep in this transition region well, indicating that the feature expressiveness is sufficient for an accurate calibration.

\FigRef{fig:elinear:tems} shows the significant deviation from linearity for the basic \topo signal \eclusem, and similarly for \eclushad, for \topo signal times \tclus reconstructed outside of the bunch crossing time window of $\SI{-12.5}{\ns} < \tclus < \SI{12.5}{\ns}$.
In particular $\median{\deleem}(\tclus) > 1$ hints on the cluster response \rclusem being dominated by out-of-time \pu signal contributions.
This leads to a slightly increased $\median{\delelcw}(\tclus) > \median{\deleem}(\tclus)$ for $|\tclus|\gtrsim \SI{12.5}{\ns}$, as the hadronic calibration from LCW is agnostic to \tclus.

Both BNN and DNN trained calibrations learn $\rclusem(\tclus)$ well, which leads to $\median{\delebnn}(|\tclus|<\SI{12.5}{\ns})\approx 0$ and $\median{\delednn}(|\tclus|<\SI{12.5}{\ns})\approx 0$.
Outside of this time window, the BNN-derived calibration shows a near-constant median deviation from linearity by underestimating \eclusdep by about \perCent{20}.
The DNN-derived calibration shows a better performance for $\tclus > \SI{12.5}{\ns}$, with a deviation from linearity within a few percent, indicating that the signal contributions from the following bunch crossings are
mitigated well.

Most of the features in \fsetclus are not involved in the reconstruction of both \eclusem and \eclushad, at least not directly as inputs.
Extending the dimensionality of the local \topo calibration, which is well supported by the machine learning applications, allows inclusion of information reflecting the cluster source and the \pp collision environment.
In addition to the results presented in \FigRef{fig:elinear:features:basics}, the devation from linearity is evaluated cluster-by-cluster for each feature in the full feature space represented in the test dataset.
\FigRef{fig:elinear:features:all} shows a selection of results for some of the features that both the EM scale and the hadronic scale from LCW are agnostic to.
The results for the remaining features are collected in \AppRef{app:calibration:central}.

A median calibration accuracy of better than \perCent{2} is achieved with both the BNN and the DNN nearly everywhere, when evaluated over the whole \clusfemc, \cluscog, \npv and \npu value ranges.
This indicates that both networks learn the response well from these features, even if those features are distributed in a complex way, as seen for \clusfemc and \cluscog in \FigRef{fig:predpower:quantitative:mfem} and \ref{fig:predpower:quantitative:mcog}, for example.
In addition, the presence of a significant \pu activity that increases the stochasticity of the \topo signal and its features is well mitigated when the corresponding \pu measures \npv and \npu are included in the (ML-learned) calibration.

\subsection{Local \topo energy resolution}
\label{sec:performance:resolution}

\begin{figure}[tt!]
\centering
\subfloat[]{\includegraphics[width=1.00\figwidth]{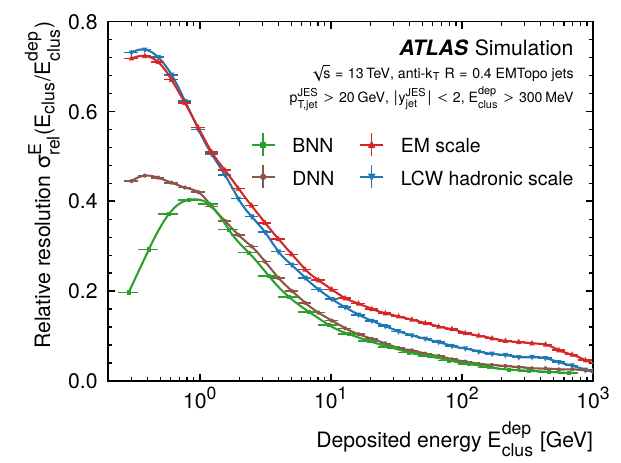}\label{fig:reso:edep}}\quad
\subfloat[]{\includegraphics[width=1.00\figwidth]{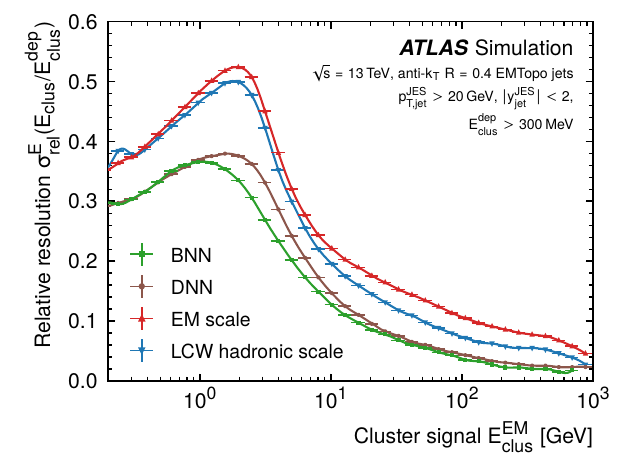}\label{fig:reso:eems}}\\
\subfloat[]{\includegraphics[width=1.00\figwidth]{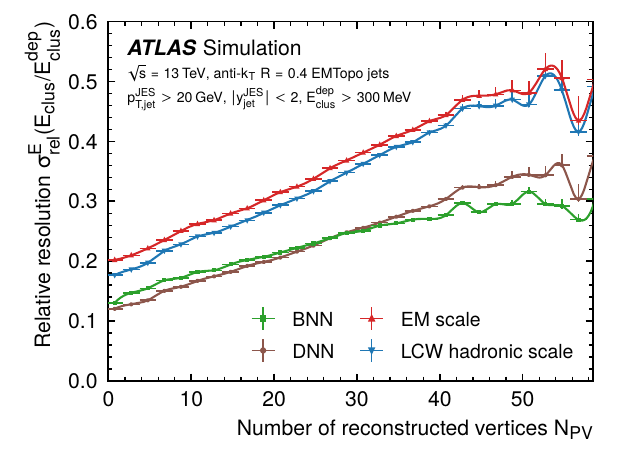}\label{fig:reso:pnpv}}\quad
\subfloat[]{\includegraphics[width=1.00\figwidth]{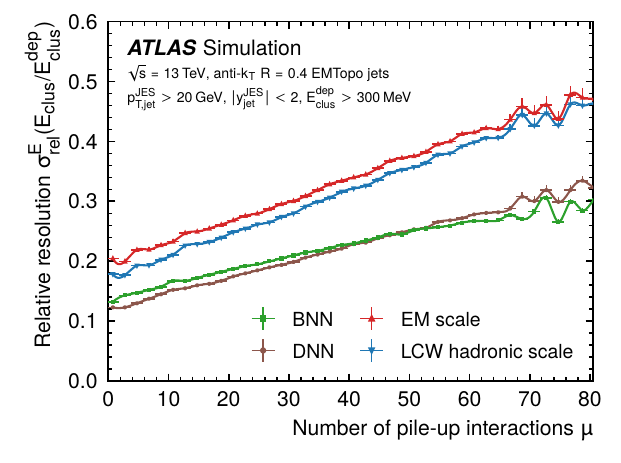}\label{fig:reso:pmmu}}
\caption[]{The relative local energy resolution of \topos at the basic EM, the LCW hadronic, the BNN-derived and the DNN-derived energy scales, measured using \EqRef{eq:reso}, as a function of
\subref{fig:reso:edep} the energy deposited in the cluster \eclusdep,
\subref{fig:reso:eems} the basic cluster signal \eclusem,
\subref{fig:reso:pnpv} the number of reconstructed vertices \npv, and
\subref{fig:reso:pmmu} the number of \pu collisions \npu in the bunch crossing.
The \topos are extracted from the test dataset as discussed in \SectRef{sec:dataset:object}.
Statistical uncertainties are indicated by vertical bars if larger than the symbol size.
}
\label{fig:reso:all}
\end{figure}

The relative local energy resolution of \topos is evaluated at the basic EM, the hadronic scale from LCW, and for the BNN- and DNN-derived calibrations.
It is measured using $\relres(\ecluskap/\eclusdep)$ introduced in \EqRef{eq:reso} in \SectRef{sec:perf-metrics:reso}, where $\kappa$ names the energy scale.

\FigRef{fig:reso:all} shows $\relres(\ecluskap/\eclusdep)$ as a function of \eclusdep, \eclusem, and the \pu measures \npv and \npu.
Both ML calibrations yield a significantly better relative resolution than the standard local hadronic calibration, together with a significant improvement relative to the EM scale resolution, on all
evaluation scales.
This is expected from the larger number of features entering the BNN and the DNN and the resulting increased exposure of these calibrations to more information concerning the \topo signal source,
in combination with the exploitation of feature correlations lost for other energy scale calibrations, and the learned smooth calibration functions.

The steeper downturn of the calibrated signal fluctuations for the BNN-derived calibration observed in \FigRef{fig:reso:edep} for $\eclusdep \lesssim \SI{1}{\GeV}$ indicates that this model best learns the transition of the signal source from inelastic hadronic interactions to more ionisation-dominated signals.\footnote{Hadrons can traverse relatively large distances in matter, if their energy
is below the typical threshold of about \SI{1}{\GeV} for inelastic interactions, according to \RefRef{Wigmans:1986pe}.
This suggests that energy deposits in \topos within $0 < \eclusdep \lesssim \SI{1}{\GeV}$ reflect a significant contribution to \eclusdep from energy loss mechanisms not involving hadronic inelastic interactions, like ionisations
by charged particles.}
This is likely associated with the use of a Gaussian mixture model for \probbayes in the BNN loss function, as described in
\EqRef{eq:bnn_gmix} of \SectRef{sec:data_net:bnn_loss}, among other possible differences between the BNN and the DNN\@.
The comparisons of the individual shapes of the \delebnn and the \delednn distributions in bins of \eclusdep show significantly suppressed tails in case of the BNN-derived calibrations.
This is expected for signals from ionisation losses, which are subject to considerably less intrinsic fluctuations than signals from energy loss mechanisms occurring in inelastic processes.
A representative selection of these comparisons is shown in \FigRef{fig:reso:spectra} in \AppRef{app:calibration:reso}.

The relative \topo energy resolution after application of the ML-based calibrations is increased, compared to the corresponding resolutions obtained for the standard energy scales, when analysed as a function of the in-time \pu measure \npv in \FigRef{fig:reso:pnpv}, and as a function of the out-of-time \pu measure \npu in \FigRef{fig:reso:pmmu}.
The increased but still partial cluster-by-cluster \pu mitigation is manifested by the reduced relative signal fluctuations observed at any given \npv and \npu when compared with the other energy scales.
The fluctuations after applying the BNN-derived \topo calibration show a more flat dependence on \npv than the residual fluctuations after application of the DNN-based calibration, with a slightly worse performance at low \npv (low \pu) but a better performance at higher \npv (higher \pu).
A similar, but less pronounced behaviour is observed in \FigRef{fig:reso:pmmu},  where the relative energy resolutions achieved with the BNN- and DNN-derived calibrations are shown as a function of \npu.

\section{Analyzing the uncertainty predictions}
\label{sec:uncertainties}

The uncertainty predictions from both the BNN and the RE are based on predictions sampled from a model representing an ensemble of trained networks.
In case of the BNN, this ensemble is formed by sampling the network parameters at inference from variational distributions with learned parameters,
as schematically laid out in \FigRef{fig:bnn_sketch} in \SectRef{sec:data_net:bnn_design}.
For the RE, the ensemble is a collection of networks interconnected by a repulsive term during training, as discussed in \SectRef{sec:data_net:re_design}.

\subsection{Predictive uncertainties}
\label{sec:uncertainties:predictions}

Learned uncertainties are usually referred to as \emph{predictive uncertainties}~\cite{Gal:2015xur,kendall2017uncertainties, Abdar_2021}.
They are divided into two major categories,
\begin{enumerate}
\item \emph{epistemic} or \emph{model uncertainties} that are related to how machines can learn and what cannot be learned, and
\item \emph{aleatoric} or \emph{data uncertainties} that are related to the residual level of stochasticity in the data that cannot be mitigated any further by the model.
\end{enumerate}
The expectation is that the epistemic uncertainties, after the final network optimisation for the \topo calibration, are dominated by contributions inherent to the data that limit what the BNN can learn.
In particular, contributions to these uncertainties are possible from a statistically insufficient coverage of the whole feature space, or certain regions of it, by \topos in the training dataset.
These contributions are thus expected to largely vanish in the limit of infinite training data and a sufficient number of epochs.
As said above, the aleatoric uncertainties reflect the lack of expressiveness of the data introduced by a relatively high level of noise predominantly introduced by \pu that cannot be overcome by increasing the training dataset or improving the network model.

\begin{table}
\centering
\caption[]{The principal qualitative relation between aleatoric and epistemic predictive uncertainties in machine-learning applications and the systematic and statistical uncertainty predictions for the \topo calibration. These relations are context-driven.}
\label{tab:unc}
\newlength{\wsp}\settowidth{\wsp}{\textbf{Alea}}
\newlength{\cola}\settowidth{\cola}{\ Residual level of stochasticity in dataAleaAlea\ }
\newlength{\colb}\settowidth{\colb}{\ Uncertainty}
\newcolumntype{P}{>{\raggedright\arraybackslash}p{\cola}}
\newcolumntype{E}{>{\raggedright\arraybackslash}p{\colb}}
\newcolumntype{M}{>{\raggedright\arraybackslash}X}
\renewcommand{\arraystretch}{1.2}
\begin{tabularx}{0.98\textwidth}{@{}P@{}@{\hspace{-0.1em}}EM@{}}
\hline\hline
Predictive uncertainty                            & Uncertainty component & Mitigation strategy \\
\hline
\emph{Epistemic (model) uncertainties} &  & \\
\hspace{\wsp} Network architecture\ \dotfill                                     & \ systematic & network configuration optimisation \\
\hspace{\wsp} Hyper-parameter choice\ \dotfill                       &  \ systematic & network configuration optimisation \\
\hspace{\wsp} Lack of expressitivity of data\ \dotfill                 &  \ systematic & feature choices \\
\hspace{\wsp} Insufficient amount of data\ \dotfill                  &   \ statistical   & sufficiently large training dataset \\
\hline
\emph{Aleatoric (data) uncertainties} &  & \\
\hspace{\wsp} Residual level of stochasticity in data\ \dotfill\  & \ systematic & none (irreducible noise) \\
\hline\hline
\end{tabularx}
\end{table}

The predictive uncertainty of a trained network model is a measure for its lack of expressive power arising from either of the two categories of uncertainties introduced above.
This ambiguity can be resolved by relating the learned uncertainties to the standard categories of statistical and systematic uncertainties that are usually associated with the application of a calibration model, or with a measurement in general.
The chosen convention here follows the general expectation that the statistical uncertainty vanishes in the limit of infinite training data, and the systematic uncertainty remains finite at this limit.
When evaluating the quality of the network, both the statistical and the systematic uncertainty contribute to the learned uncertainty as a systematic component in the sense that evaluating the network many times, without changing the training dataset size, does not reduce them.
They are thus typically considered as external nuisance parameters associated with the specific network configuration and training conditions.
The principal contributions of the predictive uncertainties to the uncertainty predictions associated with the \topo calibration are collected in \TabRef{tab:unc}.

\subsection{Uncertainty components}
\label{sec:uncertainties:components}

Separately analysing the statistical (\uncstat) and systematic (\uncsyst) component of the total learned uncertainty \unctotal requires a sufficiently large training dataset.
It is possible for both the BNN and the RE network employed here, as they use the same large training dataset introduced in \TabRef{tab:hyperparameters} of \SectRef{sec:data_net}.
This means that
\begin{align}
\unctotal = \sqrt{\uncsystsq+\uncstatsq}\;.
\label{eq:unc_total}
\end{align}
Following the previous discussion,  \uncsyst and \uncstat are characterised by their behaviour in the limit of an infinite training dataset size \DSsizeTrain,
\begin{align*}
\DSsizeTrain\rightarrow\infty\;\Rightarrow\;\left\{\begin{array}{ll}\uncsyst\rightarrow\text{const.}>0
\\ \uncstat\rightarrow 0 \end{array} \right.\;.
\end{align*}
With this understanding, \uncsyst represents the limit of the accuracy to which the model learns to describe the target with the training data available.
In case of a non-deterministic or stochastic target (noisy data)~\cite{Kasieczka:2020vlh}, like \rclusem, limited expressiveness of the network (structure uncertainty)~\cite{Badger:2022hwf}, and non-optimal choices of hyper-parameters (uncertainty in model parameters) can all be interpreted in the spirit of contributions to the systematic uncertainty inherent to the calibration model and the training data.
It has contributions to both epistemic and aleatoric uncertainty components, as qualitatively summarised in \TabRef{tab:unc}.

\begin{figure}[tt!]
\centering
\includegraphics[width=1.00\figwidth]{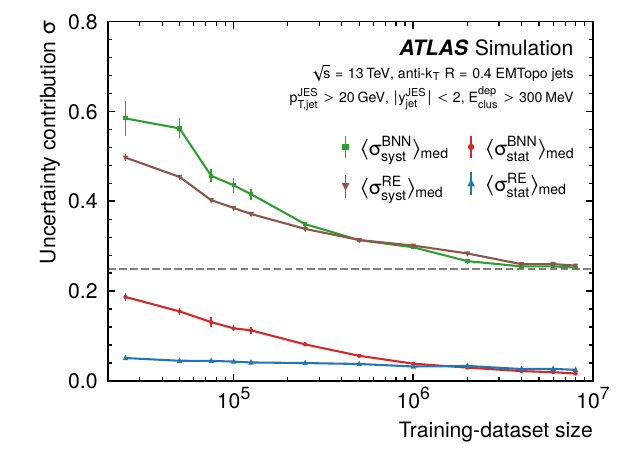}
\caption[]{The dependence of the predicted uncertainty components on the training dataset size. Shown are the median uncertainty components discussed in \SectRef{sec:uncertainties} for training datasets of
varying sizes and their specific asymptotic behaviours.
The \topos used here are collected into training datasets employing the selections shown in \SectRef{sec:dataset:object}. The statistical uncertainties on the medians are indicated as vertical bars when exceeding the symbol size.
The dashed horizontal line indicates the asymptotic limit of the systematic uncertainties.
}
\label{fig:unc:components}
\end{figure}

The statistical  uncertainty component \uncstat is constructed to become insignificantly small with increasing \DSsizeTrain, and is thus expected to become irrelevant for a sufficiently large training dataset.
Therefore, \uncstat can be considered as a significant part of the epistemic uncertainties.
A brief and more formal introduction to the calculation of \uncsyst and \uncstat can be found in \AppRef{app:metric:unc:components}.

The asymptotic behaviours of the uncertainty components as a function of the training dataset size are summarised in \FigRef{fig:unc:components}.
The figure shows the two BNN components (\uncbnnsyst, \uncbnnstat) and the two RE (\uncresyst, \uncrestat) components separately.
The results discussed in the following were achieved with a training dataset where \DSsizeTrain was sufficiently large to assume $\uncsyst \gg \uncstat \Rightarrow \unctotal = \uncsyst$.

\subsection{Comparison of the learned uncertainties}
\label{sec:uncertainties:comparisons}

\begin{figure}
\centering
\subfloat[]{\includegraphics[width=1.00\figwidth]{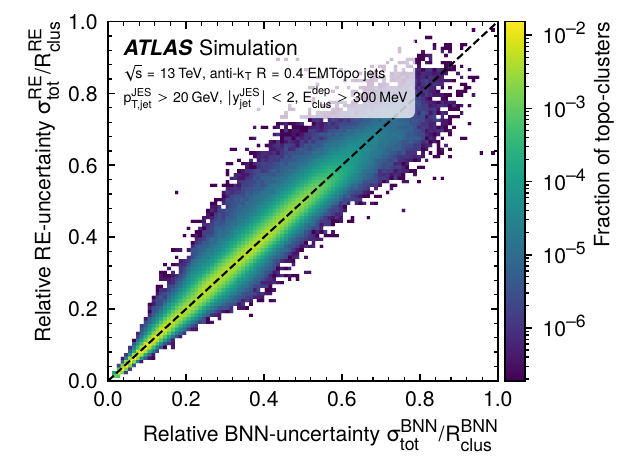}\label{fig:unc:compare:scatter}}\quad
\subfloat[]{\includegraphics[width=1.00\figwidth]{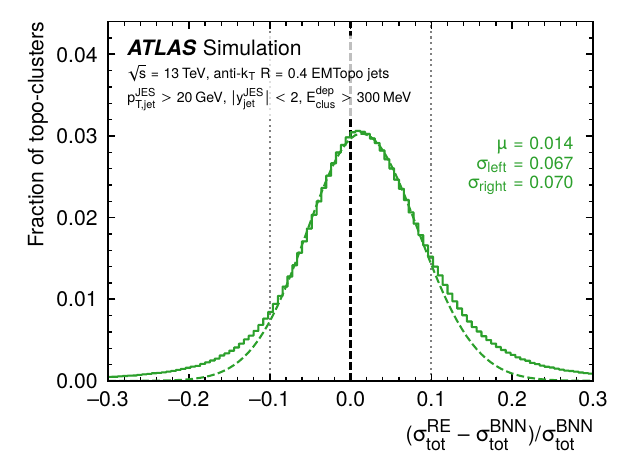}\label{fig:unc:compare:inclusive}}
\caption[]{In \subref{fig:unc:compare:scatter}, the relative total uncertainty prediction $\uncretotal/\rclusre$ from the RE is shown as a function of the relative uncertainty
$\uncbnntotal/\rclusbnn$ from the BNN\@. The distribution of the relative difference between the two uncertainties $(\uncretotal-\uncbnntotal)/\uncbnntotal$ is shown in \subref{fig:unc:compare:inclusive}.
For illustration, the dashed (centre) line indicates the completely unbiased mean of the distribution, and the dotted lines indicate the \perCent{10} range around it.
The dashed curve shows the result of a two-sided fit of Gaussian functions to the distribution, with a common mean $\mu$ and the independently fitted widths $\sigma_{\text{left}}$ and $\sigma_{\text{right}}$.
Both uncertainties are predicted for each \topo in the test dataset. These clusters are collected from the simulated jets, as described in \SectRef{sec:dataset:object}.
}
\label{fig:unc:compare}
\end{figure}

The uncertainty predictions provided along with the learned \topo calibration need to be analysed and interpreted to rule out that they are artefacts introduced by the machine learning model itself.
Rather, they are supposed to reflect limitations in the data that lower the ability of the applied network model to accurately regress on the target, in addition to limitations introduced by the model itself.

To address these concerns about specific network-introduced artefacts, the individual uncertainty predictions from the two different network models BNN and RE are directly compared cluster-by-cluster.
\FigRef{fig:unc:compare:scatter} shows the correlation between the relative learned uncertainties of the BNN ($\uncbnntotal/\rclusbnn$) and the RE ($\uncretotal/\rclusre$).
As by construction the systematic uncertainties are not deterministic, a certain spread can be expected and is observed.
Nevertheless, there is no sign of a bias in this correlation, the two predictions track each other well.

\FigRef{fig:unc:compare:inclusive} shows the distribution of the relative differences $(\uncretotal-\uncbnntotal)/\uncbnntotal$ between the (absolute) BNN and the RE uncertainty predictions.
A two-sided fit of Gaussian functions yields a near symmetric shape, indicating an agreement between the two estimates at the level of \perCent{10}.
Differences to the fitted Gaussian functions in the tails between \perCent{10} and \perCent{30}, and \perCent{-30} and \perCent{-10}, are fairly identical and thus unbiased.
They are expected for highly stochastic data like the response and the reconstructed features from \topos with significant near-random contributions from the high levels of \pu experienced in ATLAS at LHC \RunTwo operations.
This high level of comparability between the learned uncertainties shows that those predominately reflect features of the data, and not artefacts introduced by either network.

\subsection{Uncertainties and data spread}
\label{sec:uncertainties:spread}

\begin{figure}[tt!]
\centering
\subfloat[]{\includegraphics[width=1.00\figwidth]{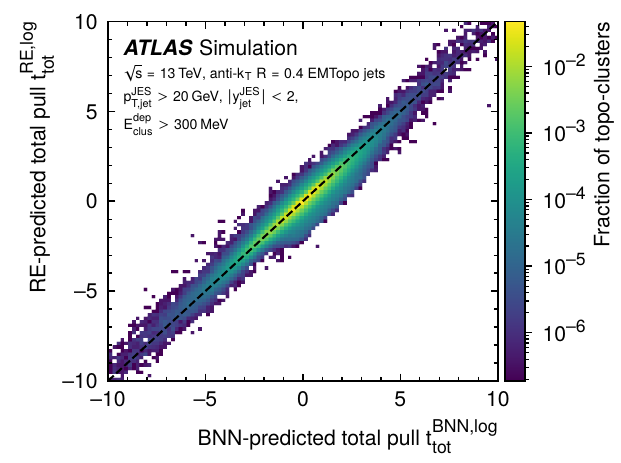}\label{fig:unc:pull:scatter}}\quad
\subfloat[]{\includegraphics[width=1.00\figwidth]{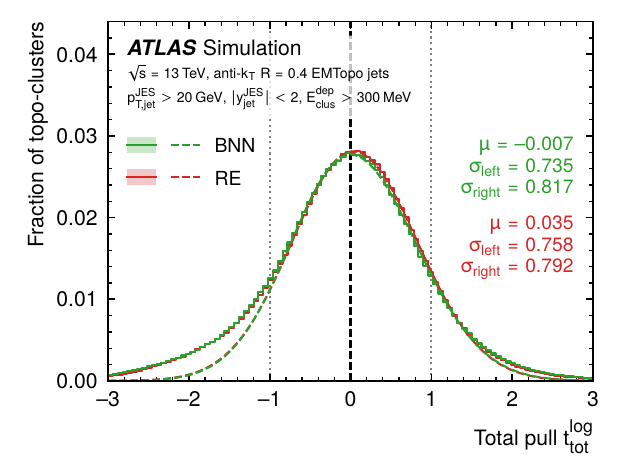}\label{fig:unc:pull:spectra}}
\caption[]{The observed pull from the RE training (\logpullre) is shown as a function of the one from the BNN training (\logpullbnn) in \subref{fig:unc:pull:scatter}. The inclusive distributions of \logpullbnn
and \logpullre are compare in \subref{fig:unc:pull:spectra}. Both \logpullbnn and \logpullre are calculated using \EqRef{eq:logpull}, for a test dataset containing \topos collected from full MC simulations,
as described in \SectRef{sec:dataset:object}.
The dashed vertical line indicates the unbiased centre of the \logpull distributions, while the dotted lines show the $\pm 1$ range around it.
The dashed curves indicate the two-sided fits of Gaussian functions for each network, with the common average $\mu$ and the independently fitted widths $\sigma_{\text{left}}$ and $\sigma_{\text{right}}$.
The corresponding fitted quantities are quoted.
}
\label{fig:unc:pull}
\end{figure}

The interpretation from the previous section are augmented with the analysis of the predicted value of the uncertainty for any given \topo.
Such an analysis can be done using the \emph{pull} \pullbnnretotal,
\begin{align}
\label{eq:pull}
\pullbnnretotal(\fsetclus) = \frac{\rclusbnnre(\fsetclus)-\rclusem}{\uncbnnretotal(\fsetclus)}\;,
\end{align}
that measures the difference between the prediction \rclusbnnre from the BNN or RE and the target \rclusem on a scale defined by the respective predicted uncertainty \uncbnnretotal.
Technically, the training is implemented such that it produces predictions for $\logten\rclusbnnre$ for the target representation  $\logten\rclusem$ and the uncertainty $\sigma_{\tagtotal}^{\text{\BNN(\RE),log}}$ at this transformed scale.
The corresponding pull \logpullbnnre is related to the pull in the linear response space given in \EqRef{eq:pull},\footnote{This relation uses $\sigma_{\tagtotal}^{\text{log}} \approx \frac{d\logten R}{dR} \cdot \unctotal  = \frac{1}{\ln 10\cdot R}\cdot\unctotal$.}
\begin{align}
\begin{split}
\logpullbnnre(\fsetclus) & = \frac{\logten\rclusbnnre(\fsetclus)-\logten\rclusem}{\sigma_{\text{tot}}^{\text{\BNN(\RE),log}}(\fsetclus)} \label{eq:logpull} \\
& \approx \ln 10\cdot\logten\frac{\rclusbnnre(\fsetclus)}{\rclusem}\cdot\frac{\rclusbnnre(\fsetclus)}{\uncbnnretotal(\fsetclus)} \\
& = \ln 10\cdot\logten\frac{\rclusbnnre(\fsetclus)}{\rclusem}\cdot\left(\pullbnnretotal(\fsetclus)+\frac{\rclusem}{\uncbnnretotal(\fsetclus)}\right)
\end{split}\;.
\end{align}

\FigRef{fig:unc:pull} compares this pull in the logarithmic response space for the \topo predictions from the BNN (\logpullbnn) with the ones from the RE (\logpullre).
Both \logpullbnn and \logpullre are highly correlated, as shown in \FigRef{fig:unc:pull:scatter}.
This indicates that both networks provide near identical pulls cluster-by-cluster, and
is consistent with $\rclusbnn\approx\rclusre$ for $\uncretotal/\rclusre\approx\uncbnntotal/\rclusbnn$ observed in the previous \SectRef{sec:uncertainties:comparisons}.

The pull distribution is expected to be close to a Gaussian distribution, with a central value $\mean{\logpull} = 0$ and a spread measured by its standard deviation of the order of unity, as motivated by the discussion in \AppRef{app:perf-metrics:pull}.
\FigRef{fig:unc:pull:spectra} shows that both the \logpullbnn and the \logpullre distributions follow Gaussian distributions in their bulk parts, with close to symmetric spreads determined by independent two-sided fits of $\sigma_{\text{left}}$ and $\sigma_{\text{right}}$ to each of these distributions.
The asymmetry observed between the tails $\logpull < \sigma_{\text{left}}$, where some deviations from a Gaussian distribution are observed, and $\logpull > \sigma_{\text{right}}$, where the distribution more closely follows a Gaussian distribution, is an artefact introduced by the logarithmic transformation of the target and the respective predictions.
The fitted spreads yield $\sigma_{\text{left}} \approx \sigma_{\text{right}} < 1$, indicating that both \uncbnntotal and \uncretotal represent a conservative uncertainty estimate safely covering the residual difference between target and prediction for each \topo.
Considering these spreads together with central values fitted to be close to $\logpull = 0$, the conclusion from the pull distributions is that both \uncbnntotal and \uncretotal are fairly unbiased and conservative uncertainties within expected value ranges.

\subsection{Sources of uncertainties}
\label{sec:uncertainties:detector}

\begin{figure}[tt!]
\centering
\includegraphics[width=1.00\figwidth]{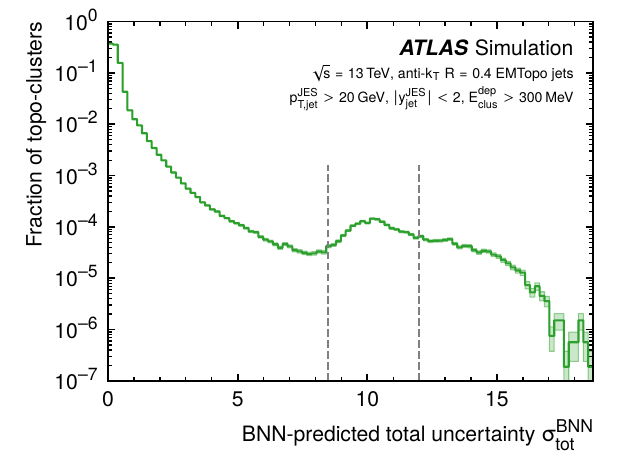}
\caption[]{The inclusive spectrum of uncertainties \uncbnntotal learned by the BNN\@. The vertical lines indicate the region $8.5 < \uncbnntotal < 12.0$ that is selected to analyse the source of the deviation
from the expected smoothly falling spectrum. The uncertainties are predicted for \topos from test dataset collected from the full MC simulations as described in \SectRef{sec:dataset:object}.
}
\label{fig:unc:anomalous}
\end{figure}

The inclusive spectrum of \uncbnntotal learned by the BNN displayed in \FigRef{fig:unc:anomalous} shows a rise of the population of \topos with large uncertainties, deviating from the expected smoothly falling spectrum.
The \topos in the region $8.5 < \uncbnntotal < 12.0$ around this emerging peak are investigated to search for possible reasons for its appearance.
They show a large response $\rclusem \gg 1$ and a small signal fraction in the electromagnetic calorimeter.
Their \cog is mostly located just in front of the \EndCap calorimeters, and
within $1\lesssim|\yclusem|\lesssim 1.4$.
This points to the instrumentation of a specific transition range between ATLAS calorimeters generating the \topo signals.

\begin{figure}[tt!]
\centering
\subfloat[]{\includegraphics[width=1.00\figwidth]{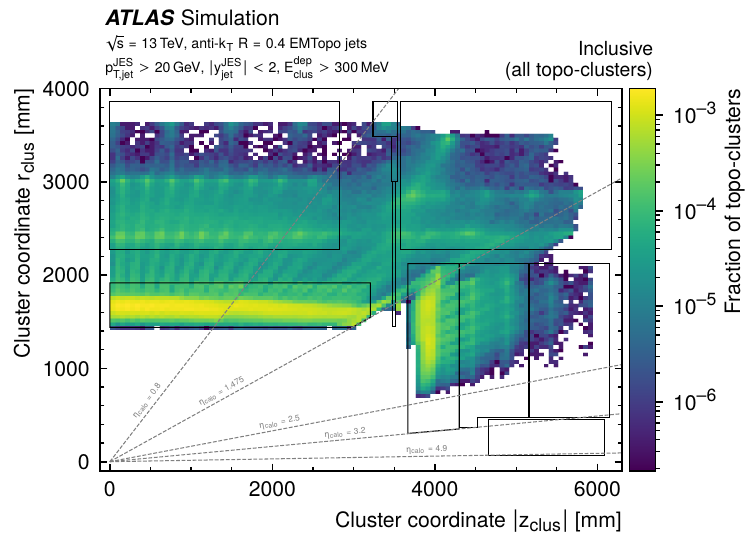}\label{fig:unc:location:all}}\quad
\subfloat[]{\includegraphics[width=1.00\figwidth]{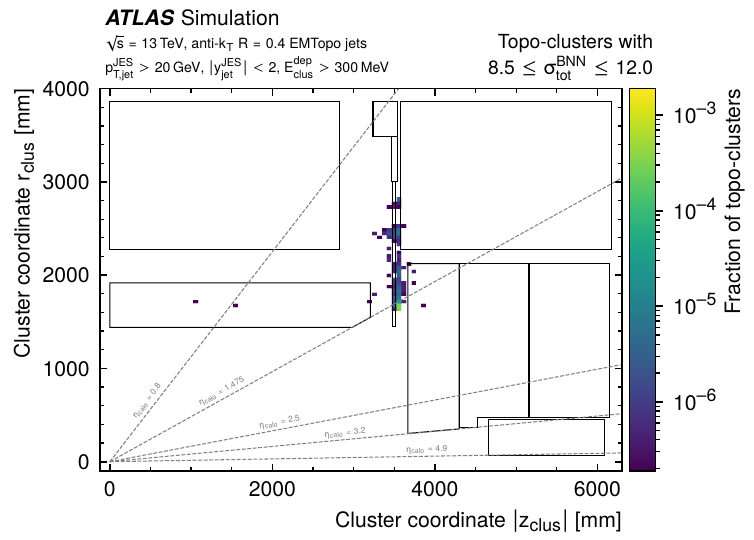}\label{fig:unc:location:sel}}
\caption[]{In \subref{fig:unc:location:all}, the \cog location projected into the $(|z_{\text{clus}}|,r_{\text{clus}})$ plane is shown for all \topos in the test dataset.
The outlined structures indicate the instrumented area covered by the various ATLAS calorimeters in this plane. The lines depict selected pseudorapidities $\eta_{\text{calo}}$ in the calorimeter pointing to
transition regions. The lack of \topos located beyond $|\eta_{\text{calo}}| \gtrsim 2.4$ is introduced by the jet selection described in \SectRef{sec:dataset:object}.
The location of the \cog of \topos with $8.5 < \uncbnntotal < 12.0$ found in the same plane is shown in \subref{fig:unc:location:sel}. These are predominantly inside,
or in close proximity to, the thin Tile gap scintillators installed in the transition region between the central and the \EndCap calorimeters.
}
\label{fig:unc:location}
\end{figure}

\FigRef{fig:unc:location:all} shows the \topo \cog projected into the $(|z_{\text{clus}}|,r_{\text{clus}})$ plane, where $z_{\text{clus}}$ is its Cartesian coordinate along the central axis of the ATLAS detector and
$r_{\text{clus}}$ is its radial distance from this central axis, for all clusters in the test dataset.
Selecting \topos with $8.5 < \uncbnntotal < 12.0$ from this sample yields the $(|z_{\text{clus}}|,r_{\text{clus}})$-plane population depicted in \FigRef{fig:unc:location:sel}.
The response and the features of these clusters are predominantly derived from signals in the Tile gap scintillator instrumenting the transition region between the central barrel and the \EndCap calorimeters.
This detector is not a regular calorimeter in that it does not generate a signal corresponding to the energy loss generated in a (regular) sampling structure of active and inactive media in a calorimeter cell.
Rather, it generates a signal reflecting the energy losses in the complex distribution of inactive material situated around it.
This environment comprises cryostat walls, mechanical support structures and inner detector services.
The energy deposits in these inactive materials are not considered in \rclusem defined in \EqRef{eq:target-response} of \SectRef{sec:data_cal} for the \topos located here, because they occur outside of calorimeter cells and are thus not accounted for in \eclusdep.

As already discussed in \SectRef{sec:data_cal:target}, the actual signal \eclusem of \topos located in the Tile gap scintillators is predominantly generated by ionisations.
Among the sources of these ionisations are secondary particles produced in the development of showers in the material around it.
These showers are generated by particles emerging from the hard scatter and the in-time \pu interactions.
The latter can also provide a second and more direct contribution to the signal when directly crossing this transition region, without previous interaction.
The BNN calibration network learns \rclusem from the feature set here, but the location of the \topo \cog suggests very little signal contributions from other calorimeters and thus to the feature reconstruction.
In particular, features related to shower shapes are expected to be reconstructed with low precision, leading to a significantly reduced expressiveness that produces predictions  with large uncertainties for \topos in this region.

The fact that the BNN learns about these regional detector signal characteristics from the training data and predicts a large total uncertainty \uncbnntotal that can be understood by relating it to the detector design, its signal formation, and its exposure to particles from the final state of the \pp collision, suggests that a meaningful interpretation of \uncbnntotal cluster-by-cluster is possible in, e.g., the context of \topo signal accuracy and precision requirements in a given final-state analysis.
Considering the previous discussions, \uncretotal is expected to reflect the same detector-induced issues.

\section{Conclusions}
\label{sec:conclusion}

The principal signal of the ATLAS calorimeters are clusters of topologically connected cell signals (\topos).
They are individually calibrated to measure the energy deposited in them using an uncertainty-aware Bayesian neural network (BNN).
This network is designed to predict the response and the corresponding uncertainty for a \topo as a function of a set of 15 features associated with it.
These features are selected under consideration of expressiveness manifested in their sensitivity to the signal source at the location of the cluster, their dependence on the detector-specific signal formation characteristics, and the proton--proton collision environment characterised by \pu.

The datasets collected for the BNN training, validation and testing comprise \topos collected from calorimeter jets in full and detailed Monte Carlo simulations of multijet production in proton--proton collisions at a centre-of-mass energy of \sqrts{13}, with the ATLAS detector configurations in place at the time of the LHC \RunTwo and with beam conditions generating levels of \pu observed in this operational period (2015--2018).

The calibrated signal performance with respect to the deviation from signal linearity
and the local relative energy resolution indicates significant
improvements on most relevant evaluation scales, if compared to the two
standard calibrations in ATLAS, the basic electromagnetic \topo calibration and the table-look-up-based hadronic calibration implemented in the LCW sequence, and to the calibration learned by a deep neural network (DNN) employing the same features as inputs as the BNN\@.
In particular, the application of the BNN calibration extends near-perfect signal linearity to a previously underperforming region populated with \topos with low-energy deposits from particles emerging from the hard scatter interaction.
This calibration also improves the relative energy resolution at cluster level when compared with the resolution achieved with the DNN-based calibration, which already shows a significant improvement with respect to the standard calibrations.

The uncertainties predicted by the BNN consist of two components that in the case of the sufficiently large training dataset used here can be distinguished and respectively interpreted as statistical and systematic uncertainties.
The BNN-learned uncertainties are confirmed using an alternative and independent calibration network featuring repulsive ensembles (RE), which predicts uncertainties that are cluster-by-cluster identical within the level of stochasticity in the dataset.
An analysis of the pull distributions finds that the uncertainties from both networks follow a Gaussian distribution in the bulk, with some small deviations observed in particular in the low pull tails.
The central values are found to be close to zero, indicating an unbiased pull.
The spread of each pull distribution is less than the expected unity, which suggests that the learned uncertainties are conservative.

Large but appropriate uncertainties of the calibrated energy derived from the BNN response predictions are observed for \topos located in the insufficiently instrumented transition region between the central and the \EndCap calorimeter modules.
This shows promise for a potential applications in which the uncertainty of the calibrated \topo signals can be used as evaluation tool for the signal quality and thus allow for selections based on accuracy (scale) and precision (resolution) requirements before the clusters enter the reconstruction of hadronic final-state objects like particles and jets, and the reconstruction of hadronic event shapes in general.

This study of the feasibility of the use of uncertainty-aware neural networks for the ATLAS calorimeter signal calibration in a realistically simulated proton--proton collision environment shows significant potential for a successful application in the context of physics analyses.
The simulation-only results presented in this paper require investigations beyond these proof-of-principle studies, including the validation and performance evaluations for the application to experimental data at present day and at significantly increased future \pu levels at the LHC\@.
The BNN model and its inputs are deliberately designed to address these challenges by adapting the derived calibrations accordingly.
This is mostly achieved by not using low-level data like calorimeter cell signals, but rather using features constructed from those as inputs.
A potentially necessary retraining of the BNN in the case of a changing feature phase space due to changing operational conditions at the LHC is thus possible without the need for major signal reconstruction campaigns.

\section*{Acknowledgements}

%
%

%
%

We thank CERN for the very successful operation of the LHC and its injectors, as well as the support staff at
CERN and at our institutions worldwide without whom ATLAS could not be operated efficiently.

The crucial computing support from all WLCG partners is acknowledged gratefully, in particular from CERN, the ATLAS Tier-1 facilities at TRIUMF/SFU (Canada), NDGF (Denmark, Norway, Sweden), CC-IN2P3 (France), KIT/GridKA (Germany), INFN-CNAF (Italy), NL-T1 (Netherlands), PIC (Spain), RAL (UK) and BNL (USA), the Tier-2 facilities worldwide and large non-WLCG resource providers. Major contributors of computing resources are listed in Ref.~\cite{ATL-SOFT-PUB-2023-001}.

We gratefully acknowledge the support of ANPCyT, Argentina; YerPhI, Armenia; ARC, Australia; BMWFW and FWF, Austria; ANAS, Azerbaijan; CNPq and FAPESP, Brazil; NSERC, NRC and CFI, Canada; CERN; ANID, Chile; CAS, MOST and NSFC, China; Minciencias, Colombia; MEYS CR, Czech Republic; DNRF and DNSRC, Denmark; IN2P3-CNRS and CEA-DRF/IRFU, France; SRNSFG, Georgia; BMBF, HGF and MPG, Germany; GSRI, Greece; RGC and Hong Kong SAR, China; ICHEP and Academy of Sciences and Humanities, Israel; INFN, Italy; MEXT and JSPS, Japan; CNRST, Morocco; NWO, Netherlands; RCN, Norway; MNiSW, Poland; FCT, Portugal; MNE/IFA, Romania; MSTDI, Serbia; MSSR, Slovakia; ARIS and MVZI, Slovenia; DSI/NRF, South Africa; MICIU/AEI, Spain; SRC and Wallenberg Foundation, Sweden; SERI, SNSF and Cantons of Bern and Geneva, Switzerland; NSTC, Taipei; TENMAK, T\"urkiye; STFC/UKRI, United Kingdom; DOE and NSF, United States of America.

Individual groups and members have received support from BCKDF, CANARIE, CRC and DRAC, Canada; CERN-CZ, FORTE and PRIMUS, Czech Republic; COST, ERC, ERDF, Horizon 2020, ICSC-NextGenerationEU and Marie Sk{\l}odowska-Curie Actions, European Union; Investissements d'Avenir Labex, Investissements d'Avenir Idex and ANR, France; DFG and AvH Foundation, Germany; Herakleitos, Thales and Aristeia programmes co-financed by EU-ESF and the Greek NSRF, Greece; BSF-NSF and MINERVA, Israel; NCN and NAWA, Poland; La Caixa Banking Foundation, CERCA Programme Generalitat de Catalunya and PROMETEO and GenT Programmes Generalitat Valenciana, Spain; G\"{o}ran Gustafssons Stiftelse, Sweden; The Royal Society and Leverhulme Trust, United Kingdom.

In addition, individual members wish to acknowledge support from Armenia: Yerevan Physics Institute (FAPERJ); CERN: European Organization for Nuclear Research (CERN DOCT); Chile: Agencia Nacional de Investigaci\'on y Desarrollo (FONDECYT 1230812, FONDECYT 1230987, FONDECYT 1240864); China: Chinese Ministry of Science and Technology (MOST-2023YFA1605700, MOST-2023YFA1609300), National Natural Science Foundation of China (NSFC - 12175119, NSFC 12275265, NSFC-12075060); Czech Republic: Czech Science Foundation (GACR - 24-11373S), Ministry of Education Youth and Sports (FORTE CZ.02.01.01/00/22\_008/0004632), PRIMUS Research Programme (PRIMUS/21/SCI/017); EU: H2020 European Research Council (ERC - 101002463); European Union: European Research Council (ERC - 948254, ERC 101089007), European Union, Future Artificial Intelligence Research (FAIR-NextGenerationEU PE00000013), Italian Center for High Performance Computing, Big Data and Quantum Computing (ICSC, NextGenerationEU); France: Agence Nationale de la Recherche (ANR-20-CE31-0013, ANR-21-CE31-0013, ANR-21-CE31-0022, ANR-22-EDIR-0002); Germany: Baden-Württemberg Stiftung (BW Stiftung-Postdoc Eliteprogramme), Deutsche Forschungsgemeinschaft (DFG - 469666862, DFG - CR 312/5-2); Italy: Istituto Nazionale di Fisica Nucleare (ICSC, NextGenerationEU), Ministero dell'Università e della Ricerca (PRIN - 20223N7F8K - PNRR M4.C2.1.1); Japan: Japan Society for the Promotion of Science (JSPS KAKENHI JP22H01227, JSPS KAKENHI JP22H04944, JSPS KAKENHI JP22KK0227, JSPS KAKENHI JP23KK0245); Norway: Research Council of Norway (RCN-314472); Poland: Ministry of Science and Higher Education (IDUB AGH, POB8, D4 no 9722), Polish National Agency for Academic Exchange (PPN/PPO/2020/1/00002/U/00001), Polish National Science Centre (NCN 2021/42/E/ST2/00350, NCN OPUS 2023/51/B/ST2/02507, NCN OPUS nr 2022/47/B/ST2/03059, NCN UMO-2019/34/E/ST2/00393, NCN \& H2020 MSCA 945339, UMO-2020/37/B/ST2/01043, UMO-2021/40/C/ST2/00187, UMO-2022/47/O/ST2/00148, UMO-2023/49/B/ST2/04085, UMO-2023/51/B/ST2/00920); Spain: Generalitat Valenciana (Artemisa, FEDER, IDIFEDER/2018/048), Ministry of Science and Innovation (MCIN \& NextGenEU PCI2022-135018-2, MICIN \& FEDER PID2021-125273NB, RYC2019-028510-I, RYC2020-030254-I, RYC2021-031273-I, RYC2022-038164-I); Sweden: Carl Trygger Foundation (Carl Trygger Foundation CTS 22:2312), Swedish Research Council (Swedish Research Council 2023-04654, VR 2018-00482, VR 2022-03845, VR 2022-04683, VR 2023-03403, VR grant 2021-03651), Knut and Alice Wallenberg Foundation (KAW 2018.0458, KAW 2019.0447, KAW 2022.0358); Switzerland: Swiss National Science Foundation (SNSF - PCEFP2\_194658); United Kingdom: Leverhulme Trust (Leverhulme Trust RPG-2020-004), Royal Society (NIF-R1-231091); United States of America: U.S. Department of Energy (ECA DE-AC02-76SF00515), Neubauer Family Foundation.

%
%


%

%
%

%
\clearpage
\appendix
\part*{Appendix}
\addcontentsline{toc}{part}{Appendix}
%

\setcounter{subsection}{0}
\renewcommand{\thesubsection}{\Alph{section}.\arabic{subsection}}
\setcounter{footnote}{1}
\renewcommand{\thefootnote}{\fnsymbol{footnote}}
\section{Local hadronic calibration sequence}
\label{app:lcw}
\begin{figure}[ht!]
\centering
\includegraphics[width=2.00\figwidth]{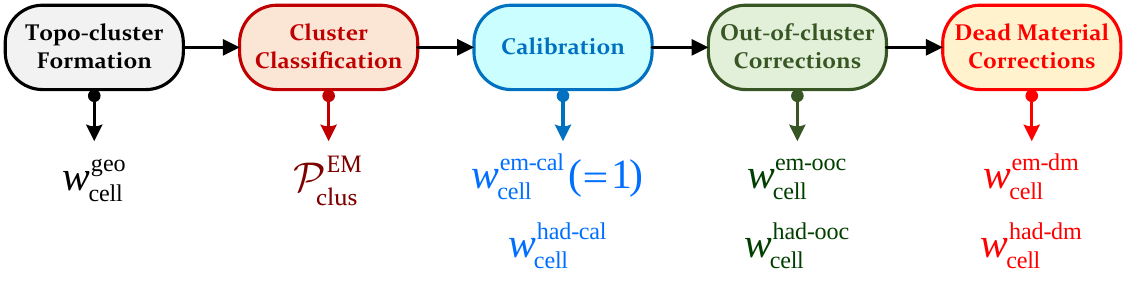}
\caption[]{The LCW calibration sequence. Taken from Ref.~\cite{PERF-2014-07}.}
\label{fig:lcw-sequence}
\end{figure}
The standard local hadronic calibration for \topos employs cell signal weighting for hadronic calibration (local cell weighting, LCW), right after the cluster classification estimating the likelihood \emlike of a cluster to be of electromagnetic origin, see sequence diagram shown in \FigRef{fig:lcw-sequence}.
The electromagnetic and hadronic scale factors are applied at cell level such that the effective \emph{cell signal weight} $w_{\cell}^{\kappa}$ at each step $\kappa$ after the classification becomes
\begin{align}
\label{eq:cellweights}
w_{\cell}^{\kappa} = \emlike w_{\cell}^{\text{em-}\kappa} + ( 1 - \emlike
) w_{\cell}^{\text{had-}\kappa}
\quad\mwith\quad
\kappa\in\{\text{cal},\text{ooc},\text{dm}\}\;.
\end{align}
Here \emlike is reconstructed in the classification step, and the `em' and `had' prefixes indicate the algorithm from which the signal weights are derived.
The calibrated \topo energies after each LCW calibration step $\kappa$ are reconstructed from sums of weighted cell energies, with weights
\begin{align*}
w_{\cell}^{\kappa}w_{\cell}^{\text{geo}}\ecell\;,
\end{align*}
applied to the cell signal \ecell at the \escale scale.
The geometrical weight $w_{\ecell}^{\text{geo}}$ is $w_{\ecell}^{\text{geo}} < 1$ for cells shared between two \topos, and $w_{\ecell}^{\text{geo}} = 1$ for cells collected into one \topo only.

The \topo energy \eclushad, which reflects the hadronic calibration applied to the \topo signal \eclusem at step $\kappa = \text{cal}$ after classification, serves as a reference for comparisons with the machine-learned calibrations and is given by
\begin{align}
\label{eq:lcw:had}
\eclushad = \sum_{i=1}^{\ncellclusall} w_{\cell,i}^{\text{geo}} \ecelli{i}
\left[\emlike w_{\cell,i}^{\text{had-em}} + \left(1-\emlike\right) w_{\cell,i}^{\text{had-had}}\right]
= \sum_{i=1}^{\ncellclusall} w_{\cell,i}^{\text{cal}} \ecelli{i}\;.
\end{align}
Here \ncellclusall is the total number of cells in the \topo.
Typical weights are $w_{\cell}^{\text{em-cal}} = 1$, as \ecell is already at EM scale, and $w_{\ecell}^{\text{had-cal}} \geq 1$ to calibrate non-compensation ($\eoverh > 1$), both used to construct the effective cell signal weight $w_{\cell}^{\text{cal}}$ at the calibration step.
The electromagnetic and hadronic weights applied in the following correction steps for out-of-cluster and inactive material energy losses are typically larger than unity.
As they depend on the nature of the underlying particle shower, they are typically also not identical for the respective electromagnetic and hadronic correction.

\setcounter{subsection}{0}
\renewcommand{\thesubsection}{\Alph{section}.\arabic{subsection}}
\section{Features}
\label{app:features}

\subsection{Feature construction}
\label{app:features:construction}

All features collected into the \topo feature set \fsetclus defined in \EqRef{eq:observables} and further described in \TabRef{tab:observables} in \SectRef{sec:datasets:features},
are documented in \RefRef{PERF-2014-07}, except for the cluster compactness measure \clusptd and the variance \tclusvar of the cell time distribution within the cluster.

The construction of \clusptd is inspired by a variable introduced in the context of quark-gluon jet tagging in \RefRef{CMS-PAS-JME-13-002}.
While preserving the naming convention introduced in the original reference, the definition of \clusptd is slightly modified for computational efficiency when considering that clustered calorimeter cells are directionally in close proximity, by replacing the transverse momenta of the jet constituents in the original definition by the cell energy \ecell such that
\begin{align}
\label{eq:ptd}
\clusptd = \frac{\sqrt{\sum_{i|\ecelli{i}>0} (\wcellij{i}{\text{geo}}\ecelli{i})^{2}}}{\sum_{i|\ecelli{i}>0}\wcellij{i}{\text{geo}}\ecelli{i}}\;.
\end{align}
Only cells with energies $\ecell > 0$ are considered in the sums.
By definition, \clusptd is numerically restricted to $0 < \clusptd \leq 1$.

A compactness of $\clusptd\rightarrow 1$ indicates that the energy \eclusem of
the \topo is increasingly contained in only one cell.
A compactness measure of $\clusptd = 1$ exact is only possible for a single cell cluster, which is not expected to happen, due to the \topo formation rules stating that at least all neighbours of a seed cell are collected~\cite{PERF-2014-07}.
Very small number of cells are possible for \topos at the edges of the calorimeter, like at the outer circumference of the \Tile, the back of the \HEC, and the front faces of the electromagnetic calorimeters \EMB and \EMEC.
A smaller value of \clusptd indicates a disperse distribution of the energy across the cells in the cluster.

The variance \tclusvar of the cell time (\tcell) distribution is given by
\begin{align}
\label{eq:timevar}
\tclusvar = \langle\tcellj{2}\rangle - \tclusj{2}\;,
\end{align}
where \tcell are the cell times and \tclus is the cluster time.
The second moment $\langle\tcellj{2}\rangle$ of the \tcell distribution for cells collected in the \topo is defined by
\begin{align}
\label{eq:timevarcalc}
\langle\tcellj{2}\rangle = \frac{\sum_{i|\scelli{i}>2}(\wcellij{i}{\taggeo}\ecelli{i})^{2}\cdot\tcellij{i}{2}}{\sum_{i|\scelli{i}>2}(\wcellij{i}{\taggeo}\ecelli{i})^{2}}\;,
\end{align}
following the definition of \tclus, which is constructed as a first moment of the \tcell distribution~\cite{PERF-2014-07}.
Only cells with a signal significance (signal-over-noise) of $\scell > 2$ are considered when calculating \tclus and \tclusvar.

\subsection{Relation of features to \topo nature}
\label{app:features:evaluation}

\begin{figure}[t!]
\centering
\subfloat[]{\includegraphics[width=0.32\textwidth]{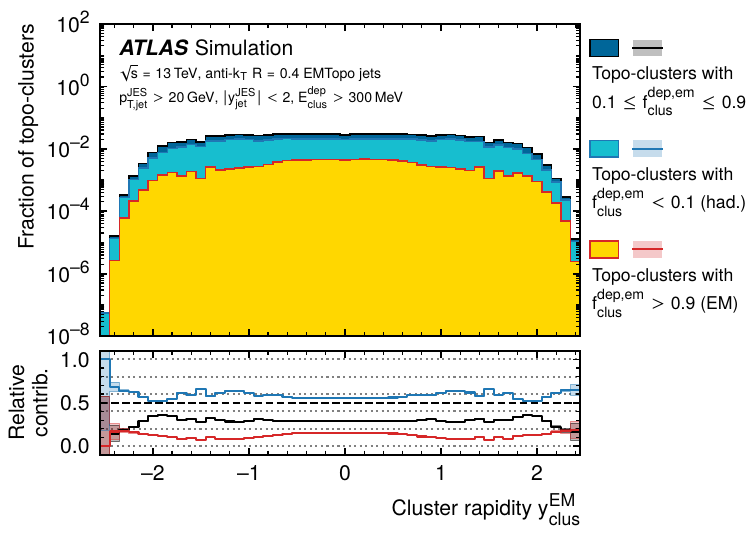}\label{fig:features_had_em_1:yems}}
\subfloat[]{\includegraphics[width=0.32\textwidth]{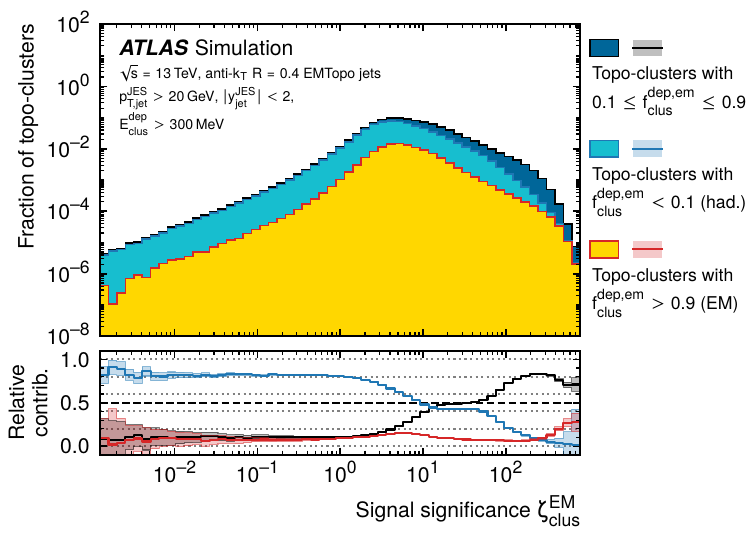}\label{fig:features_had_em_1:sems}}
\subfloat[]{\includegraphics[width=0.32\textwidth]{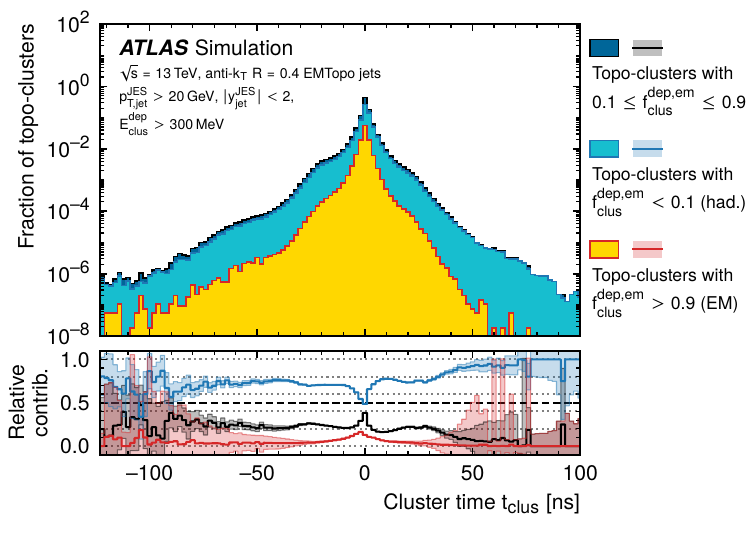}\label{fig:features_had_em_1:tems}}\\
\subfloat[]{\includegraphics[width=0.32\textwidth]{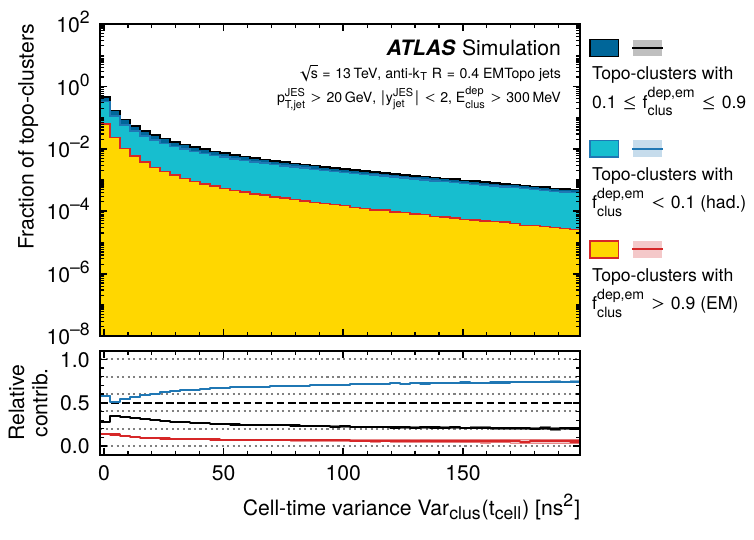}\label{fig:features_had_em_1:vart}}
\subfloat[]{\includegraphics[width=0.32\textwidth]{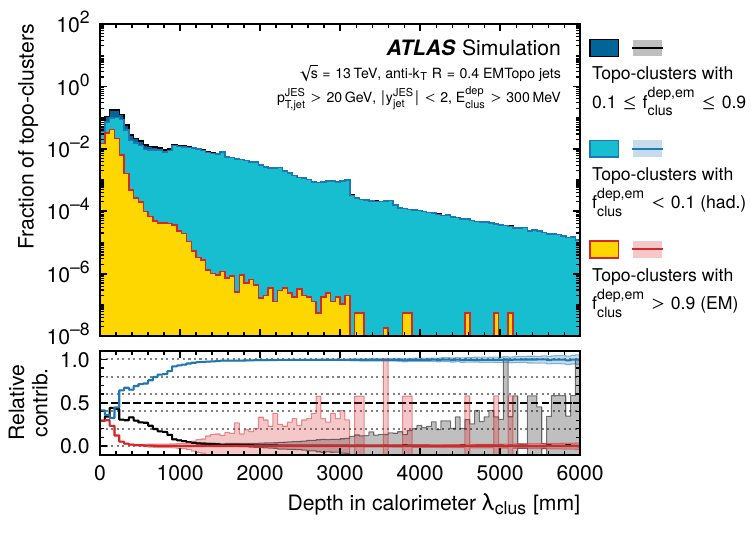}\label{fig:features_had_em_1:lamb}}
\subfloat[]{\includegraphics[width=0.32\textwidth]{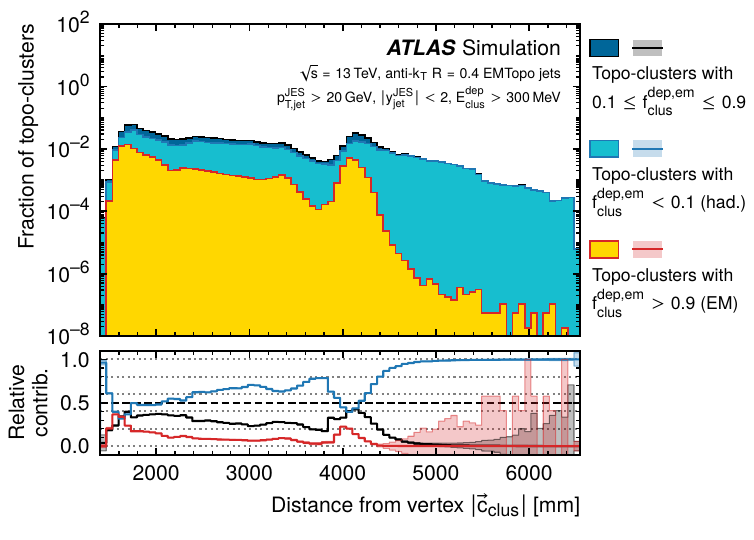}\label{fig:features_had_em_1:cogd}}\\
\subfloat[]{\includegraphics[width=0.32\textwidth]{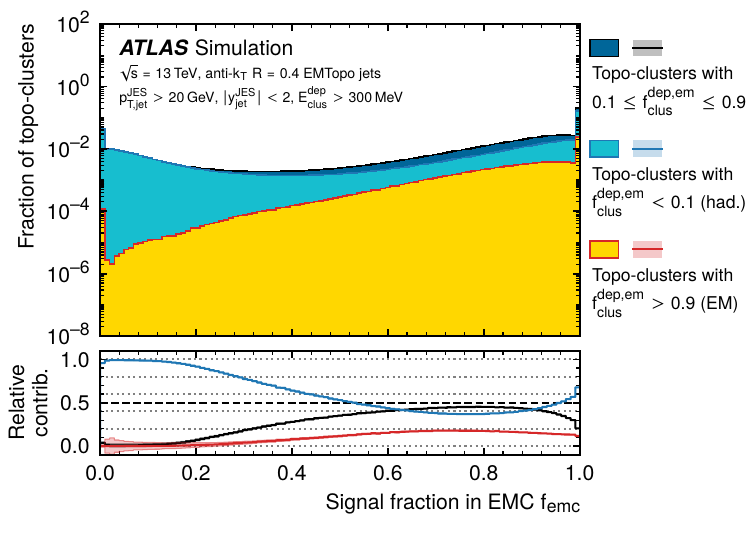}\label{fig:features_had_em_1:femc}}\quad
\subfloat[]{\includegraphics[width=0.32\textwidth]{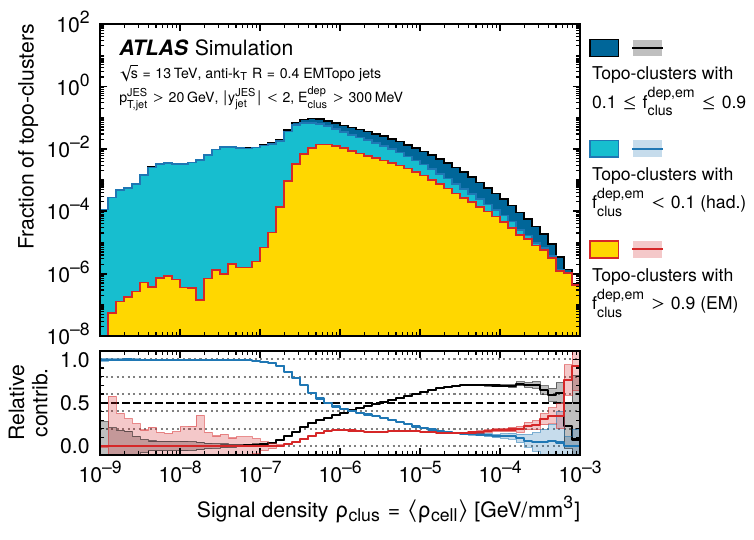}\label{fig:features_had_em_1:rhoc}}
\caption[]{Distributions of some of the features given in \EqRef{eq:observables} for \topos classified as hadronic, electromagnetic or composite following the
definitions given in \SectRef{sec:detdat:clusters:truth}. Distributions are shown for
\subref{fig:features_had_em_1:yems} the rapidity \yclusem,
\subref{fig:features_had_em_1:sems} the signal significance \sclus,
\subref{fig:features_had_em_1:tems} the time \tclus,
\subref{fig:features_had_em_1:vart} the variance of the cell time distribution \tclusvar,
\subref{fig:features_had_em_1:lamb} the distance from the calorimeter front-face \cluslam,
\subref{fig:features_had_em_1:cogd} the distance from the nominal vertex \cluscog,
\subref{fig:features_had_em_1:femc} the fraction of energy in the electromagnetic calorimeter \clusfemc, and
\subref{fig:features_had_em_1:rhoc} the signal density \clusrho.
The fractional contributions from the three \topo categories are shown in the
respective lower panels of each figure. The shaded regions show the statistical uncertainties.
More information about the definition of \tclusvar is given in \AppRef{app:features:construction}. The other features are defined in \RefRef{PERF-2014-07}. An overview on all features is given in \TabRef{tab:observables}.}
\label{fig:features_had_em_1}
\end{figure}

The features collected into the feature set \fsetclus defined in \EqRef{eq:observables} in \SectRef{sec:datasets:features} are evaluated with respect to the composition of the \topo signal source \eclusdep.
For this, feature distributions for the three categories defined in \SectRef{sec:detdat:clusters:truth}, \emTopos, \hadTopos and \compTopos, are compared in \FigsRef{fig:features_had_em_1}{fig:features_had_em_2}.
Their relative contribution to the corresponding inclusive distributions is shown in same figures.

As the \topos considered here are extracted from jets, the typical fraction of \emTopos in the sample can be estimated.
Using the canonical fraction of the jet energy carried by photons, about \perCent{25}, together with the average \topo multiplicity for hadrons shown in \RefRef{PERF-2014-07} and the assumption that each photon in a jet
generates one cluster, this fraction is about \perCent{10}.
This estimate is only given as guideline, as it ignores all full or partial merging of shower signals into one \topo and all shower signal losses introduced by the topological cell clustering algorithm and its configuration.

\begin{figure}[tt!]
\centering
\subfloat[]{\includegraphics[width=0.32\textwidth]{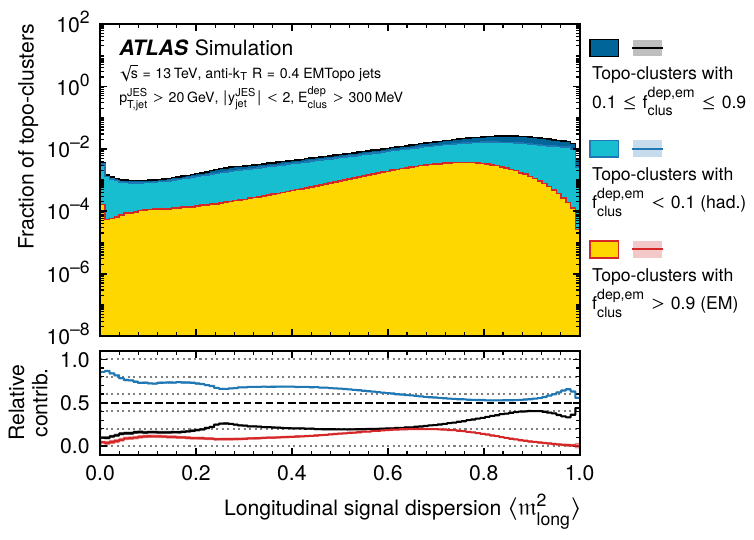}\label{fig:features_had_em_2:mlon}}
\subfloat[]{\includegraphics[width=0.32\textwidth]{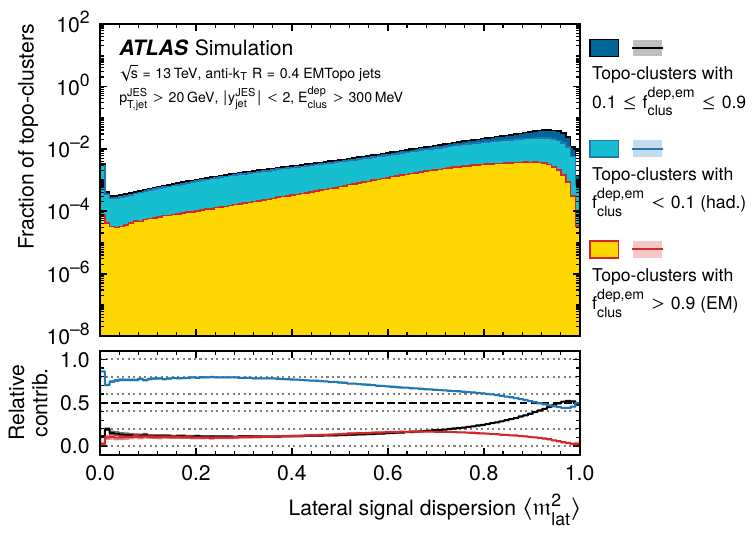}\label{fig:features_had_em_2:mlat}}
\subfloat[]{\includegraphics[width=0.32\textwidth]{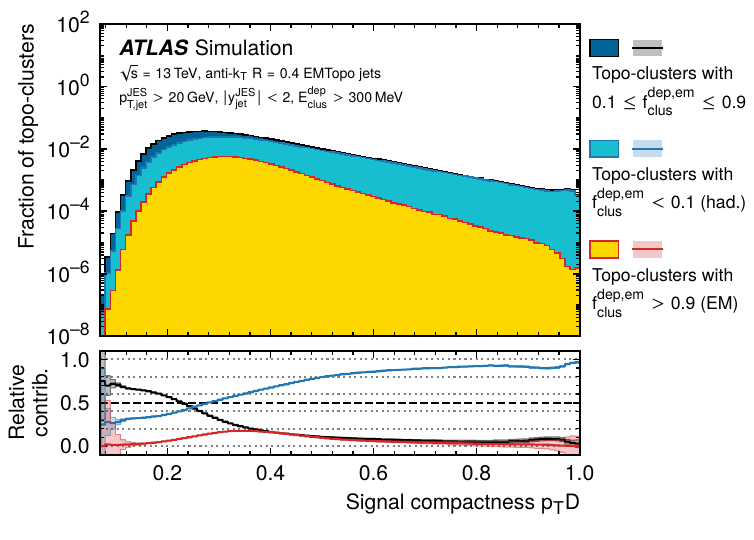}\label{fig:features_had_em_2:ptdc}}\\
\subfloat[]{\includegraphics[width=0.32\textwidth]{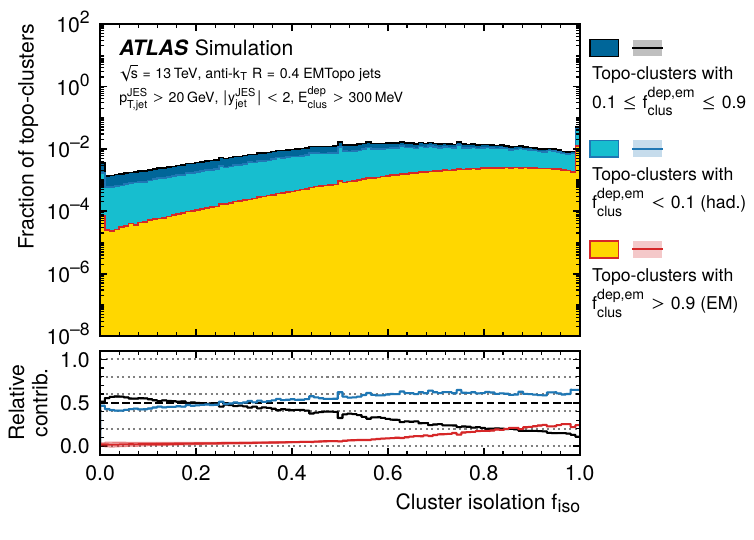}\label{fig:features_had_em_2:fiso}}
\subfloat[]{\includegraphics[width=0.32\textwidth]{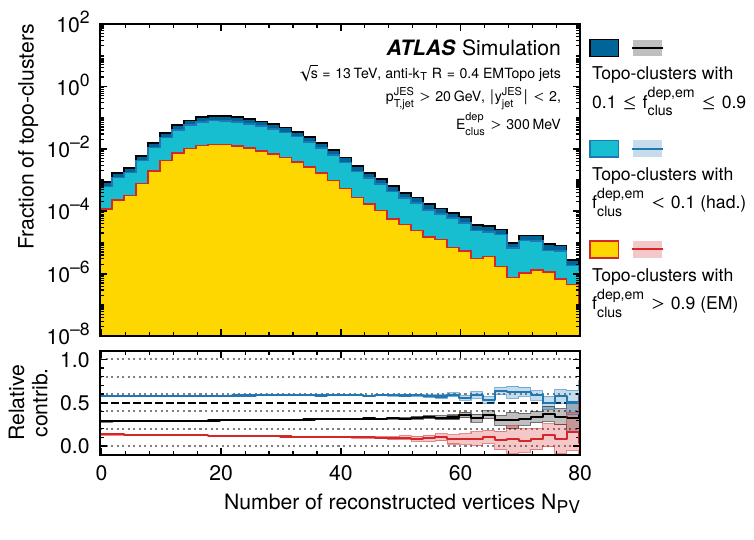}\label{fig:features_had_em_2:npv}}
\subfloat[]{\includegraphics[width=0.32\textwidth]{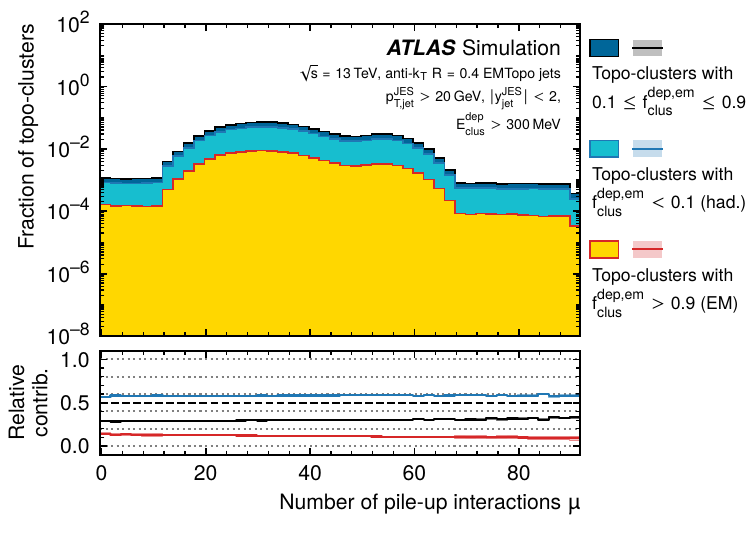}\label{fig:features_had_em_2:mu}}
\caption[]{Distributions of some of the features given in \EqRef{eq:observables}
for \topos classified as hadronic, electromagnetic or composite following the
definitions given in \SectRef{sec:detdat:clusters:truth}. Distributions are shown for
\subref{fig:features_had_em_2:mlon} the longitudinal energy dispersion \cluslong,
\subref{fig:features_had_em_2:mlat} the lateral energy dispersion \cluslat,
\subref{fig:features_had_em_2:ptdc} the compactness measure \clusptd,
\subref{fig:features_had_em_2:fiso} the isolation measure \clusfiso,
\subref{fig:features_had_em_2:npv} the number of reconstructed vertices \npv, and
\subref{fig:features_had_em_2:mu} the number of \pu interactions \npu.
The fractional contributions from the three \topo categories are shown in the
respective lower panels of each figure. The shaded regions show the statistical uncertainties.
More information about the definition of \clusptd is given in \AppRef{app:features:construction}. The other features are defined in \RefRef{PERF-2014-07}. An overview on all features is given in \TabRef{tab:observables}.}
\label{fig:features_had_em_2}
\end{figure}

Other feature distributions with near-constant contributions from \emTopos are \yclusem shown in \FigRef{fig:features_had_em_1:yems}, \npv in \FigRef{fig:features_had_em_2:npv} and \npu in \FigRef{fig:features_had_em_2:mu}.
Those also show approximately constant contributions from the other \topo categories.
This insensitivity to the cluster nature is expected, as neither \yclusem nor \npv or \npu should show any effect on the composition of the energy deposits, which to first order only depend on the fragmentation of jets from the hard-scatter interactions.
Slight variations may be introduced by the amount of inactive material in front of the active calorimeter volume that shows some variations with \yclusem.
These features are kept in \fsetclus because they provide sensitivity to changes of the detector \readout geometry, including the varying projection of the particle showers onto
the calorimeter cell geometry and thus the reconstruction of most spatial, compactness and density features, and the \pu environment.
Among the strongest indicators of the \topo nature are \sclus (\FigRef{fig:features_had_em_1:sems}), the location measures \cluslam (\FigRef{fig:features_had_em_1:lamb}) and \cluscog (\FigRef{fig:features_had_em_1:cogd}),
\clusfemc (\FigRef{fig:features_had_em_1:femc}), \clusrho (\FigRef{fig:features_had_em_1:rhoc}), the dispersion and compactness measures \cluslong, \cluslat and \clusptd in
\FigRangeRef{fig:features_had_em_2:mlon}{fig:features_had_em_2:ptdc}.
These figures all confirm expectations from electromagnetic and hadronic shower developments in the ATLAS calorimeters.
For example, the contribution of \emTopos to the \clusfemc distribution is generated by the photons in jets, nearly all of which are represented as a single cluster.
They are mostly stopped in the electromagnetic calorimeters.
This is in full agreement with the observation that \emTopos are more isolated, as they contribute more than \perCent{25} of all clusters with $\clusfiso\gtrsim 0.8$, at an inclusive contribution of about
\perCent{10}, as discussed above.

The larger contribution of \emTopos to the population of higher-density clusters seen in \FigRef{fig:features_had_em_1:rhoc} is also well within the expectation of higher-density deposits in electromagnetic showers in the
smaller volume cells of the electromagnetic calorimeters, with a higher signal yield when compared with hadronic showers.
Also as expected, \hadTopos are dominantly contributing to the population of clusters deep in the calorimeter, as seen for the \cluscog and \cluslam distributions in
\FigsRef{fig:features_had_em_1:lamb}{fig:features_had_em_1:cogd}.
Compact \topos with $\clusptd \gtrsim 0.6$ are most likely of hadronic nature, as seen in \FigRef{fig:features_had_em_2:ptdc}.
These compact \topos often have one dominant cell signal, which can occur in the course of the hadronic shower development in the hadronic calorimeters with their coarse spatial readout granularity and in the \Tile calorimeter for signals from ionisations by charged hadrons.
As already seen, \emTopos are much more likely to appear in the highly granular electromagnetic calorimeters, where despite their high energy loss density signal sharing between cells is very likely, due to the relatively small cell size, compared to typical electromagnetic shower sizes.

\FloatBarrier

\setcounter{subsection}{0}
\renewcommand{\thesubsection}{\Alph{section}.\arabic{subsection}}
\section{Network implementations and loss functions}
\label{app:loss}

\subsection{Bayesian loss function}
\label{app:loss:bnn}

As discussed in \SectRef{sec:data_net}, a Bayesian neural network (BNN) is
employed to encode the response function $\r(\fsetclus) =
\rclusbnn(\fsetclus)$ and its uncertainty over the space of all features sets
\fsetclus found for the \topos in the training dataset.
It is designed to learn $\r = \r(\fsetclus)$ as a probability distribution $p(\r)$, with the mean
\begin{align}
\mean{\r} = \int \dd\r\:\r p(\r) \;.
\label{eq:mean-value-1}
\end{align}
The network encodes the learned probability in weight configurations, conditional on the training data.
The training can be described as constructing a variational approximation~\cite{Hinton:1993vai,Blei_2017}, replacing \probbnn introduced in \SectRef{sec:data_net} with a learned $q(\theta)$
such that
\begin{align}
p(\r)
= \int \dd\theta\: p(\r | \theta) p(\theta | \trainds)
\approx \int \dd\theta\: p(\r | \theta) q(\theta) \;.
\label{eq:bnn-variational-inference}
\end{align}
The dataset \trainds is defined over the feature space containing all possible feature sets \fsetclus, following the nomenclature given in \EqRef{eq:trainds} of \SectRef{sec:data_net:bnn_loss}.
To learn $q(\theta)$, the Kullback-Leibler (KL)~\cite{KL_1951} divergence including the Bayesian transform $\probbnn\rightarrow\probbayes$
\begin{align} \label{eq:bnn-kl-divergence}
\begin{split}
\kl[q(\theta), \probbnn ]  &= \int \dd\theta\: q(\theta)
\log\frac{q(\theta)}{\probbnn} \\
&= \int \dd\theta\:
q(\theta)\log\frac{q(\theta)p(\trainds)}{\pprior(\theta)\probbayes}
\\
&\approx \kl[q(\theta), \pprior(\theta)] -\underbracket{\int
\dd\theta\: q(\theta)\log\probbayes}_{\displaystyle
\mathclap{\left\langle\log\probbayes\right\rangle_{\theta\sim
q(\theta)}}}
\end{split}
\end{align}
is minimized.
The explicit form of \probbayes is given in \EqRef{eq:bnn_gmix} in \SectRef{sec:data_net:bnn_loss}.

The first term in \EqRef{eq:bnn-kl-divergence} arises from the prior $\pprior(\theta)$ and can be viewed as a weight regularisation.
The second term samples from the negative logarithmic likelihood, where the sampling generalises the standard dropout in neural network training.
The BNN loss is then defined as given in \EqRef{eq:bnn_loss},
\begin{align}
\lbnn
=
\kl[q(\theta),\pprior(\theta)]-\big\langle\log\probbayes\big\rangle_{\theta\sim
q(\theta)}\;.
\label{eq:loss-bnn}
\end{align}
A Gaussian distribution is used as a convenient prior $\pprior(\theta)$.
Its width is a hyper-parameter that could be varied on a performance plateau to make the training more efficient~\cite{Bollweg:2019skg}.
If the weight distributions $q(\theta)$ are also chosen as Gaussian functions, the KL-divergence can be computed analytically.
Provided the network is sufficiently deep, the assumption of independent Gaussian distributions for each network parameter should not affect its expressiveness.

\subsection{Implementation of repulsive ensemble networks}
\label{app:loss:re_impl}

The principle implementation of the repulsive ensemble (RE) network training is described in \RefRef{Plehn:2022ftl}.
As already mentioned in \SectRef{sec:data_net:re_design}, RE networks modify the update rule that minimises $-\log\probbnn$ by gradient descent.
This is done by evaluating this update rule for an ensemble of $N$ networks
and introduce a repulsive term that forces the ensemble to spread out and
cover the loss around the actual minimum.
The repulsive term is designed to increase with the proximity of the network parameters $\theta$ of a given ensemble member to all other members.
It is implemented by a Gaussian kernel $k(\theta,\theta_j)$, with $j =
1,\ldots,N$, such that the update from training step $t$ to training step
$t+1$ becomes
\begin{align}
\theta^{t+1} = \theta^t + \alpha \nabla_{\theta^t} \log \probbnn - \alpha
\frac{\nabla_{\theta^t} \sum_{j=1}^N  k(\theta^t,\theta^t_j)}{\sum_{i=1}^N
k(\theta^t,\theta^t_i)} \;.
\label{eq:update_rule2}
\end{align}
The first two terms represent the standard network training by gradient descent, with the learning rate $\alpha$ and a gradient $\nabla_{\theta^{t}}$
reflecting the evolution of the network parameters $\theta^{t}$ at the training step $t\rightarrow t+1$.
The additional kernel term runs over the $N$ networks of an ensemble and
defines repulsive ensembles.

Using properties of ordinary differential equations describing the network training as a time evolution and the unique stationary solution of the Fokker-Planck equation
(see, e.g., \RefRef{doi:10.1142/4016}), it is found that this update rule leads to an ensemble of networks with parameters sampled from $\theta \sim \probbnn$.

To fit a calibration function for \topos, and similar other regression tasks, the function encoded by the network is of interest, not its latent representation $\theta$ in weight space.
For instance, two networks can encode the same function by permuting the weights, unaffected by a repulsive force in weight space.
The repulsive term therefore needs to act in the space of the actual network outputs $\rclusre(\fsetclus) = \r_{\theta}(\fsetclus)$ represented by $\theta$.
The update rule in this function space is given by
\begin{align}
\frac{\r^{t+1} - \r^t}{\alpha} = \nabla_{\r^t} \log p(\r|\trainds) - \frac{\sum_j \nabla_{\r^t} k(\r,\r_j)}{\sum_j k(\r,\r_j)}\;.
\label{eq:update_rule_fspace}
\end{align}
It is formally similar to the one in the weight space given in \EqRef{eq:update_rule2}.

The network training is still performed in the weight space, so a translation of the function-space update rule into this space is needed,
\begin{align}
\frac{\theta^{t+1}-\theta^t}{\alpha} &= \nabla_{\theta^t} \log p(\theta^t|\trainds) - \frac{\partial \r^t}{\partial \theta^t} \frac{\sum_j \nabla_{\r} k(\r_{\theta^t},\r_{\theta_j^t})}{\sum_j k(\r_{\theta^t},\r_{\theta_j^t})}\;.
\label{eq:update_rule_fspace1}
\end{align}
The repulsive kernel in the function space cannot be evaluated, so the function must be evaluated for a finite batch of \topos represented by points $x = \fsetclus$ in the feature space,
\begin{align}
\frac{\theta^{t+1} - \theta^{t}}{\alpha} \approx \nabla_{\theta^{t}} \log
p(\theta^{t}|\trainds) - \frac{\sum_{j=1}^{N} \nabla_{\theta^t}
k(\r_{\theta^t}(x),\r_{\theta_j^t}(x))}{\sum_{j=1}^{N}
k(\r_{\theta^t}(x),\r_{\theta_j^t}(x))}\;.
\label{eq:update_rule_fspace2}
\end{align}
Here $\r_{\theta^t}(x)$ in \EqsRef{eq:update_rule_fspace1}{eq:update_rule_fspace2} is understood as evaluating the function $\r$ for members $x$ of \trainds at training step $t$.

\subsection{Repulsive ensemble loss function}
\label{app:loss:re_loss}

The modified update rule of \EqRef{eq:update_rule_fspace2} can be turned into a loss function for the RE training.
Using Bayes' theorem and neglecting the evidence $p(\trainds)$, the likelihood
loss measure \probbnn can be written as
\begin{align}
\log\probbnn = \log\probbayes -
\underbracket{\frac{\theta^{2}}{2\sigma^{2}}}_{\mathclap{\text{exponential
of\ }\pprior(\theta)}} +\;\text{const.}\;.
\label{eq:re_loss_bayes}
\end{align}
The prior $\pprior(\theta)$ is chosen to be Gaussian.

For a training dataset of size \DSsizeTrain, split into batches of size \DSsizeBatch, $\r_{\theta^t}(x)$ is evaluated as the function of members $x_1,\dots,x_{\DSsizeBatch}$ for all batches of the training dataset \trainds.
The update rule from \EqRef{eq:update_rule_fspace2} can be rewritten using \EqRef{eq:re_loss_bayes},
\begin{align}
\frac{\theta^{t+1}-\theta^{t}}{\alpha}
\approx \nabla_{\theta^{t}}
\frac{\DSsizeTrain}{\DSsizeBatch}\sum_{b=1}^{\DSsizeBatch}\log p(x_{b}|\theta)
-
\frac{\sum_{j=1}^{N}\nabla_{\theta^{t}}k(\r_{\theta_i}(x),\r_{\theta_j}(x))}{\sum_{j=1}^{N}
k(\r_{\theta_i}(x), \r_{\theta_j}(x))}
- \nabla_{\theta^{t}} \frac{\theta^{2}}{2\sigma^{2}}\;.
\label{eq:update_rule_fspace3}
\end{align}

Based on this rule, the loss function is constructed by inverting the sign in front of the gradient in \EqRef{eq:update_rule_fspace3} and divide by the training dataset size \DSsizeTrain to remove the
scaling with \DSsizeTrain.
In addition, not all occurrences of $\theta$ are inside the gradient, so a \emph{stop-gradient operation} is introduced, denoted by $\overline{\r}_{\theta_j}(x)$.
With this, the RE loss function is given by
\begin{align}
\lre = \sum_{i=1}^N \Bigg[
\underbracket{-\frac{1}{\DSsizeBatch}\sum_{b=1}^{\DSsizeBatch} \log p(x_{b}|\theta_{i})}_{\mathclap{\text{(1) negative logarithmic likelihood averaged over batch}}}
+ \overbracket{\frac{1}{\DSsizeTrain} \frac{\sum_{j=1}^N
k(\r_{\theta_i}(x), \overline{\r}_{\theta_j}(x))}{\sum_{j=1}^N
k(\overline{\r}_{\theta_i}(x),
\overline{\r}_{\theta_j}(x))}}^{\mathclap{\text{(2) repulsive term in
function space}}}
+ \underbracket{\frac{1}{\DSsizeTrain}\frac{\theta_{i}^{2}}{2\sigma^{2}}}_{\mathclap{\text{(3) normalised exponent of\ }\pprior(\theta_{i})}}
\Bigg]\;.
\label{eq:loss_re}
\end{align}
Here the numbers (1), (2) and (3) refer to the discussion of the corresponding terms in \EqRef{eq:re_loss} of \SectRef{sec:data_net:re_loss}.
The normalised exponent of the prior implements a ridge ($\text{L}_{2}$) regularisation (see, for example, \RefRef{Hastie_2020}).
It controls possible over-fitting and focusses the fit on regions of relevant data.

A typical choice for the repulsive term is a Gaussian kernel, with a width given by the
\emph{median heuristic}\footnote{This is the median of all pairwise distances
between $\theta_{i}$ and $\theta_{j}$ for all $N$ ensemble members ($i =
1,\ldots,N$ and $j=1,\ldots,N$).}~\cite{liu2016stein}
between the $N$ ensemble members.
It is divided by the training dataset size \DSsizeTrain to remove any scaling with \DSsizeTrain.

\FloatBarrier

\setcounter{subsection}{0}
\renewcommand{\thesubsection}{\Alph{section}.\arabic{subsection}}
\section{Evaluation metrics}
\label{app:metric}
\subsection{Performance}
\label{app:metric:perf}

For the evaluation of the performance of the machine-learned \topo calibration, distributions of test variables like the prediction power \delr defined in \EqRef{eq:delta-r} in \SectRef{sec:perf-metrics:pred} and the deviation from signal linearity \dele defined in \EqRef{eq:delta-e} in \SectRef{sec:perf-metrics:linear} are characterised by a measure of the central tendency and a corresponding measure of their spread.
The median value of the distribution is used to express the central tendency, because it is an unbiased measure that does not require any assumption on the shape of the distribution (like the mode does) and it shows the least dependence on the sample size, compared with the statistical average of an asymmetric distribution with developing tails.

The spread is defined by the \emph{inter-quantile range} \qwidthnorm,
\begin{align}
\label{eq:qwidthnorm}
\qwidthnorm = \min(x_{k}-x_{i}) = \xqright-\xqleft \quad\text{such that}\quad \frac{\int_{\xqleft}^{\xqright}\dd x\: p(x)}{\int_{-\infty}^{+\infty}\dd x\: p(x)} = 0.68 \;,
\end{align}
which for any distribution $p(x)$ is the smallest of all $x$ value ranges that contains \perCent{68} of the collected statistics.
This range corresponds to the $[-1\sigma,+1\sigma]$ range around the mean of a Gaussian distribution, and is symmetric around any measure of central tendency in case of symmetric distributions.
Here $\xqleft$ and $\xqright$ are the respective lower and upper limit of the $x$-value range defining the \emph{inter-quantile} region $\iqrnorm$,
\begin{align}
\label{eq:iqrnorm}
\iqrnorm = \left[\xqleft,\,\xqright\right] = \left[\xqleft,\,\xqleft+\qwidthnorm\right]\;.
\end{align}

With the choices of distribution descriptors given above, a measure for the relative energy resolution \relreso can be constructed from the \dele distributions that employs both its median
$\median{\dele}$ and its spread $\qwidthnorm(\dele)$,
\begin{align}
\relreso = \frac{\qwidthnorm(\dele)}{2(\median{\dele}+1)}\;.
\label{eq:resolution}
\end{align}
If \dele is distributed following a Gaussian, $\qwidthnorm(\dele) = 2\sigma(\dele)$ such that
\begin{align*}
\relreso = \frac{\sigma(\dele)}{\ave{\dele}+1} = \frac{\sigma(\eclusbnnre)}{\ave{\eclusbnnre}} \approx \frac{\sigma(\eclusbnnre)}{\ave{\eclusdep}}
\end{align*}
is the relative energy resolution for calorimeters.
Equation~\ref{eq:resolution} can therefore be considered as an appropriate representation of this resolution.

\subsection{Uncertainties}
\label{app:metric:unc}

\label{app:metric:unc:components}

The learned uncertainties from the BNN are derived by sampling the network parameters $\theta$ and calculating the variance by
\begin{align}
\label{eq:sigma-tot}
\begin{split}
\meanr   & = \int \dd\theta\: q(\theta)\meanrw \quad\mwith\quad
\meanrw
= \int \dd\r\:\r p(\r|\theta)\;, \\
\sitotsq & = \int \dd\theta\:
q(\theta)\left[\mean{\r^{2}}_{\theta}-\meanr_{\theta}^{2}+(\meanrw-\meanr)^{2}
\right]\equiv\sigma_\text{syst}^2 + \sigma_\text{stat}^2\;.
\end{split}
\end{align}
Here $\meanr$ is the expectation value for the prediction \r considering the probability $p(\r|\theta)$ that the actual $\theta$ represents \r,
and $\sitotsq$ is the variance associated with this expectation.\footnote{$\meanr_{\theta}^{2}$ is defined in analogy to $\meanrw$.}

The total uncertainty $\sitotsq$ factorises into two terms.
The first term in \EqRef{eq:sigma-tot} vanishes in the limit of arbitrarily
well-known data, $p(\r|\theta) \to \delta(\rclusbnn(\theta) - \rclusem)$,
\begin{align}
\sigma_\text{syst}^2 = \int \dd\theta\: q(\theta)
\left[ \mean{\r^{2}}_{\theta}-\meanrw^{2} \right]  \;.
\label{eq:sistoch}
\end{align}
This uncertainty component approaches a plateau for large training data sets, and is therefore considered as a systematic uncertainty that cannot be reduced by increasing dataset sizes or the number of training attempts (epochs).
It is predicted for each individual \topo that is calibrated using \rclusbnn.

The second contribution is the $\theta$-sampled variance
\begin{align}
\sigma_\text{stat}^{2} = \int \dd\theta\: q(\theta)
\left[ \meanrw-\meanr \right]^{2} \;.
\label{eq:sipred}
\end{align}
It vanishes in the limit of perfect training, $q(\theta)\to\delta(\theta-\theta_{0})$.
This limit can be approached for large training datasets.
This uncertainty component is therefore referred to as statistical.

Contrary to the BNN, where the predictions are sampled in the variational weight space defined by $\theta$ to determine the uncertainties, the RE approach samples the predictions in function space, as briefly introduced in \AppRef{app:loss:re_loss}.
Also in this case, and for sufficient large training datasets like the one used for the \topo calibration, statistical and systematic uncertainty components can be distinguished and compared to the ones from the
BNN.

\subsection{Expectations for the pull distribution}
\label{app:perf-metrics:pull}

The pull distributions shown in \FigRef{fig:unc:pull:spectra} can be formally understood, and expectations for its width can be derived from the following considerations.

Generally, the pull of a learned distribution is defined as
\begin{align}
t(x) = \frac{f_\theta(x) - f(x)}{\sigma(x)}\;,
\end{align}
where $f_\theta(x)$ is the learned function with parameters $\theta$, $f(x)$ the true function encoded in the training data, and $\sigma(x)$ is the learned uncertainty.
For the calibration training discussed in this paper, $f_{\theta}(x) = \r_{\theta}(\fsetclus)$ for a learned set of network parameters $\theta$ and \topo feature sets \fsetclus, and
$f(x) = \rclusem(\fsetclus)$.

Ideally, $\sigma(x)$ captures the absolute value of the deviation of the learned function from the target exactly and for any $x$-value,
\begin{align}
f_\theta(x) = f(x) \pm \sigma(x) \quad\Rightarrow\quad t(x) = \pm 1\;.
\end{align}
Alternatively, $\sigma(x)$ can capture the maximum absolute value of the deviation, but the actual deviation of $f_\theta(x)$ from $f(x)$ might be smaller.
This actual deviation might also follow a flat distribution within the limits of $t(x) = \pm 1$,
\begin{align}
p(f_\theta(x)) &=
\begin{cases}
\,\text{const.} & f_{\theta}(x) \in    [f(x)-\sigma(x),f(x)+\sigma(x)] \\
\,0            & f_{\theta}(x) \notin [f(x)-\sigma(x),f(x)+\sigma(x)]
\end{cases} \notag \\
\qquad \Rightarrow \qquad
p(t(x)) &=
\begin{cases}
\,\text{const.} & t \in    [-1,1] \\
\,0            & t \notin [-1,1]
\end{cases} \;.
\end{align}
For a stochastic system like the one represented by the \topos, the learned values $f_\theta(x)$ are expected have a mean $\langle f_\theta(x) \rangle = f(x)$, but with a Gaussian smearing $\mathcal{N}(f(x),\sigma(x))$
with $x$-dependent width $\sigma(x)$ introduced by the (mostly \pu) noise,
\begin{align}
p(f_\theta(x)) &= \mathcal{N}(f(x),\sigma(x)) \propto \exp \left( - \frac{[f_\theta(x) - f(x)]^2}{2 \sigma(x)^2} \right) \notag \\
\qquad \Rightarrow \qquad
p(t(x)) &= \mathcal{N}(0,1)\;.
\end{align}
This shows that for a stochastic smearing the pull follows a standard Gaussian distribution.

\FloatBarrier

\pagebreak[4]

\setcounter{subsection}{0}
\renewcommand{\thesubsection}{\Alph{section}.\arabic{subsection}}
\section{Calibration performance}
\label{app:calibration}
\subsection{Performance in feature space}
\label{app:calibration:central}

The deviation from signal linerarity is evaluated before and after local calibrations conprising the hadronic calibration from LCW, the BNN-derived and the DNN-derived calibrations.
Before local calibration, the \topo signal \eclusem is provided on the basic EM scale.
As discussed in \SectRef{sec:perf-metrics:linear}, \eclusem is considered as uncalibrated and cannot be expected to have a direct proportionality with the deposited energy \eclusdep.
The \topo energy at the hadronic scale from LCW \ecluslcw is expected to represent a calibrated signal, thus it is expected to be calibrated.
Both of these scales do not use any of the features presented here as inputs to their respective signal reconstruction.
Both \topo energies calibrated using the BNN (\eclusbnn) or the DNN (\eclusdnn) predictions are expected to show a more linear response, as all features are used as inputs to their respective networks.

\begin{figure}[tt!]
\centering
\subfloat[]{\includegraphics[width=0.32\textwidth]{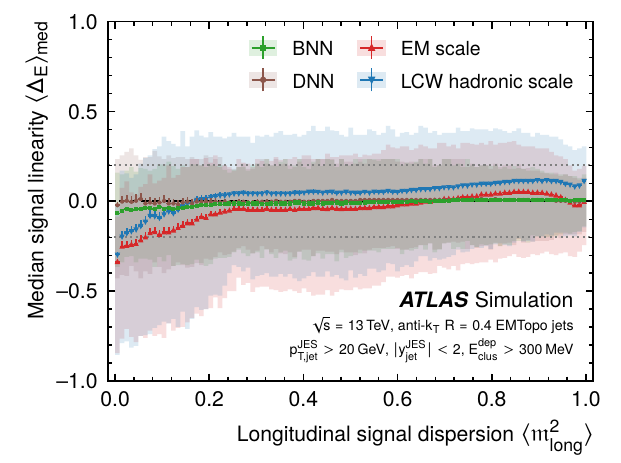}\label{fig:elinear:features:mlon}}
\subfloat[]{\includegraphics[width=0.32\textwidth]{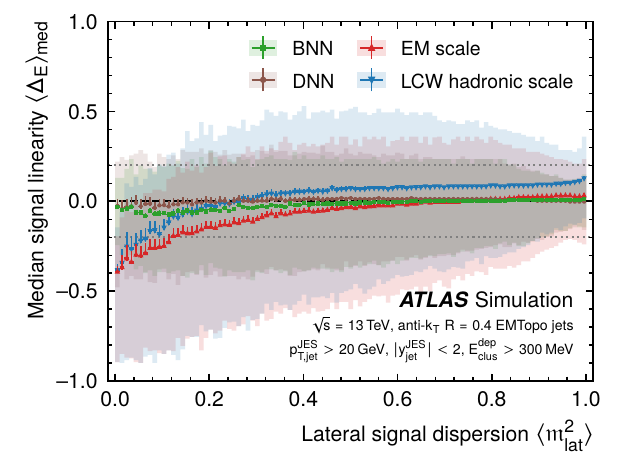}\label{fig:elinear:features:mlat}}
\subfloat[]{\includegraphics[width=0.32\textwidth]{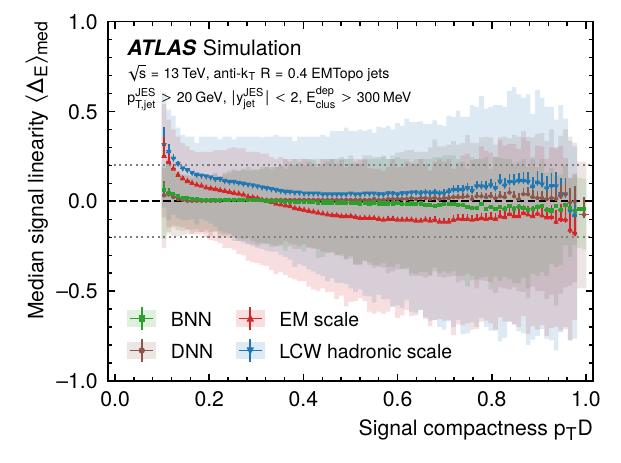}\label{fig:elinear:features:mptd}}\\
\subfloat[]{\includegraphics[width=0.32\textwidth]{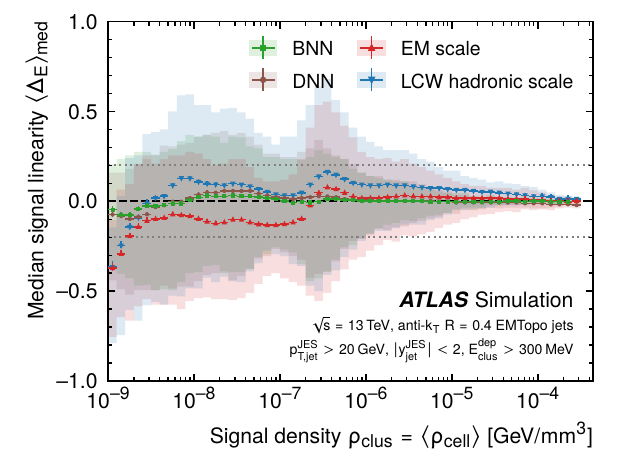}\label{fig:elinear:features:mrho}}
\subfloat[]{\includegraphics[width=0.32\textwidth]{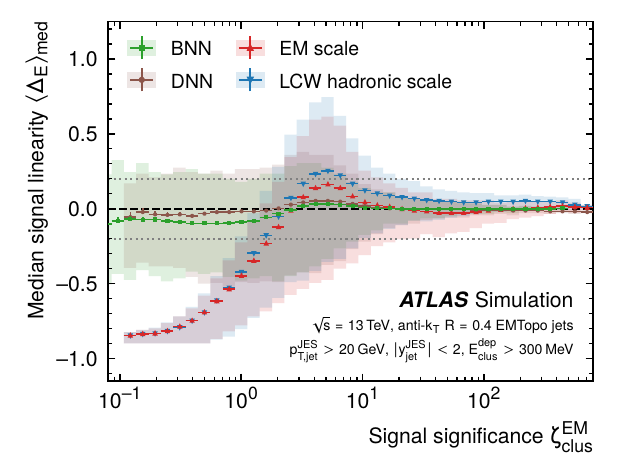}\label{fig:elinear:features:msig}}
\subfloat[]{\includegraphics[width=0.32\textwidth]{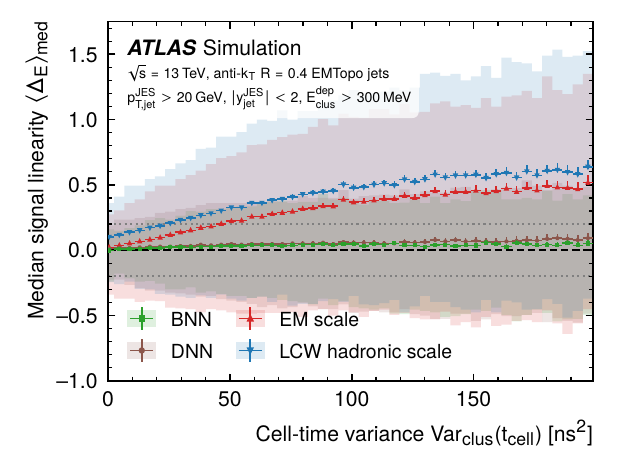}\label{fig:elinear:features:mtim}}\\
\subfloat[]{\includegraphics[width=0.32\textwidth]{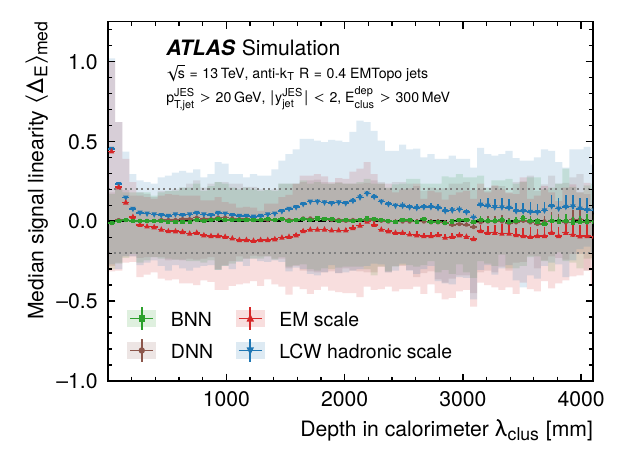}\label{fig:elinear:features:mlam}}
\subfloat[]{\includegraphics[width=0.32\textwidth]{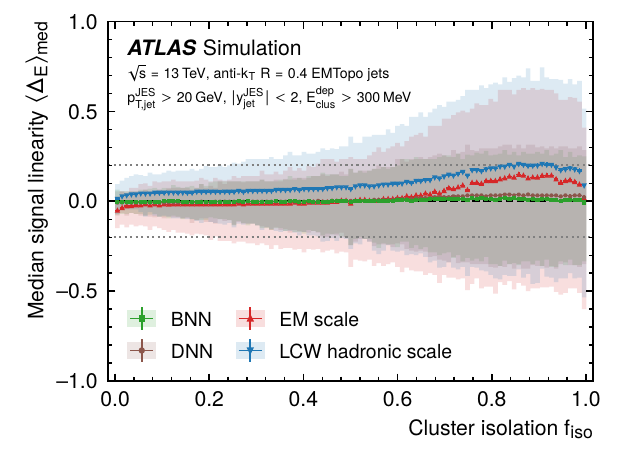}\label{fig:elinear:features:miso}}
\caption[]{The median deviation from signal linearity for the \topo at the basic EM scale (\median{\deleem}),  the hadronic calibration from LCW (\median{\delelcw}), and at the BNN-learned (\median{\delebnn}) and the DNN-learned (\median{\delednn}) calibrated energy scales, evaluated as a function of
\subref{fig:elinear:features:mlon} the longitudinal signal dispersion measure \cluslong,
\subref{fig:elinear:features:mlat} the lateral signal dispersion measure \cluslat,
\subref{fig:elinear:features:mptd} the signal compactness measure \clusptd (see \AppRef{app:features:construction}),
\subref{fig:elinear:features:mrho} the cluster signal density measure \clusrho,
\subref{fig:elinear:features:msig} the cluster signal significance \sclus,
\subref{fig:elinear:features:mtim} the second moment of the energy-squared-weighted cell time distribution \tclusvar (see \AppRef{app:features:construction}),
\subref{fig:elinear:features:mlam} the distance of the cluster \cog from the calorimeter front face \cluslam, and
\subref{fig:elinear:features:miso} the cluster isolation measure \clusfiso.
The \topos of the test sample used to produce these results are extracted as described in \SectRef{sec:dataset:object}.
}
\label{fig:elinear:features:app}
\end{figure}

\FigRef{fig:elinear:features:app} summarises the deviation from linearity for all four signal scales as a function of features only used for the BNN and DNN predictions.

\subsection{Local relative energy resolution }
\label{app:calibration:reso}

\begin{figure}[tt!]
\centering
\subfloat[]{\includegraphics[width=1.50\figwidth]{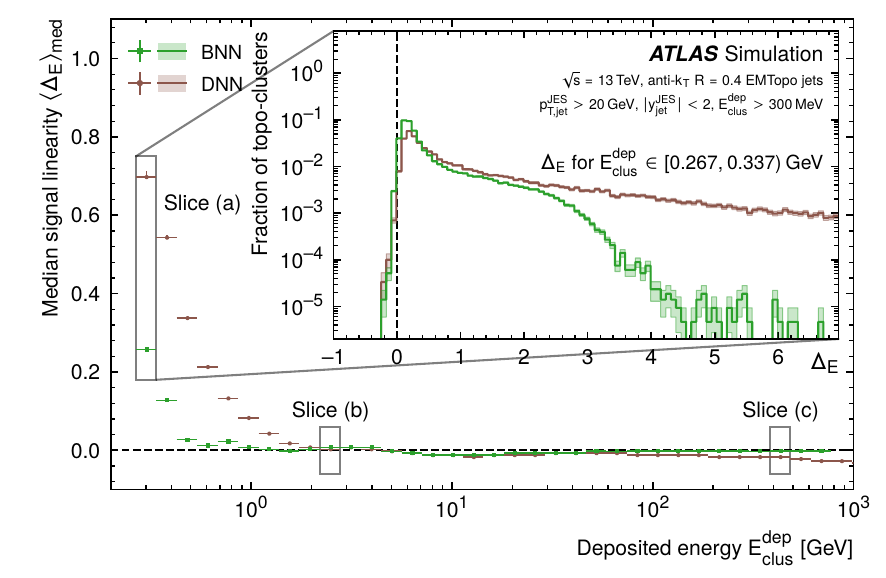}\label{fig:reso:spectra:inset}}\\
\subfloat[]{\includegraphics[width=1.00\figwidth]{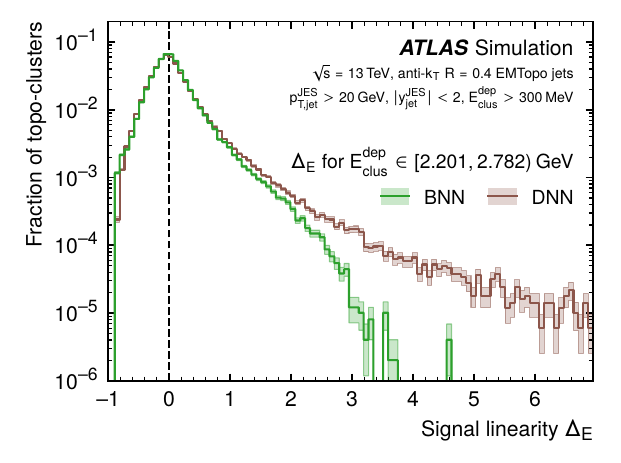}\label{fig:reso:spectra:bin23}}\quad
\subfloat[]{\includegraphics[width=1.00\figwidth]{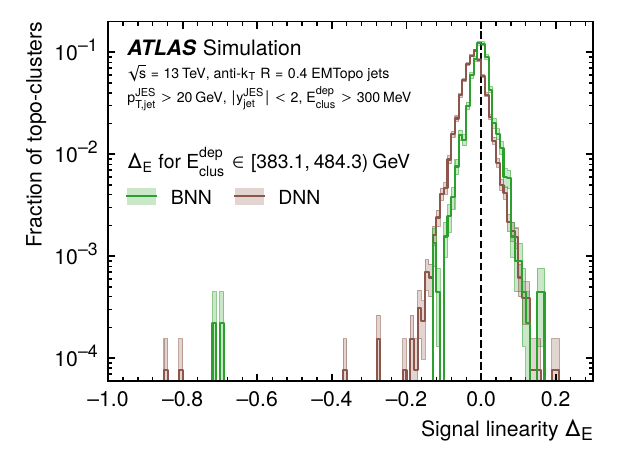}\label{fig:reso:spectra:bin45}}
\caption[]{In \subref{fig:reso:spectra:inset}, the median deviation from linearity \median{\dele} is shown as a function of the energy deposited in a \topo \eclusdep for both the BNN- and the DNN-derived calibrations.
The inset shows the shapes of the \delebnn and the \delednn distributions in Slice~(a), for \eclusdep around \SI{300}{\MeV}.
The corresponding distributions for Slice~(b) with \eclusdep around \SI{2.5}{\GeV} and Slice~(c) with \eclusdep around \SI{430}{\GeV} are shown in \subref{fig:reso:spectra:bin23} and \subref{fig:reso:spectra:bin45}, respectively.
The shaded areas indicate the statistical uncertainties.
The \topos from the test dataset used to produce these results are extracted as described in \SectRef{sec:dataset:object}.
}
\label{fig:reso:spectra}
\end{figure}

\FigRef{fig:reso:spectra} shows the median deviation from linearity \median{\dele} for the \topo energy scales after applying the calibrations learned by the BNN and the DNN, as a function of the energy
\eclusdep deposited in the cluster.
The slices (a), (b) and (c) show individual distributions of \dele in selected bins of \eclusdep, with the clear sign of significantly suppressed tails for the BNN-derived calibration at lower \eclusdep.
At higher $\eclusdep \gtrsim \SI{600}{\MeV}$ both predictions show increasingly similar distributions, with $\median{\delebnn} \approx 0$ and $\median{\delednn} < 0$ for $\eclusdep\gtrsim\SI{80}{\GeV}$ indicating a more accurate median calibration by the trained BNN.

\FloatBarrier


\FloatBarrier
\printbibliography
\clearpage
 
\begin{flushleft}
\hypersetup{urlcolor=black}
{\Large The ATLAS Collaboration}

\bigskip

\AtlasOrcid[0000-0002-6665-4934]{G.~Aad}$^\textrm{\scriptsize 104}$,
\AtlasOrcid[0000-0001-7616-1554]{E.~Aakvaag}$^\textrm{\scriptsize 17}$,
\AtlasOrcid[0000-0002-5888-2734]{B.~Abbott}$^\textrm{\scriptsize 123}$,
\AtlasOrcid[0000-0002-0287-5869]{S.~Abdelhameed}$^\textrm{\scriptsize 119a}$,
\AtlasOrcid[0000-0002-1002-1652]{K.~Abeling}$^\textrm{\scriptsize 56}$,
\AtlasOrcid[0000-0001-5763-2760]{N.J.~Abicht}$^\textrm{\scriptsize 50}$,
\AtlasOrcid[0000-0002-8496-9294]{S.H.~Abidi}$^\textrm{\scriptsize 30}$,
\AtlasOrcid[0009-0003-6578-220X]{M.~Aboelela}$^\textrm{\scriptsize 46}$,
\AtlasOrcid[0000-0002-9987-2292]{A.~Aboulhorma}$^\textrm{\scriptsize 36e}$,
\AtlasOrcid[0000-0001-5329-6640]{H.~Abramowicz}$^\textrm{\scriptsize 155}$
% ...
\bigskip
\\

% ...
$^{17}$Department for Physics and Technology, University of Bergen, Bergen; Norway.\\
% ...
$^{30}$Physics Department, Brookhaven National Laboratory, Upton NY; United States of America.\\
% ...
$^{36}$$^{(a)}$Facult\'e des Sciences Ain Chock, Universit\'e Hassan II de Casablanca;$^{(b)}$Facult\'{e} des Sciences, Universit\'{e} Ibn-Tofail, K\'{e}nitra;$^{(c)}$Facult\'e des Sciences Semlalia, Universit\'e Cadi Ayyad, LPHEA-Marrakech;$^{(d)}$LPMR, Facult\'e des Sciences, Universit\'e Mohamed Premier, Oujda;$^{(e)}$Facult\'e des sciences, Universit\'e Mohammed V, Rabat;$^{(f)}$Institute of Applied Physics, Mohammed VI Polytechnic University, Ben Guerir; Morocco.\\
% ...
$^{46}$Physics Department, Southern Methodist University, Dallas TX; United States of America.\\
% ...
$^{50}$Fakult\"{a}t Physik , Technische Universit{\"a}t Dortmund, Dortmund; Germany.\\
% ...
$^{56}$II. Physikalisches Institut, Georg-August-Universit\"{a}t G\"ottingen, G\"ottingen; Germany.\\
% ...
$^{80}$Universit\"{a}t Innsbruck, Department of Astro and Particle Physics, Innsbruck; Austria.\\
% ...
$^{104}$CPPM, Aix-Marseille Universit\'e, CNRS/IN2P3, Marseille; France.\\
% ...
$^{119}$$^{(a)}$New York University Abu Dhabi, Abu Dhabi;$^{(b)}$United Arab Emirates University, Al Ain; United Arab Emirates.\\
% ...
$^{123}$Homer L. Dodge Department of Physics and Astronomy, University of Oklahoma, Norman OK; United States of America.\\
% ...
$^{155}$Raymond and Beverly Sackler School of Physics and Astronomy, Tel Aviv University, Tel Aviv; Israel.\\
% ...
$^{a}$ Also Affiliated with an institute covered by a cooperation agreement with CERN.\\
$^{b}$ Also at An-Najah National University, Nablus; Palestine.\\
$^{c}$ Also at Borough of Manhattan Community College, City University of New York, New York NY; United States of America.\\
$^{d}$ Also at Center for High Energy Physics, Peking University; China.\\
$^{e}$ Also at Center for Interdisciplinary Research and Innovation (CIRI-AUTH), Thessaloniki; Greece.\\
$^{f}$ Also at CERN, Geneva; Switzerland.\\
% ...

\end{flushleft}


%

%
%

%

%
%
%
%

\end{document}